\documentclass[a4paper,11pt]{article}
\usepackage[english]{babel}
\usepackage[utf8]{inputenc}
\usepackage[T1]{fontenc}

\usepackage{graphicx}
\usepackage[margin=1in]{geometry}
\usepackage{amssymb,amsmath}
\usepackage{amsthm}
\usepackage{thmtools}
\usepackage{xcolor}
\definecolor{VUBlue}{rgb}{0, 0.466, 0.702}
\PassOptionsToPackage{hyphens}{url}
\usepackage[colorlinks=true, allcolors=VUBlue]{hyperref}
\usepackage{microtype}
\usepackage{comment}

\usepackage{newtxtext}
\usepackage{newtxmath}

\usepackage{enumitem}
\setlist[1]{labelindent=\parindent}
\setlist[enumerate]{label=(\arabic*)}
\setlist[itemize]{noitemsep}

\usepackage{orcidlink}

\usepackage[nottoc]{tocbibind}
\newif\ifhidetodos
\hidetodostrue

\newcommand{\vio}[1]{#1}

\newif\ifarxiv

\usepackage{mathtools}
\usepackage{bm}
\usepackage[noend,boxed]{algorithm2e}
\usepackage{xcolor}
\usepackage{placeins}
\usepackage{subcaption}
\renewcommand{\thesubfigure}{(\alph{subfigure})}
\usepackage{booktabs}
\usepackage{adjustbox}
\newcommand{\shrinktofit}[1]{\adjustbox{max width=#1}}
\usepackage{siunitx}
\usepackage{xspace}
\usepackage{tikz}
\usetikzlibrary{positioning,fit,backgrounds,matrix}
\usepackage{mdframed}
\usepackage{thmtools}
\usepackage{marginnote}
\usepackage[capitalize]{cleveref}
\def\droppanelparentaux#1(#2)\relax{(#2)}
\newcommand{\droppanelparent}[1]{\begingroup\edef\droppanelparenttmp{#1}\expandafter\droppanelparentaux\droppanelparenttmp\relax\endgroup}
\crefrangeformat{subfigure}{figures~#3#1#4\textendash#5\droppanelparent{#2}#6}
\Crefrangeformat{subfigure}{Figures~#3#1#4\textendash#5\droppanelparent{#2}#6}
\usepackage{omittedproofs}
\usepackage{amsmath}
\usepackage{graphicx}
\usepackage{float}
\usepackage[normalem]{ulem}
\usepackage{alphalph}

\definecolor{CWIBlue}   {rgb}{0.38, 0.478, 0.56}
\definecolor{CWIDGreen} {rgb}{0.372,0.494, 0.419}
\definecolor{CWILGreen} {rgb}{0.603,0.717, 0.486}
\definecolor{CWIRed}    {rgb}{0.686,0.188, 0.262}
\definecolor{VUBlue}    {rgb}{0,    0.466, 0.702}
\definecolor{VUBlue2}   {rgb}{0.364,0.678, 0.815}
\definecolor{VUGreen}   {rgb}{0,    0.502, 0.325}
\definecolor{VUGreen2}  {rgb}{0.309,0.686, 0.282}
\definecolor{VUOrange}  {rgb}{0.8,  0.254, 0}
\definecolor{VUOrange2} {rgb}{0.909,0.411, 0.176}
\definecolor{VUPurple}  {rgb}{0.231,0.129, 0.443}
\definecolor{VUPurple2} {rgb}{0.556,0.302, 0.643}
\definecolor{Cornflower}{rgb}{0.392,0.584, 0.929}

\newcommand{\smdot}[1]{\raisebox{0.26em}{\tikz[baseline=(d.center)]\node[circle,draw=#1,fill=#1!25,line width=.4pt,minimum size=1.3mm,inner sep=0pt](d){};}}

\newcommand{\algstep}[1]{%
  \vspace{2pt}%
  \textcolor{VUBlue}{\bfseries $\blacktriangleright$\ #1}\;
  \vspace{1pt}}
\DontPrintSemicolon%
\definecolor{AlgComment}{gray}{0.42}

\SetCommentSty{algcommentfont}

\newcommand{\cbox}[2][yellow]{%
  \fcolorbox{#1}{white}{\parbox{\dimexpr\linewidth-2\fboxsep}{\strut{}#2\strut{}}}%
}

\newcommand{\customtodo}[3]{\textcolor{#2}{\(\blacktriangledown\)}\marginnote{\raggedright{}\textcolor{#2}{\textbf{#1:} #3}}}
\newcommand{\customtododisplay}[3]{\noindent\cbox[#2]{\textcolor{#2}{\textbf{#1:} #3}}}
\newcommand{\customtodoinline}[3]{\textcolor{#2}{\textcolor{#2}{\textbf{#1:} #3}}}

\ifhidetodos%
  \renewcommand{\customtodo}[3]{}
  \renewcommand{\customtododisplay}[3]{}
  \renewcommand{\customtodoinline}[3]{}
\fi%

\NewDocumentCommand{\newtodoperson}{O{#2} m m}{%
  \expandafter\newcommand\csname #2\endcsname[1]{\customtodo{\MakeUppercase{#1}}{#3}{##1}}%
  \expandafter\newcommand\csname #2i\endcsname[1]{\customtodoinline{\MakeUppercase{#1}}{#3}{##1}}%
  \expandafter\newcommand\csname #2d\endcsname[1]{\customtododisplay{\MakeUppercase{#1}}{#3}{##1}}%
}

\newtodoperson[BB]{ben}{magenta}
\newtodoperson{ks}{Cornflower}
\newtodoperson[SC]{san}{red}
\newtodoperson{spp}{blue}      
\newtodoperson[GL]{greg}{VUGreen}  
\newtodoperson{jl}{VUPurple}   

\newcommand{\cO}{\mathcal{O}}
\newcommand{\Oish}{\mathcal{\widetilde{O}}}

\newcommand{\setsize}[1]{\left|#1\right|}

\newcommand{\ones}[1]{M_{#1}}
\newcommand{\zeros}[1]{\overline{\ones{#1}}}

\newcommand{\ALPAN}{$\alpha_{\mathrm{opt}}^{\mathrm{analytic}}\xspace$}
\newcommand{\ALPEX}{$\alpha_{\mathrm{opt}}^{\mathrm{exhaust}}\xspace$}
\newcommand{\ALPB}{$\alpha_{\mathrm{opt}}^{\mathrm{Brent}}\xspace$}
\newcommand{\ALPN}{$\alpha_{\mathrm{opt}}^{\mathrm{Newton}\xspace}$}

\newcommand{\RM}{\textsf{RM}\xspace}
\newcommand{\PM}{\textsf{PM}\xspace}
\newcommand{\DPM}{\textsf{DPM}\xspace}
\newcommand{\DRM}{\textsf{DRM}\xspace}

\newcommand{\SM}{\textsf{SM}\xspace}
\newcommand{\DSM}{\textsf{DSM}\xspace}

\newcommand{\FDSM}{\textsf{FDSM}\xspace}

\newcommand{\OMv}{\textsf{OMv}\xspace}
\newcommand{\BMM}{\textsf{BMM}\xspace}
\newcommand{\gOMv}{\textsf{$\gamma$-OMv}\xspace}
\newcommand{\smooth}{\alpha}
\newcommand{\smol}{\varepsilon}
\newcommand{\colors}{\Delta} 
\newcommand{\lcolors}{\colors_L}
\newcommand{\hcolors}{\colors_H}
\newcommand{\col}{c}
\newcommand{\coll}{d}
\newcommand{\freq}{f}
\newcommand{\lightBST}[1]{LT_{#1}}
\newcommand{\lightAgg}[1]{AG_{#1}}
\newcommand{\getMax}{\mathsf{get\text{-}max}}
\newcommand{\decreaseKey}{\mathsf{decrease\text{-}key}}
\newcommand{\increaseKey}{\mathsf{increase\text{-}key}}

\newcommand{\lightmode}{\mathrm{lightmode}}

\newcommand{\ie}[1]{(i.e.,~#1)}
\newcommand{\eg}[1]{(e.g.,~#1)}

\newcommand{\News}{\textsc{News}\xspace}
\newcommand{\WebKBsample}{\textsc{WebKB-sam}\xspace}
\newcommand{\Genessample}{\textsc{Genes-sam}\xspace}
\newcommand{\WebKBlarge}{\textsc{WebKB}\xspace}
\newcommand{\Geneslarge}{\textsc{Genes}\xspace}
\newcommand{\VIRlarge}{\textsc{Vir}\xspace}
\newcommand{\Files}{\textsc{Files}\xspace}
\newcommand{\Amazonlarge}{\textsc{Amazon}\xspace}
\newcommand{\sweepplot}[3]{%
\begin{subfigure}[t]{0.3286\columnwidth}
    \includegraphics[width=\linewidth]{figs/#1}
    \caption{{\footnotesize #2}}
    \label{#3}
\end{subfigure}%
}
\providecommand{\fdsmopplot}[3]{%
\begin{subfigure}[t]{0.155\textwidth}
    \centering
    \includegraphics[width=\linewidth]{figs/#1}
    \caption{\footnotesize #2}
    \label{#3}
\end{subfigure}%
}
\providecommand{\nsweepmergedplot}[5]{%
\begin{subfigure}[t]{0.155\textwidth}
    \centering
    \includegraphics[width=\linewidth]{figs/#1}
    \caption{#2}
    \label{#5}
\end{subfigure}%
}
\providecommand{\deltasweepmergedplot}[5]{%
\begin{subfigure}[t]{0.155\textwidth}
    \centering
    \includegraphics[width=\linewidth]{figs/#1}
    \caption{#2}
    \label{#5}
\end{subfigure}%
}

\declaretheoremstyle[
headfont=\bfseries\sffamily,
bodyfont=\normalfont,
headpunct={.},
postheadspace=0.5em,
mdframed={
  backgroundcolor=white,
  linecolor=gray!60,
  linewidth=0pt,
  leftline=true,
  topline=false,
  bottomline=false,
  rightline=false,
  leftmargin=0pt,
  innerleftmargin=10pt,
  innerrightmargin=10pt,
  innertopmargin=4pt,
  innerbottommargin=4pt,
  linewidth=3pt,
  linecolor=black!40,
  }
]{boxed}
\declaretheorem[numbered=no, style=boxed]{problem}
\newenvironment{ops}{\begin{list}{}{\setlength{\leftmargin}{1em}\setlength{\itemindent}{0pt}\setlength{\labelwidth}{0pt}\setlength{\labelsep}{0pt}\setlength{\topsep}{2pt}\setlength{\itemsep}{1pt}\setlength{\parsep}{0pt}\setlength{\partopsep}{0pt}\setlength{\listparindent}{0pt}}}{\end{list}}
\newcommand{\op}[1]{\item[\hspace{-1em}$\triangleright$\hspace{0.4em}\textsf{\bfseries #1}\hspace{0.5em}]}

\arxivtrue
\proofsinappendixfalse

\providecommand{\Description}[2][]{}

\makeatletter
\def\arxiv@adddotafter#1{#1\@addpunct{.}}
\renewcommand\subsubsection{\@startsection{subsubsection}{3}{\z@}%
  {3.25ex \@plus 1ex \@minus .2ex}%
  {-1em}%
  {\normalfont\normalsize\bfseries\arxiv@adddotafter}}
\makeatother

\newlength{\acmcolumnwidth}
\newlength{\acmappendixfigurewidth}
\newif\ifacmcolumnfloats
\acmcolumnfloatstrue
\makeatletter
\def\arxiv@floatbegin{%
  \centering
  \ifacmcolumnfloats
    \begin{minipage}{\acmcolumnwidth}%
    \setlength{\textwidth}{2.0996\acmcolumnwidth}
  \else
    \begin{minipage}{\acmappendixfigurewidth}%
    \setlength{\textwidth}{2.12\acmappendixfigurewidth}%
  \fi}
\def\arxiv@floatend{\end{minipage}\par}
\let\arxiv@figure\figure
\let\arxiv@endfigure\endfigure
\renewenvironment{figure}[1][tbp]%
  {\arxiv@figure[#1]\arxiv@floatbegin}{\arxiv@floatend\arxiv@endfigure}
\newif\ifarxiv@boxedfloat
\expandafter\let\expandafter\arxiv@figurestar\csname figure*\endcsname
\expandafter\let\expandafter\arxiv@endfigurestar\csname endfigure*\endcsname
\renewenvironment{figure*}[1][tbp]%
  {\arxiv@figurestar[#1]%
   \ifacmcolumnfloats\arxiv@boxedfloatfalse\else\arxiv@boxedfloattrue\arxiv@floatbegin\fi}%
  {\ifarxiv@boxedfloat\arxiv@floatend\fi\arxiv@endfigurestar}
\g@addto@macro\appendix{\acmcolumnfloatsfalse}
\makeatother


\makeatletter
\g@addto@macro\appendix{\crefalias{section}{appendix}}
\makeatother

\declaretheorem[numberwithin=section]{theorem}

\declaretheorem[sibling=theorem]{corollary}
\declaretheorem[sibling=theorem]{lemma}
\declaretheorem[sibling=theorem]{definition}
\declaretheorem[sibling=theorem]{conjecture}

\declaretheorem[sibling=theorem]{remark}
\declaretheorem[sibling=theorem]{example}
\declaretheorem[sibling=theorem]{observation}

\title{Nearly-Optimal Dynamic Subtree Mode}

\newcommand{\authorbox}[2]{%
  \parbox[t]{0.33\linewidth}{\centering #1\\[1pt]{\normalsize #2}}}
\newcommand{\authorrowbreak}{\par\vspace{2em}}
\newcommand{\authormarkspace}{\hphantom{\textsuperscript{\normalfont *}}}
\author{\parbox{0.96\textwidth}{\centering
  \authorbox{Jialong Zhou\thanks{The first two authors contributed equally to this work. All other authors are sorted alphabetically.}\authormarkspace~\orcidlink{0000-0001-7972-6418}}
            {King's College London\\London, United Kingdom}\hfill
  \authorbox{Ben Bals\footnotemark[1]\authormarkspace~\orcidlink{0009-0009-1054-8444}}
            {CWI\\Amsterdam, The Netherlands\\Vrije Universiteit\\Amsterdam, The Netherlands}\hfill
  \authorbox{Sankardeep Chakraborty~\orcidlink{0000-0002-2395-4160}}
            {University of Tokyo\\Tokyo, Japan}\authorrowbreak
  \authorbox{Grigorios Loukides~\orcidlink{0000-0003-0888-5061}}
            {King's College London\\London, United Kingdom}\hfill
  \authorbox{Solon P.\ Pissis~\orcidlink{0000-0002-1445-1932}}
            {The Cyprus Institute\\Nicosia, Cyprus}\hfill
  \authorbox{Kunihiko Sadakane~\orcidlink{0000-0002-8212-3682}}
            {University of Tokyo\\Tokyo, Japan}%
}}
\hypersetup{
  pdftitle={Nearly-Optimal Dynamic Subtree Mode},
  pdfauthor={Jialong Zhou, Ben Bals, Sankardeep Chakraborty, Grigorios Loukides, Solon P. Pissis, Kunihiko Sadakane},
}
\date{\vspace{1em}\today}

\begin{document}

\RestyleAlgo{ruled}

\providecommand{\vio}[1]{\textcolor{violet}{#1}}
\providecommand{\flag}[1]{\textcolor{red}{\textbf{[#1]}}}

\ifarxiv\pagenumbering{roman}\maketitle\fi

\begin{abstract}
Computing the \emph{mode} (i.e., the most frequent element) in a range of an array, or in a path of a colored tree, has been studied extensively in both the static and dynamic setting. 
However, the problem of computing the mode in a \emph{subtree} of a leaf-colored tree, referred to as the Subtree Mode (SM), has only been studied in the static setting, despite its applications in domains ranging from file systems, to e-commerce, and bioinformatics. Here, we study the SM problem as the tree undergoes changes. We define Dynamic Subtree Mode (DSM), which asks for a data structure supporting leaf recolorings in a fixed tree, and Fully Dynamic Subtree Mode (FDSM), which additionally supports leaf insertions and deletions and subtree moves.

For DSM on a tree of $n$ nodes, we propose a linear-space data structure with $\cO(n)$ preprocessing and, for any $\alpha \in [0,1]$, query and update times $\Oish(n^\alpha)$ and $\Oish(n^{1-\alpha})$, respectively. This trade-off is polynomially optimal under the Online Matrix-Vector Multiplication conjecture. For trees of height $h$, query and update times improve to $\Oish(1)$ and $\Oish(h)$, respectively. For FDSM, a logarithmic-overhead reduction from Dynamic Range Mode gives a conditional lower bound. For a color set $\colors$, we obtain a linear-space data structure supporting queries and subtree moves in $\Oish(|\colors|)$ time and recolorings and leaf insertions and deletions in $\Oish(1)$ time. 
Thus, while \DSM exploits the tree structure to achieve
query--update trade-offs that are conditionally
impossible for Dynamic Range Mode, \FDSM
inherits its conditional hardness but admits efficient
solutions when $|\colors|$ is small. Experiments on datasets from different domains, as well as case studies using workloads from e-commerce and file systems, show speedups of orders of magnitude over natural baselines. 
\end{abstract}

\ifarxiv
\thispagestyle{empty}
\clearpage
\tableofcontents
\thispagestyle{empty}
\clearpage
\pagenumbering{arabic}
\setcounter{page}{1}
\else
\maketitle
\fi

\section{Introduction}\label{sec:intro}

A key summary statistic of a collection of values is its  \emph{mode} (i.e., its most frequent value)~\cite{DBLP:books/mk/HanKP2011}. In most applications, however, users require finding the mode of specific parts of the data. For example, a market analyst queries e-commerce transaction logs to identify the most frequently purchased item over a specific promotional window~\cite{sea23}, a network engineer queries a routing path to detect the prevailing traffic type~\cite{DBLP:conf/esa/Gao022}, and a biologist queries a phylogenetic tree to determine the dominant mutation within a specific evolutionary clade~\cite{static-paper}. Due to the large number of queries and volume of data in such applications, constructing  a data structure that supports fast querying of arbitrary parts of data is crucial. 

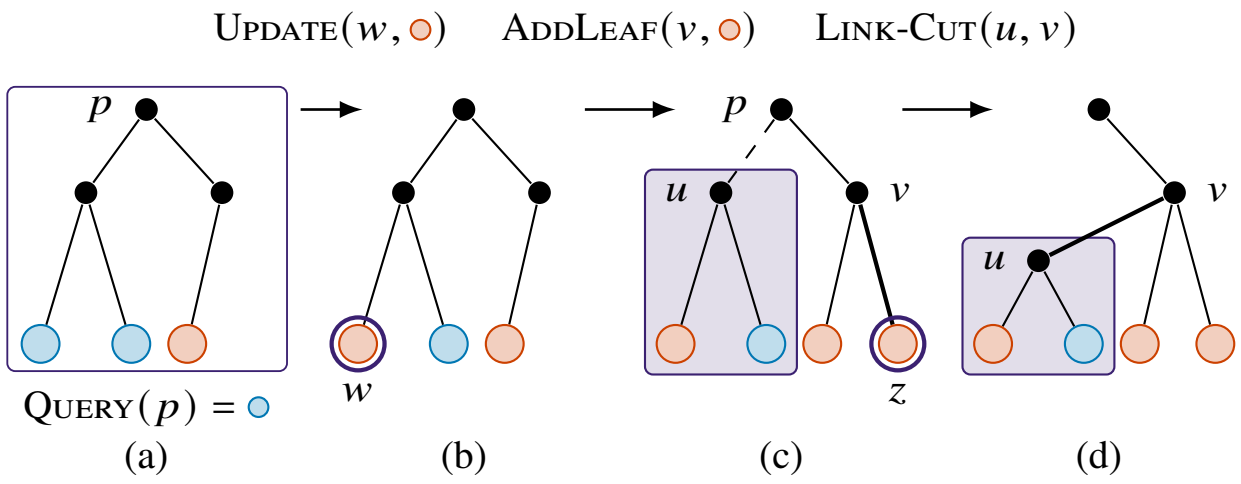
\begin{figure}[t]
  \centering
  \captionsetup{skip=3pt}
\captionsetup[subfigure]{skip=0pt}

  \begingroup
  \phantomsubcaption\label{fig:subtree-mode-operations-query}%
  \phantomsubcaption\label{fig:subtree-mode-operations-update}%
  \phantomsubcaption\label{fig:subtree-mode-operations-addleaf}%
  \phantomsubcaption\label{fig:subtree-mode-operations-linkcut}%
  \endgroup
  \shrinktofit{\columnwidth}{%
\begingroup
\begin{tikzpicture}[
  x=1cm,y=1cm,font=\fontsize{7.5}{9}\selectfont,
  line width=.4pt,
  sm inner/.style={circle,fill=black,minimum size=1.5mm,inner sep=0pt},
  sm leaf/.style={circle,draw=black,minimum size=2.5mm,inner sep=0pt},
  sm blue/.style={sm leaf,draw=VUBlue,fill=VUBlue!25},
  sm orange/.style={sm leaf,draw=VUOrange,fill=VUOrange!25},
  sm changed/.style={circle,draw=VUPurple,thick,minimum size=3.5mm,inner sep=0pt},
  sm region/.style={draw=VUPurple,fill=VUPurple!15,rounded corners=1.5pt,inner sep=2pt},
  sm query/.style={draw=VUPurple,line width=.35pt,rounded corners=1.5pt},
  sm label/.style={inner sep=1pt},
  sm call/.style={align=center,inner sep=0pt},
  sm arrow/.style={->,>=latex,line width=.5pt}
]
\path[use as bounding box] (0,-3.12) rectangle (8.3,.12);

\node[sm call] at (.95,-2.62)
  {$\textsc{Query}(p)=\smdot{VUBlue}$};
\node[sm call,anchor=south] at (2.175,-.25)
  {$\textsc{Update}(w,\smdot{VUOrange})$};
\node[sm call,anchor=south] at (4.15,-.25)
  {$\textsc{AddLeaf}(v,\smdot{VUOrange})$};
\node[sm call,anchor=south] at (6.25,-.25)
  {$\textsc{Link\mbox{-}Cut}(u,v)$};

\begin{scope}
  \node[sm inner] (a-p) at (.95,-.65) {};
  \node[sm inner] (a-u) at (.55,-1.20) {};
  \node[sm inner] (a-v) at (1.45,-1.20) {};
  \node[sm blue] (a-w) at (.25,-2.20) {};
  \node[sm blue] (a-b) at (.85,-2.20) {};
  \node[sm orange] (a-o) at (1.22,-2.20) {};
  \begin{scope}[on background layer]
    \draw[sm query] (.03,-2.38) rectangle (1.88,-.50);
    \draw (a-p)--(a-u) (a-p)--(a-v);
    \draw (a-u)--(a-w) (a-u)--(a-b) (a-v)--(a-o);
  \end{scope}
  \node[sm label,anchor=east] at (.78,-.65) {$p$};
\end{scope}

\begin{scope}[xshift=2.1cm]
  \node[sm inner] (b-p) at (.95,-.65) {};
  \node[sm inner] (b-u) at (.55,-1.20) {};
  \node[sm inner] (b-v) at (1.45,-1.20) {};
  \node[sm orange] (b-w) at (.25,-2.20) {};
  \node[sm blue] (b-b) at (.85,-2.20) {};
  \node[sm orange] (b-o) at (1.22,-2.20) {};
  \node[sm changed] at (b-w) {};
  \begin{scope}[on background layer]
    \draw (b-p)--(b-u) (b-p)--(b-v);
    \draw (b-u)--(b-w) (b-u)--(b-b) (b-v)--(b-o);
  \end{scope}
  \node[sm label,anchor=north] at (.25,-2.42) {$w$};
\end{scope}

\begin{scope}[xshift=4.2cm]
  \node[sm inner] (c-p) at (.95,-.65) {};
  \node[sm inner] (c-u) at (.55,-1.20) {};
  \node[sm inner] (c-v) at (1.45,-1.20) {};
  \node[sm orange] (c-w) at (.25,-2.20) {};
  \node[sm blue] (c-b) at (.85,-2.20) {};
  \node[sm orange] (c-o) at (1.22,-2.20) {};
  \node[sm orange] (c-z) at (1.72,-2.20) {};
  \node[sm changed] at (c-z) {};
  \begin{scope}[on background layer]
    \node[sm region,fit=(c-u)(c-w)(c-b)] {};
    \draw[dashed] (c-p)--(c-u);
    \draw (c-p)--(c-v) (c-u)--(c-w) (c-u)--(c-b) (c-v)--(c-o);
    \draw[line width=.8pt] (c-v)--(c-z);
  \end{scope}
  \node[sm label,anchor=east] at (.78,-.65) {$p$};
  \node[sm label,anchor=east] at (.38,-1.20) {$u$};
  \node[sm label,anchor=west] at (1.62,-1.20) {$v$};
  \node[sm label,anchor=north] at (1.72,-2.42) {$z$};
\end{scope}

\begin{scope}[xshift=6.3cm]
  \node[sm inner] (d-p) at (.95,-.65) {};
  \node[sm inner] (d-v) at (1.45,-1.20) {};
  \node[sm inner] (d-u) at (.55,-1.65) {};
  \node[sm orange] (d-w) at (.25,-2.20) {};
  \node[sm blue] (d-b) at (.85,-2.20) {};
  \node[sm orange] (d-o) at (1.22,-2.20) {};
  \node[sm orange] (d-z) at (1.72,-2.20) {};
  \begin{scope}[on background layer]
    \node[sm region,fit=(d-u)(d-w)(d-b)] {};
    \draw (d-p)--(d-v) (d-v)--(d-o) (d-v)--(d-z);
    \draw (d-u)--(d-w) (d-u)--(d-b);
    \draw[line width=.8pt] (d-v)--(d-u);
  \end{scope}
  \node[sm label,anchor=east] at (.38,-1.65) {$u$};
  \node[sm label,anchor=west] at (1.62,-1.20) {$v$};
\end{scope}

\draw[sm arrow] (1.97,-.65)--(2.38,-.65);
\draw[sm arrow] (3.85,-.65)--(4.45,-.65);
\draw[sm arrow] (5.95,-.65)--(6.55,-.65);

\foreach \x/\l in {.95/a,3.05/b,5.15/c,7.25/d}
  \node[sm label,anchor=base] at (\x,-3.02) {(\l)};
\end{tikzpicture}
\endgroup
  }  
  \caption{A tree under successive operations.}
  \label{fig:subtree-mode-operations}
\end{figure}

This leads to three fundamental problems, depending on how the data are modeled. When the data are modeled as an array, querying for the mode of a subarray (e.g., a contiguous window in sequential data, or a range of attribute values in a database) leads to the well-known \emph{Range Mode} (\RM) problem~\cite{DBLP:journals/njc/KrizancMS05,DBLP:journals/corr/abs-1101-4068,DBLP:conf/icalp/GreveJLT10,DBLP:conf/soda/WilliamsX20}.  When the data are modeled as a tree, querying for the mode of the labels along a tree path leads to the \emph{Path Mode} (\PM) problem~\cite{DBLP:conf/esa/Gao022}, 
and querying for the mode of the labels of the leaves in the subtree of a node leads to the \emph{Subtree Mode} (\SM) problem~\cite{static-paper}.
For example, consider the tree in \Cref{fig:subtree-mode-operations-query}. The answer to a subtree mode query for node $p$ is $\smdot{VUBlue}$, as this is the most frequent color among the leaves in its subtree.

There are efficient data structures for all these three problems in the \emph{static} setting (see \Cref{sec:related}). 
Yet, in most applications, the data are updated; purchased items are added into e-commerce transaction logs, network routing  topologies adapt dynamically to link failures, and genome databases undergo new annotations. In this setting, it is  impractical to build the (static) data structures for \RM, \PM, or \SM from scratch after each update. Also, these data structures cannot be readily adapted to deal with updates. Thus, designing  \emph{dynamic} data structures that support both efficient queries and updates is necessary. 
The \emph{Dynamic Range Mode} (\DRM) problem, where an update modifies, inserts, or deletes an entry of an array of $n$ entries, has received significant attention~\cite{DBLP:journals/corr/abs-1101-4068,DBLP:conf/esa/El-Zein0MS18,DBLP:conf/icalp/SandlundX20,DBLP:conf/icalp/GuPVX21,DBLP:conf/stoc/JinX22}. The state-of-the-art combinatorial data structure~\cite{DBLP:conf/esa/El-Zein0MS18} supports queries and updates in $\cO(n^{2/3})$ time using linear space. 
Using fast matrix multiplication, the query and update time can be improved to $\cO(n^{0.6524})$ using $\cO(n^{1.3262})$ space~\cite{DBLP:conf/icalp/GuPVX21}.  
For the \emph{Dynamic Path Mode} (\DPM) problem, where an update changes the label of a node of a tree with $n$ nodes, the state-of-the-art is a dynamic data structure achieving $\Oish(n^{0.6581})$ query and update time via fast matrix multiplication~\cite{DBLP:conf/esa/Gao022}. 

As dynamic hierarchical data are ubiquitous and \SM is important in many applications~\cite{static-paper}, we introduce and study two dynamic versions of \SM: the \emph{Dynamic Subtree Mode} (\DSM) and the \emph{Fully Dynamic Subtree Mode} (\FDSM) problem.  We consider both problems for a leaf-colored tree, but the case where internal nodes are also colored can be reduced to the leaf-colored case easily; see~\cite{static-paper}.

\subsection{The Dynamic Subtree Mode}

\begin{problem}[Dynamic Subtree Mode (\DSM)]
  Preprocess a rooted tree $\mathcal{T}$ on $n$ nodes whose leaves are colored from a fixed finite universe $\colors=\{0,\ldots,|\colors|-1\}$, to support:
  \begin{ops}
    \op{Query}Given a node $v$, return the most frequent color among the leaves in the subtree of $v$.
    \op{Update}Given a leaf $v$ and $c\in\colors$, recolor $v$ with $c$.
  \end{ops}
\end{problem}

\vspace{+1mm}
\DSM models the common setting in which the tree is fixed but the attributes associated with the leaves (i.e., the colors) evolve. 
For example, in \Cref{fig:subtree-mode-operations-update} an update changes the color of the leftmost leaf from $\smdot{VUBlue}$ to $\smdot{VUOrange}$.

\vspace{+1mm}
\noindent{\bf Motivation.}~From a theoretical standpoint, \DSM can be motivated as follows. \DSM can be easily reduced to \DRM and by using the state-of-the-art data structure for the latter problem, we obtain a baseline with $\cO(n^{2/3})$ query and update time~\cite{DBLP:conf/esa/El-Zein0MS18}. In addition, there is strong  theoretical evidence that these bounds cannot be improved for general \DRM by a combinatorial data structure~\cite{DBLP:conf/stoc/JinX22}.
This gives rise to two fundamental questions: 

\begin{itemize}
    \item[\textbf{(Q1)}] \textit{Can we exploit the tree structure, in which there are $n$ instead of $n^2$ possible queries (ranges), to solve \DSM faster than solving it via \DRM?} 
\item[\textbf{(Q2)}] \textit{Can we solve \DSM even faster in practical parameter regimes?} 
\end{itemize}

From a practical standpoint, \DSM is motivated by several applications in domains such as computational biology and e-commerce. 

{\it Annotated phylogenetic trees.}~
In biology, a phylogenetic tree models evolutionary relationships among organisms: leaves represent sampled organisms, while internal nodes represent their common ancestors~\cite{phylobook}. Leaves can be annotated with categorical metadata such as SNP states, geographic origin, or antibiotic-resistance status~\cite{10193,plospathogens, Arora2017OriginSyphilis}. A subtree-mode query at an internal node returns the most prevalent annotation within the corresponding clade, supporting studies of pathogen spread and resistance evolution, as discussed in~\cite{static-paper}. As new laboratory results or revised classifications become available, \DSM updates the corresponding leaf colors without changing the phylogenetic tree. 

{\it E-Commerce product taxonomies.}~Retailers organize products in a taxonomy tree (e.g., a root-to-leaf path can be \emph{Product $\to$ Electronics $\to$ Audio $\to$ Headphones $\to$ Apple Airpods 3})~\cite{DBLP:books/mk/HanKP2011}. 
Internal nodes represent product categories, and leaves represent products, each having an attribute, such as rating tier, shipping speed (e.g., \emph{Same-Day} vs. \emph{Standard}), supplier, or price tier. By finding the subtree mode of a node $v$ corresponding to a category, a retailer can identify the most frequent attribute value of the  products in this category. In e-commerce systems, the taxonomy tree is typically curated by experts and is stable, so as to maintain consistent website navigation and SEO indexing, but the product attribute values change frequently~\cite{LaudonTraver2023,MorvilleRosenfeld2006}. Thus, {\DSM} allows a retailer to update the attribute values and query for the subtree mode, as the attribute values are updated.

\begin{example}
A retailer wishes to identify the most frequent
product rating category within each category of
an e-commerce taxonomy and maintain this information
as product ratings evolve.
To this end, they construct our \DSM data structure
on a tree $\mathcal{T}$ derived from the Amazon Reviews
dataset~\cite{mcauley2015image}, containing
$n=\num{863317}$ nodes and \num{859478} 
leaves (products), with $|\colors|=5$ colors (rating categories). 
As new reviews arrive, the retailer uses the data
structure to maintain the subtree modes under
\num{1000} rating-category updates, while answering
\num{5845} subtree mode queries.
Processing this workload with \DSM is about $71{,}000$ times faster than with the \DRM-based baseline.\end{example}

\subsection{The Fully Dynamic Subtree Mode}

\begin{problem}[Fully Dynamic Subtree Mode (\FDSM)]
Preprocess a rooted tree $\mathcal{T}$ on $n$ nodes whose leaves are colored from a fixed finite universe $\colors=\{0,\ldots,|\colors|-1\}$, to support:
\begin{ops}
\op{Query}Given a node $v$, return the most frequent color among the leaves in the subtree of $v$.
\op{Update}Given a leaf $v$ and $c\in\colors$, recolor $v$ with $c$.
\op{Add Leaf}Given a node $p$ and $c\in\colors$, add a new child $v$ of $p$ with color $c$.
\op{Delete Leaf}Given a non-root leaf $v$, delete $v$.
\op{Link-Cut Update}Given a non-root node $u$ with parent $p$ and a node $v$ outside the subtree of $u$, replace edge $(p,u)$ by $(v,u)$.
\end{ops}
If a deletion or link-cut turns an internal node into a leaf, the operation also specifies a color $c\in\colors$ for it.
\end{problem}

\vspace{+1mm}
\FDSM models the setting where, in addition to queries  and updates as in \DSM, the tree structure is updated. Specifically, a node can be added into $\mathcal{T}$ as a leaf, a leaf can be deleted from $\mathcal{T}$, or the subtree rooted at a node $u$ of $\mathcal{T}$ can be detached and appended under $v$. The latter is performed with Link-Cut Update; we do not separate the deletion of the edge $(p,u)$ and the addition of $(v,u)$ to keep the tree connected.
In our example \Cref{fig:subtree-mode-operations-addleaf}, an update adds a leaf colored $\smdot{VUOrange}$ below $v$. Finally, in \Cref{fig:subtree-mode-operations-linkcut}, a whole subtree is moved by first cutting $u$ from its parent and then making $v$ its new parent.
Also, there is \emph{no} ``Delete Internal Node'' operation in \FDSM, as it can be performed by the allowed operations of \FDSM.\footnote{To delete an internal node $u$ without its descendants, we can use Link-Cut Update to move the subtree of each child to $p$ and then delete $u$ which has become a leaf. To delete both $u$ and its  descendants, we can apply Delete Leaf to the subtree of $u$ bottom-up.}

\vspace{+1mm}
\noindent{\bf Motivation.}~From a theoretical standpoint, \FDSM can be motivated as follows: Unlike \DSM, where a fixed tree topology allows leaf subtrees to map directly to static array intervals yielding a baseline via \DRM, the tree structure updates in \FDSM may reorder and split subtrees, making reductions to array range queries inapplicable. At the same time, rebuilding the \SM data structure~\cite{static-paper} based on the updated tree after each update incurs a prohibitive $\mathcal{O}(n)$ update time. This gives rise to two fundamental questions:
\begin{itemize}
    \item[\textbf{(Q3)}] \textit{Can we maintain efficient query and update times when the tree structure is dynamic?}
    \item[\textbf{(Q4)}] \textit{Can we solve \FDSM efficiently in practical parameter regimes?} 
\end{itemize}

From a practical standpoint, \FDSM is motivated by applications in file-system management and network monitoring. 

{\it File-system hierarchies.}~ 
Consider a file system in which folders need to be tagged with the most frequent file type (e.g., image, document, source code, etc.) contained in them~\cite{static-paper}. The file system can be modeled as a tree $\mathcal{T}$, in which each non-empty directory is an internal node, its subdirectories are its children, and each file is a leaf that is attached to the node of the containing directory. Furthermore, each leaf is colored according to the type of the file it represents. As discussed in~\cite{static-paper}, by finding the subtree mode for a node $v$ of $\mathcal{T}$, an administrator can find the most frequent file type in the directory of $v$, obtaining useful visual clues about the prevalent contents in the directory. In practice, files are often created or deleted, and files or entire subdirectories may be moved between directories. These changes correspond to leaf insertions, leaf deletions, and subtree moves (link-cut updates), respectively. Thus, \FDSM can build a data structure for answering subtree mode queries as the file system evolves.

\begin{example}
A file-system administrator wishes to identify the most
frequent file type in each directory of a software repository
and maintain this information as the repository evolves.
To this end, they construct our \FDSM data structure
on a snapshot of the LLVM Project repository~\cite{llvm-project},
containing $n=\num{182079}$ nodes, with files assigned
to $|\colors|=14$ types. 
As the next \num{1000} commits arrive, the administrator
uses our data structure to maintain the subtree modes
under $695$ file and directory additions, deletions,
and moves, while answering \num{12806} queries on
affected directories and their ancestors.
Processing this workload with \FDSM is about \emph{$80$ times faster} compared to when rebuilding the static
\SM{} data structure of~\cite{static-paper} after updates.
\end{example}

{\it Prefix aggregation trees.}~Prefix aggregation trees~\cite{rfc4632,ruiz2001survey} organize IP addresses; internal nodes represent routing prefixes (subnetworks), while leaves represent IP addresses (network endpoints), each colored by a traffic attribute value 
 (e.g., healthy, degraded, or offline, for the service status attribute). Finding the subtree mode for a node $v$ allows engineers to identify the predominant traffic attribute value (e.g., service status) in the subnetwork corresponding to the subtree of $v$, enabling traffic engineering and automated failover detection. In practice, endpoint subnets join and leave frequently, and subnets are dynamically reassigned to different gateways~\cite{DBLP:conf/imc/KrencF16, DBLP:conf/sigcomm/KochYABK23}. These operations correspond to leaf additions, deletions, and link-cut updates, respectively. Thus,    
\FDSM can build a data structure enabling the tracking of  predominant attribute values as the network topology evolves.

\subsection{Contributions}

In addition to introducing the \DSM and \FDSM problems, our work makes the following contributions.

\vspace{+1mm}
\noindent{\bf 1.}~\emph{Optimal and Smooth Trade-off for \DSM.}~We develop an $\cO(n)$-space data structure for \DSM offering a \emph{smooth} query/update trade-off: for any trade-off parameter $\alpha \in [0, 1]$, our data structure supports subtree mode queries in $\Oish(n^{\alpha})$ time and leaf color updates in $\Oish(n^{1-\alpha})$ time, after $\mathcal{O}(n)$-time preprocessing. 
We complement this upper bound with a matching conditional lower bound, proving that our trade-off is (conditionally) optimal. This answers  question Q1 above in the affirmative and provides strong evidence that the trade-off cannot be improved. 
Note that setting $\alpha = 1/2$ yields $\Oish(\sqrt{n})$ query and update time. This outperforms the aforementioned \DRM-based baseline, which cannot be improved for general \DRM by a combinatorial data structure~\cite{DBLP:conf/stoc/JinX22}. See \Cref{sec:dsm_ds}.

\vspace{1mm} \noindent{\bf 2.}~\emph{Height-Parameterized Analysis of the \DSM Data Structure.}~We prove that our \DSM data structure admits an analysis parameterized by the height $h$ of the tree $\mathcal{T}$. In this setting, it achieves $\Oish(1)$ query time, $\Oish(h)$ update time, and $\cO(n)$ construction time and space. This is important because in several \DSM applications $h$ is much smaller than $n$. 
For example, 
large-scale commercial e-commerce product taxonomies have $h$ up to $8$~\cite{DBLP:conf/sigir/LinDD18,DBLP:conf/bigdataconf/LinDF18}.  
See \Cref{sec:parameterized_dsm}. 

\vspace{+1mm}
\noindent{\bf 3.}~\emph{Hardness of \FDSM.}~We show that dynamic updates to the tree structure fundamentally limit the achievable efficiency. Specifically, we reduce \DRM to \FDSM with $\cO(\log n)$ overhead per operation, so any (combinatorial) \FDSM data structure with $\mathcal{O}(n^{2/3-\varepsilon})$ query and update time contradicts the $n^{2/3-o(1)}$ conditional lower bound of~\cite{DBLP:conf/stoc/JinX22}.  
Thus, \FDSM is inherently as hard as \DRM, and question Q3 is answered in the negative for general $|\colors|$; Contribution~4 shows what is possible when $|\colors|$ is small. See \Cref{sec:hardness_fdsm}. 

\vspace{+1mm} 
\noindent{\bf 4.}~\emph{Color-Parameterized Data Structure  for \FDSM.}~We design an \FDSM data structure that is 
parameterized by the number $|\colors|$ of colors. It supports subtree mode queries and link-cut updates in $\Oish(|\colors|)$ time, while color updates and leaf additions/deletions take $\Oish(1)$ time. The data structure can be constructed in $\Oish(n)$ time using $\cO(n)$ space. This parameterization is important in practice because in many \FDSM applications, $|\colors|$ is much smaller than $n$. For example, in file-system hierarchies the number of file types or extensions is clearly much smaller than the number of files.
We also show a direct application of our data structure to the
Top-$1$ Document Retrieval problem~\cite{DBLP:conf/soda/Muthukrishnan02} under \emph{document additions}~\cite{DBLP:journals/siamcomp/NavarroN17}. See \Cref{sec:parameterized_fdsm}.


\vspace{+1mm} 
\noindent{\bf 5.}~\emph{Experimental Evaluation.}
We extensively evaluate our data structures on large, real-world datasets
from different domains. For our \DSM data structure, we develop an efficient and effective method
for configuring the trade-off parameter $\alpha$. 
Our results show that this data structure outperforms the aforementioned \DRM{}-based baseline by
up to \emph{four orders of magnitude}  and can be constructed faster
on large datasets while maintaining linear space. A case study using data from Amazon~\cite{mcauley2015image} further shows a $71{,}000$-fold speedup of our \DSM data structure over the \DRM-based baseline, and an LLVM case study shows an $80$-fold speedup of our \FDSM data structure over rebuilding the static \SM{} data structure. See \Cref{sec:experimentw}.

The rest of the paper is organized as follows. \Cref{sec:related} discusses related work and \Cref{sec:preliminaries} provides the necessary background.  \Cref{sec:conclusion} concludes the paper. 
\ifproofsinappendix
\emph{Proofs of statements marked with~\omittedproofmark{} are deferred to \Cref{sec:omitted-proofs}.}
\fi

\section{Related Work}\label{sec:related}
\noindent{\bf Holistic aggregates in data management.}~Gray et al.~\cite{DBLP:journals/datamine/GrayCBLR97} classify aggregates as \emph{distributive} (e.g., $\textsf{count}$, $\textsf{sum}$, $\textsf{min}$, $\textsf{max}$), \emph{algebraic} (e.g., $\textsf{average}$), and \emph{holistic} (e.g., the median and the mode).
\emph{Distributive} and \emph{algebraic} aggregates can be computed by composing constant-size summaries of disjoint data subsets. Thus, standard data structures can maintain such aggregates efficiently, e.g., segment trees over window frames~\cite{DBLP:journals/pvldb/LeisKK015} and sliding-window aggregation for associative operators~\cite{DBLP:journals/pvldb/TangwongsanHSW15}. For example, \emph{distributive/algebraic} path and subtree aggregates (e.g., sum, min/max, diameter) on trees can be maintained in $\cO(\log n)$ time per operation via \emph{top trees}~\cite{DBLP:journals/talg/AlstrupHLT05} or, for path aggregates, \emph{link-cut trees}~\cite{DBLP:conf/stoc/SleatorT81} and, for subtree aggregates, Euler-tour trees~\cite{DBLP:journals/jacm/HenzingerK99}.

On the other hand, \emph{holistic} aggregates, such as the mode and percentiles, \emph{cannot} be determined from bounded-size summaries of subproblems \eg{a summary from which the mode of a union of subsets can be determined must essentially store the frequency of every color}. Thus, such aggregates are fundamentally harder to maintain~\cite{DBLP:journals/datamine/GrayCBLR97}, and, in practice, they are often evaluated from scratch.
Database research has studied holistic aggregates over window frames, their incremental view maintenance, and hierarchical data with subtree moves and roll-ups~\cite{DBLP:journals/pvldb/WesleyX16,DBLP:conf/sigmod/VogelsgesangNLK22,DBLP:conf/vldb/PalpanasSCP02,DBLP:conf/icde/BrunelFFMK0F15,DBLP:journals/pvldb/FinisBK0MF15,DBLP:journals/pvldb/BrunelMK16}, but none of these works maintains a holistic aggregate over all subtrees of a hierarchy in sublinear time per update.

\noindent{\bf Static mode queries.}~As mentioned in \Cref{sec:intro}, there has been much work on problems for mode queries in the static setting due to their importance in areas as diverse as databases (e.g., OLAP roll-up aggregations over dimension hierarchies and data profiling)~\cite{DBLP:books/mk/HanKP2011,DBLP:conf/sigir/LinDD18}, information retrieval~\cite{sea23}, network engineering~\cite{DBLP:conf/esa/Gao022,rfc4632}, and bioinformatics~\cite{static-paper}.
Krizanc et al.~\cite{DBLP:journals/njc/KrizancMS05} introduced \RM and \PM.
For \RM, Chan et al.~\cite{DBLP:journals/corr/abs-1101-4068} gave a linear-space data structure with $\Oish(\sqrt{n})$ query time and $\Oish(n^{3/2})$ construction time.
They also showed that, under the combinatorial \BMM (Boolean Matrix Multiplication) hypothesis, no \emph{combinatorial} data structure has both $\cO(n^{3/2-\smol})$ construction time and $\cO(n^{1/2-\smol})$ query time; using fast matrix multiplication, $n$ range mode queries can be answered in $\Oish(n^{1.4854})$ total time~\cite{DBLP:conf/soda/WilliamsX20}.

Approximate range mode queries too have received attention~\cite{DBLP:conf/stacs/BoseKMT05,DBLP:conf/icalp/GreveJLT10,sea23}.
For \PM, Durocher et al.~\cite{DBLP:journals/algorithmica/DurocherSST16} gave a linear-space data structure with $\Oish(\sqrt{n})$ query time. Recently, the query time was improved by a $\log\log n$ factor, with $\Oish(n^{3/2})$ construction time~\cite{DBLP:conf/swat/Rata26}.
Since \PM generalizes \RM, the conditional lower bound for \RM carries over to \PM. For \SM, on the other hand, the subtree modes of all nodes can be computed in $\cO(n)$ time~\cite{static-paper}, which gives an optimal data structure with $\cO(n)$ construction time and $\cO(1)$ query time.


\noindent{\bf Dynamic mode queries.}~
\DSM is a special case of \DRM, since we can keep the leaves of the tree in DFS order, so that the leaf-descendants of every node form a contiguous array range. A subtree-mode query then becomes a range-mode query, while a leaf recoloring becomes a point update.
For \DRM, Chan et al.~\cite{DBLP:journals/corr/abs-1101-4068} gave a linear-space data structure with $\Oish(n^{3/4})$ query time and $\Oish(n^{3/4})$ amortized expected update time.
El-Zein et al.~\cite{DBLP:conf/esa/El-Zein0MS18} improved this to $\cO(n^{2/3})$ worst-case query and update time, using $\cO(n)$ space and $\cO(n^{4/3})$ construction time, and additionally support insertions and deletions.
Using fast matrix multiplication, Sandlund and Xu~\cite{DBLP:conf/icalp/SandlundX20} broke the $n^{2/3}$ barrier, and Gu et al.~\cite{DBLP:conf/icalp/GuPVX21} further improved the query and update time to $\cO(n^{0.6524})$ using $\cO(n^{1.3262})$ space.
These algebraic data structures \emph{are impractical} due to the large constants in fast matrix multiplication.
Thus, in our experiments, we use the combinatorial data structure of~\cite{DBLP:conf/esa/El-Zein0MS18} as our \DRM baseline.

On the lower-bound side, Chan et al.~\cite{DBLP:journals/corr/abs-1101-4068} reduced the multiphase problem of P\u{a}tra\c{s}cu~\cite{DBLP:conf/stoc/Patrascu10} to \DRM, which gives a polynomial lower bound under the 3SUM hypothesis.
Jin and Xu~\cite{DBLP:conf/stoc/JinX22} showed that, under the combinatorial $4$-Clique hypothesis, the $\cO(n^{2/3})$ bound of El-Zein et al. is optimal for combinatorial data structures up to subpolynomial factors.
This shows that the trade-offs we give for \DSM with $1/3 < \smooth < 2/3$ are conditionally impossible for combinatorial \DRM data structures. They also observed that the reduction of Chan et al. can be adapted to show an $n^{1/2-o(1)}$ lower bound under the Online Matrix-Vector Multiplication (\OMv) conjecture, which also applies to algebraic data structures.
Our \Cref{cor:laminar-drm} strengthens this: the full query--update trade-off of \Cref{thm:dsm-lb} holds even if all query ranges come from a fixed laminar family; our \DSM data structure (\Cref{thm:dsm_ds}) shows that this is tight for such families.


\section{Preliminaries}\label{sec:preliminaries}
We work in the word RAM model with $\Theta(\log n)$-bit words.
All trees are rooted, and every node is its own ancestor and descendant.
The \emph{subtree of} a node $v$ consists of $v$ and its descendants, whereas a \emph{subtree} of a tree is any connected subgraph, rooted at its shallowest node.
For a node $v$ of $\mathcal{T}$ and a color $\col \in \colors$, $\freq_v(\col)$ is the number of leaf-descendants of $v$ colored $\col$, and $\freq(\col)$ is this number at the root.

\begin{theorem}[Tree covering~{\cite[Theorem~1]{DBLP:journals/algorithmica/FarzanM14}}]\label{thm:farzan-munro}
  For any integer $1 \le L \le n$, any tree on $n$ nodes can be decomposed in $\cO(n)$ time into $\cO(n/L)$ subtrees of size $\cO(L)$ each, which are pairwise disjoint aside from their roots.
  For each subtree, aside from edges from its root to its children, at most one edge leads from a node of the subtree to a child outside of it.
\end{theorem}

\noindent{\bf Data Structures.}~A static tree on $n$ nodes supports LCA (lowest common ancestor) queries in $\cO(1)$ time after $\cO(n)$-time preprocessing~\cite{DBLP:conf/latin/BenderF00}. 
Link-cut trees~\cite{DBLP:conf/stoc/SleatorT81} maintain a forest of rooted trees on $n$ nodes with node values under the following $\cO(\log n)$-time operations: making a root a child of a node in another tree, cutting a node from its parent, adding a number to the values of all ancestors of a node, reading a value, and LCA queries.

\noindent{\bf Fine-grained Complexity Hypotheses.}~An algorithm is \emph{combinatorial} if it does not use fast matrix multiplication-style techniques; see~\cite{DBLP:conf/stoc/HenzingerKNS15} for a discussion.
The following hypothesis generalizes the combinatorial \BMM hypothesis~\cite{DBLP:journals/jacm/WilliamsW18}.

\begin{conjecture}[Combinatorial $k$-Clique Hypothesis~\cite{DBLP:journals/siamcomp/AbboudBW18}]\label{conj:k-clique}
  For any constants $k \ge 3$ and $\smol > 0$, no combinatorial algorithm detects a $k$-clique in an $n$-node graph in $\cO(n^{k-\smol})$ time.
\end{conjecture}


\begin{definition}[\gOMv~\cite{DBLP:conf/stoc/HenzingerKNS15}]
  For a constant $\gamma > 0$ and $n_1 = \lfloor n_2^\gamma \rfloor$, preprocess a Boolean matrix $M \in \{0,1\}^{n_1 \times n_2}$; then, for Boolean vectors $v^1, \ldots, v^{n_3} \in \{0,1\}^{n_2}$ arriving one by one, output $M v^t$ before $v^{t+1}$ arrives.
\OMv is the special case $n_1 = n_2 = n_3 = n$.
\end{definition}

\begin{conjecture}[\OMv Conjecture~\cite{DBLP:conf/stoc/HenzingerKNS15}]\label{conj:omv}
  For any $\smol > 0$, no algorithm solves \OMv in $\cO(n^{3-\smol})$ time with error probability $\le 1/3$.
\end{conjecture}

Unlike \Cref{conj:k-clique}, \Cref{conj:omv} is not restricted to combinatorial algorithms.
By a theorem of \cite{DBLP:conf/stoc/HenzingerKNS15}, \Cref{conj:omv} implies hardness for \gOMv, meaning that for any $\gamma > 0$, no algorithm with $\mathrm{poly}(n_1, n_2)$ preprocessing time solves \gOMv in total time $\cO(n_1^{1-\smol} n_2 n_3 + n_1 n_2^{1-\smol} n_3 + n_1 n_2 n_3^{1-\smol})$, for a constant $\smol > 0$, with error probability at most $1/3$.


\section{\DSM: An Optimal Smooth Trade-off}\label{sec:dsm_ds}

We show that there is a smooth trade-off between the query and update times for the \DSM problem.
That is, \emph{any} (polynomial) query time can be achieved, but for smaller query times the update needs to be larger. In \Cref{sec:upper-bound-static-connectivity}, we show how to achieve the precise trade-off and in \Cref{sec:lower-bound-static-connectivity} that it cannot be improved by a polynomial, by providing a conditional lower bound for \DSM.

\subsection{The Data Structure}\label{sec:upper-bound-static-connectivity}

\Cref{thm:dsm_ds}, which we prove in this section, presents the construction and operations of our data structure for \DSM.

\begin{theorem}
  For any $0\le \smooth \le 1$, there is a data structure for the $\DSM$ problem with query time $\Oish(n^\smooth)$, update time $\Oish(n^{1-\smooth})$, construction time $\cO(n)$ and space $\cO(n)$.\label{thm:dsm_ds}
\end{theorem}

\subsubsection{Proof Overview} 
Fix $0 \le \smooth \le 1$. 
We partition the colors into heavy and light colors based on their frequency such that  
there are at most $n^\smooth$ heavy colors, so their frequencies can be
considered explicitly during a query. 

For every color, we maintain its \emph{single-color tree}, containing its leaves and only the branching nodes needed to connect them; this permits us to access the frequency of a color at a node of the tree $\mathcal{T}$ without storing a color $\col$ at every ancestor of each leaf colored $\col$. 
For light colors, we additionally maintain local frequency information and decompose $\mathcal{T}$ into rooted subtrees of size $\cO(n^\smooth)$. This decomposition handles light colors that are not represented at the queried node: 
if all occurrences of such a color lie in one child subtree, we determine 
its contribution by local work in a rooted subtree of size
$\cO(n^\smooth)$, rather than inspecting all light colors.
Below the rooted subtree containing the queried node, we rely on a heap stored at every subtree root, which maintains the most frequent light color below that root.
During an update, we explicitly repair these structures only for light colors.
Since a light color has frequency less than $n^{1-\smooth}$, its single-color tree has $\cO(n^{1-\smooth})$ nodes, which bounds the update time.

\vspace{+1mm}
We first describe the maintained information and preprocessing.
We then give the update and query algorithms, prove their correctness,
and finally provide the construction time and space complexity of our data structure.  


\subsubsection{Maintained Information}
\label{sec:dsm-maintained-information} 

We call a color $\col$ heavy if $\freq(\col) \ge n^{1-\smooth}$ and light otherwise.
Write $\hcolors$ and $\lcolors$ for the heavy and light colors, respectively.
Observe that there are at most $n^\smooth$ heavy colors.

Write $T_\col$ for the \emph{single-color} tree for color $\col$: it is the tree containing all leaves of color $\col$ and all inner nodes that have leaves of color $\col$ in at least two of their subtrees \ie{the virtual tree of the leaves colored $\col$, which is their closure under lowest common ancestors}.
By a result from~\cite{static-paper}, we can construct \emph{all} of the single-color trees at once in $\cO(n)$ time. We will later show how to update them.

For every node $v$, we will maintain a binary search tree, BST for short, $\lightBST{v}$ that will contain all pairs $(\col, \freq_v(\col))$ for all light colors $\col$ such that $v \in T_\col$, that is, a BST of all light colors that exist in at least two subtrees of $v$ (or, for a leaf $v$, its own color if it is light). Heavy colors are not stored in these BSTs; they are handled separately during queries.
The sorting of this BST will be 
in increasing order of frequency (breaking ties with $\col$), meaning we support the operations $\getMax$, $\decreaseKey$, and $\increaseKey$ in $\cO(\log n)$ time.\footnote{There are many standard data structures that can give us these operations, one example being the ``humble''  heap~\cite[Section~6.5]{cormen2022introduction}.}

Additionally, to maintain information on the light colors, we use the tree covering of \Cref{thm:farzan-munro} with $L = \lceil n^\smooth \rceil$, that is, $\cO(n^{1-\smooth})$ subtrees of size $\cO(n^\smooth)$.

We assume that we have such a decomposition of $\mathcal{T}$ into subtrees and that we store, for every node, the subtrees of the decomposition that contain it (note from the theorem statement that the subtrees overlap only in their roots).
In particular, we will maintain $\lightmode(u)$, the most common light color in the descendants of $u$, for all roots of all these subtrees.

To this end, we let the subtree roots cut $\mathcal{T}$ into regions.
The \emph{region} $S(r)$ of a subtree root $r$ consists of $r$ and all descendants of $r$ that are not subtree roots and are reached from $r$ without passing through another subtree root.
The \emph{child roots} of $r$ are the first subtree roots below $r$ \ie{the subtree roots $r' \neq r$ such that the path from $r$ to $r'$ contains no further subtree root}.
Thus, the subtree of $r$ splits into the region $S(r)$ and the subtrees of the child roots of $r$.

Intuitively, the heap $\lightAgg{r}$ summarizes the subtree of $r$ by combining the best light candidate of every node in its region with the summaries of its child roots.
Concretely, $\lightAgg{r}$ contains (1) one entry $\getMax(\lightBST{u})$ for every $u \in S(r)$ and (2) one entry $\getMax(\lightAgg{r'})$ for every child root $r'$ of $r$, and it supports the same operations as the $\lightBST{v}$. 
Let $\lightmode(r)$ be the color of $\getMax(\lightAgg{r})$.
Every node lies in exactly one region and every subtree root is a child root of at most one subtree root, so these heaps take $\cO(n)$ space in total.

Now, $\lightmode(r)$ is indeed a most frequent light color below $r$: as the subtree of $r$ splits into $S(r)$ and the subtrees of its child roots, induction over the child roots shows that $\getMax(\lightAgg{r})$ is exactly this entry.
The heaps can be initialized bottom-up in linear time once the $\lightBST{v}$ are built.

\subsubsection{Update}
\label{sec:dsm-updates}

We state the update procedure in \Cref{alg:update}. 

\begin{algorithm}[t]\small 
  \caption{Update($v$, $\col \to \coll$): recolor leaf $v$ from $\col$ to $\coll$ and restore all maintained information.\label{alg:update}}
  \KwIn{Leaf $v$ whose color changes from $\col$ to $\coll$}

  \algstep{Step 1: update the single-color trees}
  Update $T_\col$ and $T_\coll$ and the frequencies of $\col$ and $\coll$ at their nodes using link-cut trees\;

  \algstep{Step 2: maintain the BSTs of light colors per node}

  \If{$\col \in \lcolors$}{
    \ForEach{ancestor $u$ of $v$ in $T_\col$}{
      $\decreaseKey$ for $\col$ in $\lightBST{u}$\;
    }
    Delete $\col$ from $\lightBST{u}$ for the $\cO(1)$ nodes removed from $T_\col$\;
  }

  \If{$\coll \in \lcolors$}{
    Insert $(\coll, \freq_u(\coll))$ into $\lightBST{u}$ for the $\cO(1)$ new nodes in $T_\coll$\;
    \ForEach{ancestor $u$ of $v$ in $T_\coll$ not added in Step 1}{
      $\increaseKey$ for $\coll$ in $\lightBST{u}$\;
    }
  }

  \BlankLine
  \If{$\col$ becomes light}{
    Remove $\col$ from $\hcolors$, add to $\lcolors$\;
    \ForEach(\tcp*[h]{$\cO(\freq(\col)) \subseteq \cO(n^{1-\smooth})$ many}){$u$ in $T_\col$}{
      Insert $(\col, \freq_u(\col))$ into $\lightBST{u}$\;
    }
  }

  \If{$\coll$ becomes heavy}{
    Add $\coll$ to $\hcolors$, remove from $\lcolors$\;
    \ForEach(\tcp*[h]{$\cO(\freq(\coll)) \subseteq \cO(n^{1-\smooth})$ many}){$u$ in $T_\coll$}{
      Delete $\coll$ from $\lightBST{u}$\;
    }
  }

  \algstep{Step 3: update the heaps of subtree roots}
  Update the entry of every node $u$ whose $\lightBST{u}$ changed in Step 2 in the heap $\lightAgg{r}$ with $u \in S(r)$\;
  \ForEach{subtree root $r'$ whose heap $\lightAgg{r'}$ changed, in order of decreasing depth}{
    Update the entry of $r'$ in $\lightAgg{r}$, for every child root $r'$ of $r$\;
  }
  \end{algorithm}

We should explain how to update $T_\col$ and $T_\coll$, the single-color trees of the old color $\col$ and the new color $\coll$,  respectively.  
Let us look at the update of $T_\col$ first.
Here, $v$ is a leaf that is now not colored $\col$ anymore, thus we simply delete it from $T_\col$.
Additionally, this might make the parent $p$ of $v$ in $T_\col$ not be branching anymore, forcing us to delete it from $T_\col$.
This cannot cascade: if $p$ is deleted, the subtree of $p$ still contains leaves colored $\col$, so every ancestor of $p$ in $T_\col$ still has leaves colored $\col$ in the same subtrees as before.
We also need to decrease the value of $\freq_u(\col)$ for all ancestors $u$ of $v$ in $T_\col$.
This can be done via a path-update in the link-cut tree.
The operation for updating $T_\coll$ is almost symmetric; the only difference is that the counterpart of deleting the parent of $v$ in $T_\col$ is slightly more difficult.
In particular, a new internal node might appear in $T_\coll$ (as the parent of $v$) if there is another leaf $v'$ s.t. the lowest common ancestor of $v$ and $v'$ is not already in $T_\coll$. 
 To find it, we maintain, for every color, its leaves in DFS order in a BST. Let $v'$ and $v''$ be the predecessor and successor of $v$ among the leaves colored $\coll$ in this order. The only candidate for the new internal node (and the parent of $v$ in $T_\coll$) is the deeper of the lowest common ancestors (LCA) of $v, v'$ and of $v, v''$, so two LCA queries suffice.
Thus, $T_\col$ and $T_\coll$ change by only $\cO(1)$ nodes.
If the candidate $x$ is not yet in $T_\coll$, let $y$ be the topmost node of $T_\coll$ in the subtree of $x$, found as in \Cref{sec:dsm-queries} before $v$ is inserted into the DFS-order BST; we splice $x$ in between $y$ and the parent of $y$ in $T_\coll$ with $\freq_x(\coll) \coloneqq \freq_y(\coll)$, make $v$ a child of $x$, and add $1$ to $\freq_u(\coll)$ for all ancestors $u$ of $v$ in $T_\coll$ by a path-update.
This takes $\cO(1)$ link-cut operations.

For the running time, note that Steps 2 and 3 touch only light colors; heavy colors only need the $\cO(\log n)$-time path-updates of Step 1.
A light color $\col$ satisfies $\freq(\col) < n^{1-\smooth}$, so $T_\col$ has $\cO(n^{1-\smooth})$ nodes. The same holds when a color crosses the heavy/light threshold, as its frequency then is at most $n^{1-\smooth}+1$.
Hence, Step 2 changes $\cO(n^{1-\smooth})$ BSTs, each in $\cO(\log n)$ time.
In Step 3, every one of the $\cO(n^{1-\smooth})$ subtree roots is handled at most once, after all changes to its heap. 
Thus, an update takes $\Oish(n^{1-\smooth})$ time.

\subsubsection{Query}
\label{sec:dsm-queries} 

For the query procedure, see \Cref{alg:query}.            %
First, note how we can access $\freq_v(\col)$.
If $v \in T_\col$, we have directly stored the value.
Otherwise, let $u$ be the topmost node of $T_\col$ in the subtree of $v$; by the definition of $T_\col$, $\freq_v(\col) = \freq_u(\col)$, or $\freq_v(\col) = 0$ if $u$ does not exist.
We find $u$ as the LCA of the first and the last leaf colored $\col$ in the subtree of $v$, which two searches in the DFS-order BST of \Cref{sec:dsm-updates} provide, in $\cO(\log n)$ time.
To complete the explanation, we must explain how to compute $\lightmode$.
For a subtree root $v$, we directly take the value $\lightmode(v)$ we maintained during updates.
For any non-root node $u$, it will be computed recursively. That is, we take the maximum of $\getMax(\lightBST{u})$ and $\lightmode(w)$ over all children $w$ of $u$.
Note that this will then be executed at most once per node that is in the subtree of $u$ (which is unique as $u$ is not a subtree root).
By \Cref{thm:farzan-munro}, this recurses at most $\cO(n^{\alpha})$ times in total.
Moreover, each call takes $\cO(1)$ time per child of its node.
All children of these nodes lie in the same subtree of the decomposition, except for at most one child in total by \Cref{thm:farzan-munro}; as the parent of this child is not a subtree root, the child must be the root of its subtree, so we take its stored value in $\cO(1)$ time.
Hence, computing $\lightmode(v)$ takes $\cO(n^{\alpha})$ time.

It remains to see why the query returns a most frequent color below $v$.
For the heavy colors, this is immediate, as we check all of them explicitly.
For the light colors, recall from \Cref{sec:dsm-maintained-information} that a most frequent light color $x$ below $v$ is stored with its full frequency at the topmost node $u$ of $T_x$ in the subtree of $v$, and that no entry exceeds the true frequency of its color below $v$.
Thus, it suffices that the query accounts for $\lightBST{u}$.
Indeed, the computation of $\lightmode(v)$ descends from $v$ towards $u$ and reaches $u$ or stops at a subtree root on the path from $v$ to $u$ (possibly $v$), whose heap accounts for $\lightBST{u}$.
Note that this includes the case $u = v$, where the leaves colored $x$ below $v$ branch at $v$: then $\lightBST{v}$ is taken into account either directly by the recursion or, if $v$ is a subtree root, through $\lightAgg{v}$, as $v \in S(v)$.
Thus, we do not need to consult $\lightBST{v}$ separately.

\begin{algorithm}[ht]\small 
  \KwIn{Node $v$}

  \algstep{Step 1: best light color, via the tree decomposition}
  $a \gets \lightmode(v)$\tcp*{$\cO(n^{\smooth})$}

  \algstep{Step 2: best heavy color, by checking all of them}
  $b \gets \operatorname*{arg\,max}_{\col \in \hcolors} \freq_v(\col)$ \tcp*{$\cO(\setsize{\hcolors}) \subseteq \cO(n^{\smooth})$}

  \Return{the one of $a$ and $b$ with larger $\freq_v$}\;
  \caption{Query($v$): report a most frequent color among the leaves below $v$.\label{alg:query}}
\end{algorithm}

\subsubsection{Construction Time and Space}

Preprocessing begins by running the algorithm from~\cite{static-paper} on all colors.
This also gives us the initial $T_c$'s and the LCA data structure (see \Cref{sec:preliminaries}).
Then we transform these $T_\col$'s into link-cut trees (see~\cite{DBLP:conf/stoc/SleatorT81}), build the DFS-order BSTs of \Cref{sec:dsm-updates} by sorting the leaves of each color by DFS number, build the heaps $\lightBST{v}$ and $\lightAgg{r}$ bottom-up, and compute the decomposition from \Cref{thm:farzan-munro}.
All of this can be done in $\cO(n)$ time using $\cO(n)$ space~\cite{static-paper,DBLP:conf/stoc/SleatorT81}.  

\vspace{-2mm}
\subsection{The Lower Bound}\label{sec:lower-bound-static-connectivity}
Techniques from fine-grained complexity allow us to show that we cannot significantly improve on the data structure we presented in \Cref{sec:upper-bound-static-connectivity}, unless central hardness assumptions fail. Specifically, we provide the following theorem, which rules out a polynomial improvement to the tradeoff between update and query time.
 
\begin{theorem}\label{thm:dsm-lb}
  For any~$\smooth\in(0,1)$, if there is a data structure for the \DSM
  problem with polynomial preprocessing, query time
  $\Oish(n^{\smooth-\smol_Q})$ and update time
  $\Oish(n^{1-\smooth-\smol_U})$ for some $\smol_Q,\smol_U\ge 0$ with
  $\smol_Q+\smol_U>0$, then the \OMv{} conjecture fails.
\end{theorem}
\begin{appendixproof}
We reduce \gOMv{} to \DSM and assume that we have access to a $\DSM$ data structure violating \Cref{thm:dsm-lb}.
If the input matrix is $M \in \{0,1\}^{n_1 \times n_2}\relax$, then the reduction builds a single tree $T$ of size $n=\Theta(n_1 n_2)\relax$ 
during
preprocessing and, in each \OMv{} round, performs a controlled number of
updates and queries.
We choose $\gamma$ based on $\alpha$ and the alleged running time improvements $\smol_U,\smol_Q$ and thus derive a contradiction for
every $\smooth$.
Concretely, we pick any $\gamma$ such that
\begin{align*}
  \smooth - \smol_Q \;<\; \frac{1}{1+\gamma} \;<\; \smooth + \smol_U.
\end{align*}
As one of $\smol_Q, \smol_U$ is greater than $0$ and $\smooth \in (0,1)$, such a $\gamma > 0$ exists.

Recall that by definition $(M v)_i = \bigvee_{j} (M_{i,j} \land v_j{})\relax$.

\begin{figure}[t]
  \centering
\begin{tikzpicture}[
    x=1cm, y=1cm,
    font=\normalsize,
    leaf/.style={draw, fill=white, rectangle, minimum size=0.42cm, inner sep=0pt},
    inner/.style={circle, fill, inner sep=0pt, minimum size=3.5pt},
    set/.style={anchor=base},
    region/.style={draw=#1, fill=#1!15, thin},
    ph/.style={leaf, draw=black!40, text=black!40},
  ]

  \node[leaf] (a1) at (0.00,0) {1};  \node[leaf] (a2) at (0.48,0) {2};  
  \node[leaf] (b1) at (1.29,0) {1};  \node[leaf] (b3) at (1.77,0) {3};  
  \node[leaf] (c2) at (2.47,0) {2};                                     
  \node[leaf] (d2) at (3.28,0) {2};                                     
  \node[leaf] (e1) at (3.98,0) {1};  \node[leaf] (e3) at (4.46,0) {3};  
  \node[ph] (u1) at (5.27,0) {4};                                       
  \node[leaf, draw=VUOrange, very thick] (u2) at (5.75,0) {2};          
  \node[ph] (u3) at (6.23,0) {6};                                       

  \node[inner] (wM3) at (0.24,0.65) {};
  \node[inner] (wZ2) at (1.53,0.65) {};
  \node[inner] (wM2) at (2.47,0.65) {};
  \node[inner] (wZ1) at (3.28,0.65) {};
  \node[inner] (wM1) at (4.22,0.65) {};

  \node[inner, label={right:$w_1$}]       (w1) at (5.20,1.25) {};
  \node[inner, label={above right:$w_2$}] (w2) at (3.80,1.80) {};
  \node[inner, label={above:$w_3$}]       (w3) at (2.10,2.35) {};

  \draw (wM3) -- (a1) (wM3) -- (a2)
        (wZ2) -- (b1) (wZ2) -- (b3)
        (wM2) -- (c2)
        (wZ1) -- (d2)
        (wM1) -- (e1) (wM1) -- (e3);
  \draw (w1) -- (wM1) (w1) -- (u1) (w1) -- (u2) (w1) -- (u3);
  \draw (w2) -- (wM2) (w2) -- (wZ1) (w2) -- (w1);
  \draw (w3) -- (wM3) (w3) -- (wZ2) (w3) -- (w2);

  \path (0.24, -0.65) node[set] {$\ones{3}$}
        (1.53, -0.65) node[set] {$\zeros{2}$}
        (2.47, -0.65) node[set] {$\ones{2}$}
        (3.28, -0.65) node[set] {$\zeros{1}$}
        (4.22, -0.65) node[set] {$\ones{1}$}
        (5.27, -0.65) node[set] {$u_1$}
        (5.75, -0.65) node[set] {$u_2$}
        (6.23, -0.65) node[set] {$u_3$};

  \matrix (M) [matrix of math nodes, anchor=north west, inner sep=1pt, left delimiter=(, right delimiter=),
               nodes={inner sep=0pt, minimum size=0.36cm}] at (0.45,3.05)
    { 1 & 0 & 1 \\ 0 & 1 & 0 \\ 1 & 1 & 0 \\ };
  \node[anchor=east] at ([xshift=-5pt]M.west) {$M =$};
  \node[anchor=north west, inner sep=0pt, text=VUOrange] at (4.15,3.0) {$v = (0, 1, 0)$};
  \node[anchor=north west, inner sep=0pt] at (4.15,2.5) {$Mv = (0, 1, 1)$};

  \begin{scope}[on background layer]
    \draw[region=VUPurple] (-0.33,-0.78) rectangle (0.81,0.82);
    \draw[region=VUGreen]  ( 0.96,-0.78) rectangle (2.80,0.82);
    \draw[region=VUBlue]   (2.95,-0.78) rectangle (4.79,0.82);
    \draw[region=VUOrange] (4.94,-0.78) rectangle (6.56,0.82);
    \fill[VUBlue!15]   (M-1-1.north west) rectangle (M-1-3.south east);
    \fill[VUGreen!15]  (M-2-1.north west) rectangle (M-2-3.south east);
    \fill[VUPurple!15] (M-3-1.north west) rectangle (M-3-3.south east);
  \end{scope}
\end{tikzpicture}
  \caption{The tree of the reduction for the shown $M$ and $v$, after loading $v$. Leaves show their colors; placeholder colors are gray and the recolored leaf $u_2$ is highlighted. The shaded regions mark the leaves encoding the rows of $M$ and the vector $v$.}
  \label{fig:omv-reduction}
\end{figure}
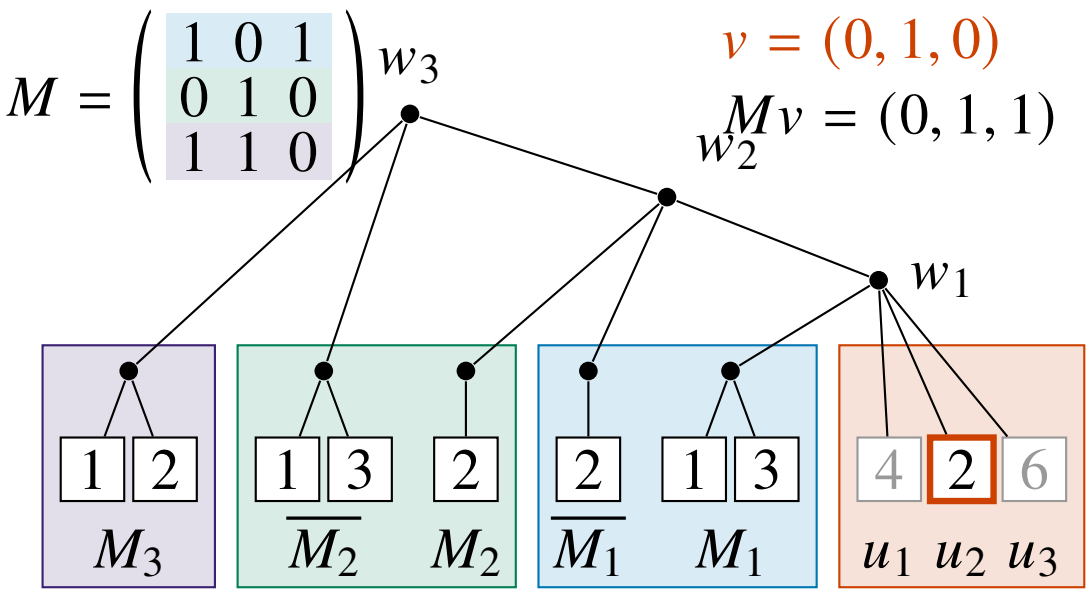

\paragraph{Construction}
First, let us define, for each $i \in [n_1]$, the sets $\ones{i}$ and $\zeros{i}$ as the sets of indices $j \in [n_2]$ such that $M_{i,j} = 1$ and the complement of that set, respectively.
That is, the indices of the one and zero bits in row $i$, respectively.

We create leaves for each element of each of the sets $\ones{i}$ and $\zeros{i}$.\footnote{Technically, we omit this for $\zeros{n_1}$.}
Call these leaves the \emph{matrix leaves} (we will later also describe vector leaves to encode the queried vector $v$).
Each of these sets has a special parent node $w_{\ones{i}}$ and $w_{\zeros{i}}$ (which we omit if the set is empty).
For each row $i \in [n_1] \setminus \{1\}$, create a node $w_i$ that has as children: (1) the node $w_{\ones{i}}$, (2) the node $w_{i-1}$, (3) the node $w_{\zeros{i-1}}$.
The node $w_1$ only has the first of these and the vector leaves we describe later.
The root of the tree is $w_{n_1}$.
See \Cref{fig:omv-reduction} for an example.
Now, the following observation is crucial for our analysis.

\begin{observation}\label{obs:row-gadget}
  For each $i \in [n_1]$ and $j \in [n_2]$, the number of descendants of $w_i$ colored $j$ among the matrix leaves is $\freq_{w_i}(j) = i-1+M_{i,j}$.
\end{observation}


Additionally, we create $n_2$ children of $w_1$, we name them $u_j$ (for each $j \in [n_2]$).
These are the \emph{vector leaves} and, initially, these are each colored with a unique color \ie{one that appears nowhere else in the tree}, for example $n_2 + j$.

On this leaf-colored tree, call the preprocessing of the alleged $\DSM$ data structure.

\paragraph{\OMv{} Query}
When we get a query $v \in \{0,1\}^{n_2}\relax$, we do the following for each $j \in [n_2]$ such that $v_j = 1$: we color the node $u_j$ with color $j$. 
If $v_j = 0$, we color the node with its placeholder color $n_2 + j$.
We call this operation \emph{loading $v$}; it is possible using at most $n_2$ updates to our alleged \DSM{} data structure.

Now, this together with the previous observation yields the following result.
\begin{corollary}
  After loading $v$, and for every $i \in [n_1]$, the frequency of the subtree mode of $w_i$ is $i+1$ iff there is a $j \in [n_2]$ such that (1) $M_{i,j} = 1$, and (2) $v_j = 1$.
  Otherwise, that frequency is $i-1$ or $i$.
  Note that this is exactly the condition for $(Mv)_i = 1$.
\end{corollary}

From this corollary, we immediately get that we can extract the vector $Mv$ using $n_1$ queries to our alleged \DSM{} data structure.
Call this operation \emph{reading $Mv$}.


\paragraph{Size and Running Times}
It is easy to see that we have at most $n_1 n_2 + n_2$ leaves in this tree.
As there are only $\cO(n_1)$ inner nodes, there are at most $\cO(n_1n_2)$ nodes in total.
Write $n$ for this number of nodes.
As this is polynomial in $n_1 n_2$ and the preprocessing of our alleged \DSM{} data structure is polynomial in its input, the total \OMv{} preprocessing is also polynomial.

Now, let us consider the total time to process $n_3$ \OMv{} queries.
While servicing each \OMv{} query, we make $n_2$ updates to the alleged \DSM{} data structure and $n_1$ queries.
By assumption, this takes a total of
\begin{equation*}
  \begin{split}
    &\cO(n_3 n_1 n^{\alpha-\smol_Q} + n_3 n_2 n^{1-\alpha-\smol_U})
    \\ = &\cO(n_3 n_2^{(1+\gamma) (\alpha - \smol_Q) + \gamma}
          \; + n_3 n_2^{(1+\gamma) (1 - \alpha - \smol_U) + 1}).
  \end{split}
\end{equation*}

\paragraph{Deriving the contradiction}
Computing $n_3$ products of type $Mv$ naively costs $\Theta(n_1 n_2 n_3) =
\Theta(n_3\, n_2^{1+\gamma})$; by the hardness of \gOMv (\Cref{sec:preliminaries}), no algorithm with
polynomial preprocessing beats this by a polynomial factor. Simple rearrangement shows our
reduction does exactly that, \ie{that both exponents above are strictly
  below $1+\gamma$}.

For the query term,
\begin{align*}
  (1+\gamma)(\alpha-\smol_Q) + \gamma \;<\; 1+\gamma
  \quad\Longleftrightarrow\quad
  \alpha - \smol_Q \;<\; \frac{1}{1+\gamma},
\end{align*}
and for the update term,
\begin{align*}
  (1+\gamma)(1-\alpha-\smol_U) + 1 \;<\; 1+\gamma
  \quad\Longleftrightarrow\quad
  \frac{1}{1+\gamma} \;<\; \alpha + \smol_U.
\end{align*}

Thus, the existence of the alleged data structure violates the \OMv{} conjecture.

\begin{remark}
  This reduction uses that the \DSM{} data structure returns the frequency of the subtree mode when queried.
  This is not a restriction as such a data structure can be derived from one that only reports the color of the subtree mode with only \emph{additive} extra logarithmic running time in both the query and update.

    To achieve this, maintain the single-color trees and the DFS-order BSTs of \Cref{sec:dsm-updates}, which takes $\cO(\log n)$ time per update, and compute $\freq_v(\col)$ for the reported color $\col$ as in \Cref{sec:dsm-queries}, in $\cO(\log n)$ time.
\end{remark}

\end{appendixproof}

\vspace{-2mm}
\begin{corollary}\label{cor:laminar-drm}
  \Cref{thm:dsm-lb} also holds for \DRM{} with array length $n$, where point updates change the color of one element, and all query ranges come from a fixed laminar family of intervals.
\end{corollary}
\begin{appendixproof} 
  Order the leaves of the tree constructed above in DFS order and let $A$ be the array of their colors, so that $|A| = \Theta(n)$.
  The leaf-descendants of every node $v$ then form an interval $[L_v, R_v]$ of $A$, and these intervals form a laminar family, as two subtrees are either disjoint or nested.
  A leaf recoloring is a point update of $A$ and a subtree mode query at $v$ is a range mode query on $[L_v, R_v]$.
  Thus, a \DRM{} data structure violating the bounds of \Cref{thm:dsm-lb} on this family yields a \DSM{} data structure violating them on our instances, contradicting the \OMv{} conjecture.
\end{appendixproof}

\vspace{-2mm}
\section{Low Tree Height Can Improve \DSM}\label{sec:parameterized_dsm}

In several of the applications for \DSM, 
the underlying trees have very small height (see \Cref{sec:intro}). 
For instance, in the applications presented in \cite{static-paper}, the input trees are suffix trees of string collections, and suffix trees are known to have a small height for certain string distributions~\cite{DBLP:journals/siamcomp/DevroyeSR92}. It is thus natural to ask if our \DSM data structure can be improved in this case.
Interestingly, the construction from \Cref{sec:upper-bound-static-connectivity} can be reanalyzed in this case to give tighter bounds.

\begin{theorem}
  \label{thm:height-upper-bound}
  There is a data structure for the $\DSM$ problem for trees of height $h$ with query time $\Oish(1)$, update time $\Oish(h)$, construction time $\cO(n)$ and space $\cO(n)$.
\end{theorem}

\begin{appendixproof}
  This follows directly from our data structure in \Cref{sec:upper-bound-static-connectivity} when we treat all colors as light (this means we set $\smooth=0$; as a tree with $n \ge 2$ nodes has fewer than $n$ leaves, no color is heavy).
  It directly follows from that section that the query time then becomes $\Oish(1)$. 
  For the update time, we examine \Cref{alg:update} again.
  First observe that the conditional operations for when a color changes between light and heavy will never trigger as all colors are always light.
  Second, observe that the updates in Step 1 only require $\Oish(1)$ time anyways.
  Lastly, observe that the remaining loops (in Step 2 and 3) run for each ancestor of the changed node, thus at most $h$ times, as only BSTs and heaps of ancestors of the changed node change.
  Each operation performed inside a loop iteration is logarithmic, giving the desired bound.
\end{appendixproof}

In \Cref{sec:parameterized_fdsm}, we show an application that specifically utilizes suffix trees in another special case: \FDSM with small $|\colors|$.

\begin{remark}
  The instances of \Cref{sec:lower-bound-static-connectivity} have height $\Theta(n_1)$.
  Hence, a data structure with $\Oish(1)$ query time and $\Oish(h^{1-\smol})$ update time, for some $\smol > 0$, processes each \OMv round, that is, $n_2$ updates and $n_1$ queries, in $\Oish(n_2 n_1^{1-\smol} + n_1)$ time. By the hardness of \gOMv (\Cref{sec:preliminaries}), this violates the \OMv conjecture for any $\gamma > 0$.
\end{remark}

This means that the update time of \Cref{thm:height-upper-bound} cannot be improved by a polynomial unless \OMv fails.

\section{Why Is the Connectivity Not Dynamic?}\label{sec:hardness_fdsm}

The allowed update operations of the data structure presented in the previous section do not allow the structure of the tree to change. 
It is thus natural to ask if there is a data structure that allows, for example, the insertion and deletion of edges. 
To ensure that the tree remains connected, such updates are frequently modeled as the addition and deletion of leaves together with the Link-Cut operation described in the definition of \FDSM.
The following theorem shows that if we allow such operations, we run into a well-known barrier: such a data structure would directly yield a data structure for the dynamic range mode problem \DRM.

\begin{theorem}
  If there is a data structure for the $\FDSM$ problem with query and update time $\Oish(T(n))$, then there is such a data structure for $\DRM$.
  If the $\FDSM$ data structure is combinatorial, then so is the derived $\DRM$ one.\label{thm:fdsm-drm-reduction}
\end{theorem}

\begin{appendixproof}[thm:fdsm-drm-reduction]
  Consider a \DRM instance with an array $A$ of length $n$ over the color set $\colors$.
  For initialization, create $n$ leaves $a_1$, \dots, $a_n$ and color each $a_i$ with $A[i]$.
  Also create $n$ inner nodes $u_1, \dots, u_{n}$, where $u_{n}$ is the parent of $a_n$ only and all other $u_i$ are parent of $u_{i+1}$ and $a_i$.
  Thus, for every $i \in [n]$, the leaf descendants of $u_i$ are exactly $a_i, \dots, a_n$.
  Finally, create a root $v_1$ as the parent of $u_1$.

  For a \DRM query $i,j$ with $i \le j$, map it to a constant number of \FDSM operations as follows: if $j < n$, cut $u_{j+1}$ from its parent $u_j$ and attach it to $v_1$.
  Now (also if $j = n$) the leaf descendants of $u_i$ are exactly $a_{i}, \dots, a_j$, so that the subtree mode of $u_i$ is exactly the desired range mode.
  Afterwards, undo the link-cut operation so that the topology of the tree is the same as initially.

  For a \DRM update that changes an entry, recolor $a_i$. To insert or delete an entry, create or delete the corresponding $a_i$ and $u_i$; as $u_i$ lies on the path $u_1, \dots, u_n$, this takes a constant number of leaf additions/deletions and Link-Cut Updates. Note that the indices of leaves and internal nodes change, since inserting or deleting an entry shifts all later positions.
  To find the nodes belonging to a position, we keep the pairs $(a_i,u_i)$ in array order in a balanced search tree, which costs $\cO(\log n)$ time per operation.

  In conclusion, every \DRM operation boils down to a constant number of \FDSM operations on a tree with $\cO(n)$ nodes, plus $\cO(\log n)$ time for the search tree.
  This gives query and update time $\Oish(T(n))$.
  Initialization only adds linear time to the construction of the \FDSM data structure.
  Finally, the reduction uses the \FDSM data structure as a black box, so it is combinatorial if the \FDSM data structure is.
\end{appendixproof}

Note that \Cref{thm:fdsm-drm-reduction} yields hardness for \FDSM only in combination with a lower bound for \DRM.
For combinatorial data structures, a tight conditional lower bound is due to Jin and Xu~\cite{DBLP:conf/stoc/JinX22}.
This matches the $\cO(n^{2/3})$ upper bound of El-Zein et al.~\cite{DBLP:conf/esa/El-Zein0MS18} up to subpolynomial factors.
Setting $T(n)=n^{2/3-\smol}$ in \Cref{thm:fdsm-drm-reduction}, together with these conditionally tight bounds, gives the following.

\begin{corollary}\label{cor:fdsm-lb}
  Under the combinatorial $4$-Clique hypothesis (\Cref{conj:k-clique} with $k=4$), there is no combinatorial $\FDSM$ data structure with polynomial preprocessing time and $\cO(n^{2/3-\smol})$ query and update time for any $\smol>0$.
\end{corollary}

Since $\Oish(n^{2/3-\smol})\subseteq\cO(n^{2/3-\smol/2})$, \Cref{cor:fdsm-lb} also rules out $\Oish(n^{2/3-\smol})$ query and update time.
The restriction to combinatorial data structures is essential for this argument.
Using fast matrix multiplication, Gu et al.~\cite{DBLP:conf/icalp/GuPVX21} obtain a \DRM data structure with $\cO(n^{0.6524})$ query and update time, so \Cref{thm:fdsm-drm-reduction} does not preclude an algebraic \FDSM data structure with $\cO(n^{2/3-\smol})$ query and update time.
Still, any \FDSM data structure with $\cO(n^{0.6524-\smol})$ query and update time would improve the state of the art for \DRM.

\section{Few Colors Allow \FDSM}\label{sec:parameterized_fdsm}

Motivated by the hardness result from the previous section, we ask if there can 
exist efficient data structures for \FDSM for important special cases. As discussed in \Cref{sec:intro}, in several applications of \FDSM, $|\colors|$ is much smaller than $n$. 
In this section, we will tolerate algorithms that have linear overhead in $|\colors|$.
We show that there is a linear-space \FDSM data structure where every operation costs $\cO(\log^2 n)=\Oish(1)$ or $\cO(|\colors|\log^2 n)=\Oish(|\colors|)$ worst-case time. 
The structure consists of one dynamic Euler sequence and per-color
search trees ordered through it. There is no polynomial dependence on $n$, which helps the aforementioned applications, where $|\colors| \ll n$. 

\begin{theorem}
  There is an \FDSM data structure of size $\cO(n)$ which can be constructed in $\cO(n\log n)$ time, with the following worst-case operation times:
  \begin{itemize}
     \item {\bf Query}~(query the subtree mode of a node): $\cO(|\colors|\log^2 n)$ time 
    \item {\bf Update}~(change the color of a leaf): $\cO(\log^2 n)$ time 
    \item {\bf Link-Cut Update} (link and cut a subtree, moving all its vertices): $\cO(|\colors|\log^2 n)$ time    
    \item {\bf Add/Delete Leaf} (adding or deleting a leaf):  $\cO(\log^2 n)$ time  
  \end{itemize}\label{thm:fdsm_few_colors}
\end{theorem}

\begin{appendixproof}
\emph{Representation and order comparisons.}~We choose an arbitrary order of the children of every node and maintain the corresponding Euler sequence $E$. For every vertex $v$, the sequence contains an entry token $e_v$ and an exit token $x_v$. The interval from $e_v$ through $x_v$ consists exactly of the tokens of the subtree rooted at $v$. We store $E$ in a worst-case balanced sequence tree with parent pointers and subtree sizes, supporting insertion, deletion, splitting, concatenation, and the rank of a token in $\cO(\log n)$ time. Each vertex has handles to its two tokens.

For every color $c$ currently present on a leaf, we maintain a balanced search tree $B_c$ containing the entry tokens of exactly the leaves of color $c$, in their order in $E$. Each $B_c$ stores subtree sizes and supports order-statistic searches, insertion, deletion, splitting at a search position, and concatenation of ordered trees. A balanced dictionary maps the currently present colors to their trees; it takes $\cO(n)$ space and $\cO(\log n)$ time per lookup. Empty color classes need no permanent record. We also store parent pointers and child counts in the underlying rooted tree.

To compare two keys of $B_c$, we compare their current ranks in $E$. This costs $\cO(\log n)$ time, so a search or a split located by such comparisons takes $\cO(\log^2 n)$ time. Once the relevant handles or split positions are known, the structural balanced-tree operations take $\cO(\log n)$ time. In particular, we do not assume constant-time comparisons of positions in a sequence undergoing subtree moves. The order of a color tree remains valid whenever the relative Euler order of its keys is preserved; absolute ranks need not be stored or updated.

\emph{Query.}~For a token $a$, let $r_c(a)$ be the number of keys of $B_c$ strictly preceding $a$ in $E$. The leaf-color frequency in the subtree of $v$ is
\[
  \freq_v(c)=r_c(x_v)-r_c(e_v).
\]
Indeed, precisely the colored leaf-entry tokens of that subtree lie in this interval, including $e_v$ itself when $v$ is a leaf. Two order-statistic searches compute this difference in $\cO(\log^2 n)$ time. An absent color has frequency zero. Iterating over the nonempty color classes and returning one of maximum frequency therefore takes $\cO(|\colors|\log^2 n)$ time.

\emph{Recoloring.}~To recolor a leaf $w$ from $c$ to $d$, delete $e_w$ from $B_c$ and insert it into $B_d$, creating or removing dictionary records when necessary. The Euler sequence is unchanged, and no other color tree is affected. This takes $\cO(\log^2 n)$ time.

\emph{Add/Delete Leaf.}~To attach a new leaf $w$ to $p$, insert $e_w,x_w$ immediately before $x_p$, choosing $w$ as the last child of $p$, and insert $e_w$ into the search tree for its color. If $p$ was a leaf, first remove $e_p$ from the search tree for its old color: $p$ is now internal and contributes no leaf color. To delete a non-root leaf $w$, remove its color record and its two Euler tokens. If its parent becomes a leaf, insert the parent's entry token into the tree for the color specified by the operation. Each operation changes only constantly many color records. Insertion or deletion of Euler tokens preserves the relative order of all surviving tokens, so all other color trees remain valid without modification. Both operations take $\cO(\log^2 n)$ time, independently of $|\colors|$.

\emph{Link-Cut Update.}~Let $p$ be the old parent of $u$, and let $v$ lie outside the subtree of $u$. If $v=p$, the update changes nothing.
If $v$ is a leaf, it becomes internal, so we first delete $e_v$ from the search tree of its color in $\cO(\log n)$ time.
Let $I=[e_u,x_u]$ be the Euler interval to be moved.
For each currently represented color $c$, use two searches and splits to factor $B_c$ into the keys before $I$, the keys in $I$, and the keys after $I$. Save the middle tree $C_c$ and concatenate the other two trees to obtain $R_c$. Keep the temporary color records until the update is complete, even if $R_c$ is empty.

All these extractions are performed \emph{before changing $E$}. Next, cut $I$ out of $E$ and insert it immediately before $x_v$, making $u$ the last child of $v$. This uses constantly many sequence splits and concatenations and takes $\cO(\log n)$ time. The keys of each $R_c$ retain their relative order, as do the keys of each $C_c$, so both remain valid search trees under the new Euler order. For each $c$, split $R_c$ at $x_v$ in this new order and concatenate its first part, $C_c$, and its second part. Every key of $C_c$ now belongs between those two parts: the moved interval immediately precedes $x_v$, and no key of $R_c$ lies inside that interval. This reconstructs $B_c$ in $\cO(\log^2 n)$ time per color.

Finally update the underlying parent pointers and child counts. If $p$ has become a leaf, insert $e_p$ with the color supplied by the operation; discard any empty color records. These boundary changes cost $\cO(\log^2 n)$ time in total. Thus a link-cut update takes $\cO(|\colors|\log^2 n)$ time. The two-phase procedure is essential: searching unprocessed color trees after moving $I$ could otherwise use an invalid key order.

\emph{Unary-node insertion and removal.}~We will also use a simple extension in the applications below. To subdivide an edge leading to $u$ with a new unary node $z$, insert $e_z$ immediately before $e_u$ and $x_z$ immediately after $x_u$. The new node is internal, and no existing leaf changes status or relative Euler order, so no color tree changes. Removing a unary node is the reverse operation. Each costs $\cO(\log n)$ time. The same argument allows a new unary root to be inserted or removed.

\emph{Space, construction, and worst-case bounds.}~The sequence tree has $2n$ tokens, and the total number of records in all color trees is exactly the number of leaves, at most $n$. The dictionary, underlying-tree records, and temporary trees used during a move also occupy $\cO(n)$ space: splitting and concatenation rearrange existing color records rather than copying them. Hence the total space is $\cO(n)$.

For construction, traverse the initial tree in Euler order, build $E$, and append each leaf-entry token to the list for its color, locating these lists through the balanced color dictionary. This takes $\cO(n\log n)$ time; each list is already sorted in Euler order and can be converted into its balanced search tree in linear time in its length. All stated operation bounds are worst-case when worst-case balanced sequence and search trees are used.
\end{appendixproof}

In Appendix~\ref{sec:1DR}, we show a direct application of \Cref{thm:fdsm_few_colors}
to the Top-1 Document Retrieval problem under document additions.

The same representation directly supports the frequency of a specified color. The two order-statistic searches in the proof of \Cref{thm:fdsm_few_colors} give the following corollary.

\begin{corollary}
  There is a data structure of size $\cO(n)$ that returns the frequency of a specified color $c$ among the leaf descendants of a queried node $v$ in $\cO(\log^2 n)$ worst-case time. Recolorings and leaf insertions/deletions take $\cO(\log^2 n)$ time, while link-cut updates take $\cO(|\colors|\log^2 n)$ time.
\end{corollary}

To compare our two data structures, note that $\log^2 n=o(n^{1-\smooth})$ for every fixed $0<\smooth<1$. 
  
\begin{corollary}
    The data structure of \Cref{thm:fdsm_few_colors} has
  asymptotically faster color updates than the data structure of
  \Cref{thm:dsm_ds} for any fixed $0 <  \smooth < 1$. It also has asymptotically
  faster queries whenever $|\colors|\log^2 n = o(n^{\smooth})$. 
\end{corollary}

\begin{corollary}\label{cor:drm-few-colors}
  Combined with the reduction of \Cref{thm:fdsm-drm-reduction} and the unary-node extension in the proof of \Cref{thm:fdsm_few_colors}, our construction yields a linear-space combinatorial $\DRM$ data structure with query time $\cO(|\colors|\log^2 n)$ and update time $\cO(\log^2 n)$. These update bounds are asymptotically smaller than the $\cO(n^{2/3})$ bound of the state-of-the-art combinatorial \DRM data structure~\cite{DBLP:conf/esa/El-Zein0MS18} for every $\colors$; the query bound is asymptotically smaller if and only if $|\colors|=o(n^{2/3}/\log^2 n)$. 
\end{corollary}

\begin{appendixproof}
  Use the tree representation from \Cref{thm:fdsm-drm-reduction} on $\cO(n)$ nodes. Each range query requires constantly many subtree moves and one subtree-mode query. An entry recoloring requires one leaf recoloring. Inserting or deleting an entry additionally requires constantly many leaf insertions/deletions and insertion or removal of a unary node. The unary-node extension proved above costs $\cO(\log n)$ time and changes no color search tree. The balanced search tree mapping array positions to vertices adds $\cO(\log n)$ time per operation. Hence queries take $\cO(|\colors|\log^2 n)$ time and updates take $\cO(\log^2 n)$ time, in $\cO(n)$ space. The construction uses only comparisons and balanced-tree operations, so it is combinatorial. Finally, $\log^2 n=o(n^{2/3})$, and comparing $|\colors|\log^2 n$ with $n^{2/3}$ gives the stated query condition.
\end{appendixproof}

Per-color search trees are also used for \DRM~\cite[proof of Theorem~2]{DBLP:conf/icalp/SandlundX20}. The data structure of \Cref{thm:fdsm_few_colors} extends this approach to dynamic subtrees by ordering the color trees through a shared Euler sequence. Comparing two positions in that sequence takes $\cO(\log n)$ time instead of $\cO(1)$ time for array positions, which accounts for the additional logarithmic factor.

\section{Experimental Evaluation}\label{sec:experimentw}
\subsection{Datasets and Methods}
We used $5$ real-world, publicly available datasets in our scalability experiments, as well as $2$ sample datasets derived from them; see \Cref{tab:scalability_datasets}. Following~\cite{static-paper}, we derived all datasets except \Amazonlarge~\cite{mcauley2015image} by constructing a generalized suffix tree for each original dataset (a string collection) and coloring each leaf with the id of the string in the collection, to derive large-scale datasets. \Amazonlarge is a tree dataset, representing a product taxonomy, whose product leaves are colored by sales-rank bands; see Appendix~\ref{app:dataset-details} for details.  
In the \FDSM experiments and \Cref{sec:case_studies}, we used a dataset, called \Files{}, which is derived from the \emph{LLVM directory hierarchy}~\cite{llvm-project}. The latter is a tree formed by directories and files in the source-code repository of an open-source compiler infrastructure project~\cite{llvm-project}.

\begin{table}
\centering
\caption{(a) Characteristics of the datasets used in the scalability
experiments. (b) \Files dataset used in the \FDSM experiments and case study.}
\label{tab:scalability_datasets}

\setlength{\tabcolsep}{2pt}
\renewcommand{\arraystretch}{1.05}

\begin{subtable}{\linewidth}
\centering
\vspace{-1mm}
\caption{String-derived instances and \Amazonlarge.}
\label{tab:scalability-datasets}
\shrinktofit{\linewidth}{%
\begin{tabular}{@{}ll
*{2}{S[table-format=9.0]}
S[table-format=4.0]
S[table-format=6.0]@{}}
\toprule
\textbf{Dataset} &
\textbf{Domain} &
${n}$ &
{$\#~\textbf{Leaves}$} &
$h$ &
{$|\colors|$} \\\midrule
\News~\cite{zhou2015pattern}
    & News & 728307 & 478100 & 59 & 4781 \\
\WebKBlarge~\cite{craven1998learning}
    & Web & 35429531 & 26068175 & 167 & 790340 \\
\Geneslarge~\cite{ncbi_handbook_chapter6}
    & Biology & 519275197 & 499580789 & 569 & 887249 \\
\VIRlarge~\cite{ncbi_genome_2025}
    & Biology & 287825267 & 249925868 & 1466 & 143588 \\
\Amazonlarge~\cite{mcauley2015image}
    & E-commerce & 863317 & 859478 & 9 & 3802 \\
\midrule
\Genessample~\cite{ncbi_handbook_chapter6}
    & Biology & 3723300 & 2000000 & 76 & 10000 \\
\WebKBsample~\cite{craven1998learning}
    & Web & 4693596 & 3103000 & 101 & 31030 \\
\bottomrule
\end{tabular}%
}
\end{subtable}
\vspace{3pt}

\begin{subtable}{\linewidth}
\centering
\caption{\Files dataset derived from the LLVM Project repository.}
\label{tab:files_dataset}
\shrinktofit{\linewidth}{%
\begin{tabular}{@{}ll
*{4}{S[table-format=6.0]}
S[table-format=2.0]
S[table-format=3.0]@{}}
\toprule
\textbf{Source} &
\textbf{Snapshot} &
{$\#$~\textbf{Directories}} &
{$\#$~\textbf{Files}} &
{$n$} &
{$\#~\textbf{Leaves}$} &
{$h$} &
{$|\colors|$} \\
\midrule
LLVM Project &
10 Nov.\ 2025 &
15637 &
166441 &
182079 &
166442 &
16 &
399 \\
\bottomrule
\end{tabular}%
}
\end{subtable}
\end{table}

We compare our \DSM data structure to: (I) A baseline based on the reduction to Dynamic Range Mode (\DRM)~\cite{DBLP:conf/esa/El-Zein0MS18}, outlined in the introduction. Specifically, this baseline orders the leaves of $\mathcal{T}$ in the DFS order. This reduces a subtree-mode query to a range-mode query, and a leaf update to a point update query. This baseline uses our implementation of the state-of-the-art practical data structure for \DRM~\cite{DBLP:conf/esa/El-Zein0MS18}, which takes $\cO(n^{4/3})$ time to build, occupies $\cO(n)$ space, and answers queries and updates in $\cO(n^{2/3})$ time. The implementation defaults to a block size of $\lceil m^{2/3}\rceil$, where $m$ is the number of leaves of $\mathcal{T}$. 
(II) \textsc{SCM-rebuild}, which builds from scratch  the static SCM data structure~\cite{static-paper} after each update. This approach was chosen, as it significantly outperformed other baselines for \SM~\cite{static-paper} which are omitted. Also, we omit a baseline that stores at every internal node the frequency of each color occurring below it and updates the counts at all ancestors of a changed leaf, 
due to its $\cO(\#\text{leaves}\cdot h)$ space  and costly updates. For example, on a \WebKBlarge{} subtree with $n=\num{989035}$, $h=157$, and $|\colors|=\num{65536}$, under $Q/U=1/10$, its updates were $4$ and $40$ times slower than \DSM{} and \FDSM{}, respectively, while its space was $90\%$ larger than that of \FDSM{}, and prohibitive for \VIRlarge{}.



\vspace{-2mm}
\subsection{Setup}\label{sec:exp-setup}
To our knowledge, there are no publicly available
query logs for \SM. Thus, for our experiments (except the case studies), we generate synthetic workloads as follows. A query samples an internal node uniformly at random and returns the mode color in its subtree. 
An update samples a structural leaf (i.e., a node with no children in the tested tree)  uniformly at random (with replacement) and replaces its current color $c$ with a color sampled uniformly from $\colors\setminus\{c\}$; hence, no update leaves the color unchanged. 

We use workloads $\mathcal{W}_{Q/U}$ with $Q/U\in\{100,50,25,5,1,1/100\}$, where $U$ and $Q$ are the numbers of color updates and queries in each cycle, respectively. Each cycle performs $U$ color updates followed by $Q$ \SM queries. Let $\mathbf{U}$ and $\mathbf{Q}$ denote the total numbers of updates and queries in a complete workload, respectively. For the small datasets and \Files{}, $\mathbf{U}+\mathbf{Q}$ equals the number of structural leaves in the tested tree. At full scale, this gives \num{3103000}, \num{478100}, \num{2000000}, and \num{166442} operations on \WebKBsample{}, \News{}, \Genessample{}, and \Files{}, respectively. For the  large datasets, the total operation count equals the number of structural leaves divided by $100$ and rounded down, giving \num{260681}, \num{4995807}, \num{2499258}, and \num{8594} operations on \WebKBlarge{}, \Geneslarge{}, \VIRlarge{}, and \Amazonlarge{}, respectively. 
When varying $\alpha$, we also use a query-only workload $\mathcal{W}_{\mathbf{Q}}$ and an update-only workload $\mathcal{W}_{\mathbf{U}}$. To examine the impact of $n$, we use nested instances $25\%$, $50\%$, $75\%$, and $100\%$ of the nodes of the full tree. To examine the impact of $|\colors|$, we varied the color assignment of the full tree.

We report \emph{average operation time} (i.e., the average time over all the operations of a workload), \emph{construction time}, and \emph{space} (i.e., peak memory for construction and answering the workload). 

All experiments were conducted on compute nodes equipped with an AMD EPYC 7702 64-Core Processor (2.00 GHz), 1 TB of RAM, and Ubuntu 22.04.5 LTS.

\emph{Our code and data} can be found at \ifarxiv\url{https://github.com/JialongZhou666/dynamic-subtree-mode}\else\url{https://anonymous.4open.science/r/dynamic-subtree-mode-93C5/}\fi. 

\subsection{\DSM Data Structure}

\subsubsection{Workload-aware Selection of $\alpha$} 

Recall from \Cref{thm:dsm_ds} that $\alpha\in [0,1]$ affects both the \SM query and update time. When a workload contains both \SM queries and updates, $\alpha$ controls the trade-off, and it is thus important to find a value $\alpha_{\mathrm{opt}}$ that minimizes time for answering the workload. In the following, we examine four different methods for finding this value. 

\noindent{\bf Analytic.}
For each dataset and workload $\mathcal{W}_{Q/U}$, we first evaluate
$11$ fixed values for $\alpha$, namely $\{0.0, 0.1, \dots, 1.0\}$. Let $f_Q = Q/(Q+U)$ and
$f_U = U/(Q+U)$ denote the fractions of queries and updates,
respectively. Motivated by the theoretical trade-off, we model the
average mixed-operation time as
\begin{equation}
    T(\alpha)
    =
    f_Q A n^\alpha
    +
    f_U B n^{1-\alpha},
    \label{eq:t-alpha}
\end{equation}
where $n$ is the number of nodes of  $\mathcal{T}$, and $A,B$ are
implementation-dependent constants for  query and update cost, 
respectively. We estimate $A$ and $B$ by least-squares fitting of 
Equation~\eqref{eq:t-alpha} to the $11$ fixed values for $\alpha$.  
Minimizing Equation~\eqref{eq:t-alpha} w.r.t.  $\alpha$ yields
the predictor
    $\alpha_{\mathrm{opt}}^{\mathrm{analytic}}
    =
    \operatorname{clip}_{[0,1]}
    \left[
        \frac{1}{2}
        \left(
            1 +
            \frac{\ln\bigl((f_U B)/(f_Q A)\bigr)}
                 {\ln n}
        \right)
    \right]$, 
where $\operatorname{clip}_{[0,1]}$ projects the value onto
$[0,1]$. \ALPAN decreases when queries
dominate and increases when updates dominate. It is an
approximation used to test how well the model of \Cref{eq:t-alpha} predicts $\alpha_{\mathrm{opt}}$, as it ignores logarithmic factors suppressed by $\Oish(\cdot)$ in the bounds and changes caused by the value of $\alpha$ in 
the data structure configuration.

\begin{table*}[t]
\centering
\caption{Comparison of the $\alpha$ selection methods on different datasets and workloads $\mathcal{W}_{Q/U}$. The constants $A, B$ in the Analytic method are measured in nanoseconds (ns). The Exhaustive-search and Analytic method take the same time, measured in seconds. $\#$ Eval. denotes the number of evaluations performed. The Error of every method is relative to \ALPEX.}
\label{tab:alpha_opt_unified}
\vspace{-2mm}
\footnotesize
\setlength{\tabcolsep}{3pt}
\begin{adjustbox}{max width=\textwidth}
\begin{tabular}{@{}l l S[table-format=1.1] S[table-format=1.3] S[table-format=1.3] S[table-format=1.3] S[table-format=2.1] S[table-format=5.0] S[table-format=1.3] S[table-format=2.1] S[table-format=2.0] S[table-format=3.0] S[table-format=1.3] S[table-format=2.1] S[table-format=2.0] S[table-format=4.0]@{}}
\toprule
& & {\textbf{Exhaust.}} & \multicolumn{4}{c}{\textbf{Analytic}} & {\textbf{Exh.\ \& Ana.}} & \multicolumn{4}{c}{\textbf{Brent's Method}} & \multicolumn{4}{c}{\textbf{Newton's Method}} \\
\cmidrule(lr){4-7} \cmidrule(lr){9-12} \cmidrule(l){13-16}
{\textbf{Dataset}} & {$\bm{Q/U}$} & {${\alpha_{\mathrm{opt}}^{\mathrm{exhaust}}}$} & {$\bm{A}$ \textbf{(ns)}} & {$\bm{B}$ \textbf{(ns)}} & {${\alpha_{\mathrm{opt}}^{\mathrm{analytic}}}$} & {\textbf{Error (\%)}} & {\textbf{Time (s)}} & {${\alpha_{\mathrm{opt}}^{\mathrm{Brent}}}$} & {\textbf{Error (\%)}} & {\textbf{\# Eval.}} & {\textbf{Time (s)}} & {${\alpha_{\mathrm{opt}}^{\mathrm{Newton}}}$} & {\textbf{Error (\%)}} & {\textbf{\# Eval.}} & {\textbf{Time (s)}} \\
\midrule
\WebKBsample         & $100$   & 0.4 & 2.054 & 0.865 & 0.322 & 19.5 & 77760 & 0.369 &  7.7 &  8 &  37 & 0.355 & 11.2 & 15 &   69 \\
($n=\num{4.69e6}$)   & $1$     & 0.5 & 2.292 & 0.880 & 0.469 &  6.2 & 56820 & 0.500 &  0.0 & 11 & 747 & 0.575 & 15.0 & 12 & 1694 \\
                     & $1/100$ & 0.5 & 2.350 & 0.879 & 0.618 & 23.6 & 16620 & 0.562 & 12.4 &  8 & 106 & 0.564 & 12.8 &  9 &   94 \\
\midrule
\News                & $100$   & 0.4 & 0.791 & 0.532 & 0.315 & 21.3 &   900 & 0.323 & 19.2 &  8 &   5 & 0.450 & 12.5 & 15 &   10 \\
($n=\num{7.28e5}$)   & $1$     & 0.4 & 0.993 & 0.585 & 0.480 & 20.1 &   699 & 0.553 & 38.3 & 12 &   9 & 0.550 & 37.5 &  9 &   95 \\
                     & $1/100$ & 1.0 & 1.043 & 0.571 & 0.648 & 35.2 &   301 & 0.784 & 21.6 & 11 &  10 & 0.888 & 11.2 & 18 &   22 \\
\midrule
\Genessample         & $100$   & 0.3 & 0.915 & 0.920 & 0.348 & 16.0 & 22440 & 0.277 &  7.7 &  9 &  27 & 0.450 & 50.0 &  6 &   18 \\
($n=\num{3.72e6}$)   & $1$     & 0.5 & 1.012 & 0.868 & 0.495 &  1.0 & 16500 & 0.560 & 12.1 &  7 & 327 & 0.550 & 10.0 &  6 &   34 \\
                     & $1/100$ & 0.5 & 1.077 & 0.890 & 0.646 & 29.2 &  8700 & 0.601 & 20.3 & 11 &  93 & 0.550 & 10.0 &  9 &   71 \\
\bottomrule
\end{tabular}
\end{adjustbox}
\end{table*}

\noindent{\bf Brent's method.}
We use Brent's derivative-free bounded minimization~\cite{brent1973}
to select $\alpha\in[0,1]$ that minimizes the measured average
operation time. For each dataset and $Q/U$ ratio, we generate
a pilot workload and candidate values
are evaluated on the same pilot workload, containing one operation
per ten structural leaves. The selected value
$\alpha_{\mathrm{opt}}^{\mathrm{Brent}}$ is then used on the full workload.

\noindent{\bf Newton's method.}
We also minimize the measured average operation time using
Newton's method~\cite{nocedal2006,dennis1996}, estimating its
first and second derivatives by finite differences. The method uses the same pilot workload as Brent's method. 
Starting from $\alpha_0=0.5$, we restrict each iterate to $[0,1]$
and terminate when the improvement or step size is below $0.01$,
or after six iterations. The resulting
$\alpha_{\mathrm{opt}}^{\mathrm{Newton}}$ is used on the full workload.

\noindent{\bf Exhaustive-search.}
For each dataset and workload $\mathcal{W}_{Q/U}$, we evaluate the \DSM data
structure at every candidate value $\alpha \in \{0.0, 0.1,$ $\dots, 1.0\}$ 
and record the average
operation time. We then select the value 
\ALPEX that yields the minimum
observed time, and compare it to the
$\alpha$ values selected by the other methods.

\noindent{\bf Comparison of the $\alpha$ selection methods.}
\Cref{tab:alpha_opt_unified} reports the fitted constants $A,B$,
the analytic prediction \ALPAN, and the exhaustive-grid reference \ALPEX.
Analytic models the trade-off in \Cref{thm:dsm_ds}, achieving
$19.1\%$ average relative error, but requires the same $11$-point
sweep as Exhaustive-search, costing up to $21.6$ hours.
Brent's and Newton's reduce calibration to $5$--$747$\,s and
$10$--$1694$\,s, respectively. Brent's achieves lower average
($15.5\%$ versus $18.9\%$) and maximum ($38.3\%$ versus $50.0\%$)
errors, making it our default heuristic.
These errors measure agreement with the best tested grid value
of $\alpha$, rather than runtime optimality.

\begin{figure}[htbp]
\centering
\captionsetup{skip=3pt}
\captionsetup[subfigure]{skip=0pt}

\begin{subfigure}[b]{0.32\columnwidth}
    \includegraphics[width=\linewidth,trim={0 4mm 0 0},clip]{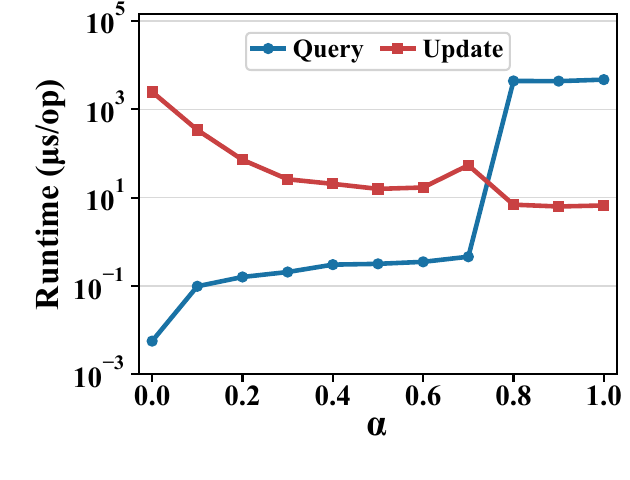}
    \caption{\WebKBsample}
    \label{fig:alpha_webkb_query_time}
    \label{fig:alpha_webkb_update_time}
\end{subfigure}\hfill
\begin{subfigure}[b]{0.32\columnwidth}
    \includegraphics[width=\linewidth,trim={0 4mm 0 0},clip]{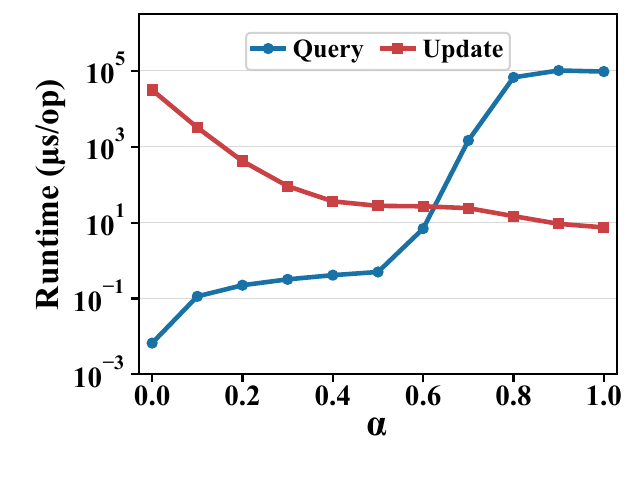}
    \caption{\WebKBlarge}
    \label{fig:webkb_large_alpha_complexity}
\end{subfigure}\hfill
\begin{subfigure}[b]{0.32\columnwidth}
    \includegraphics[width=\linewidth,trim={0 4mm 0 0},clip]{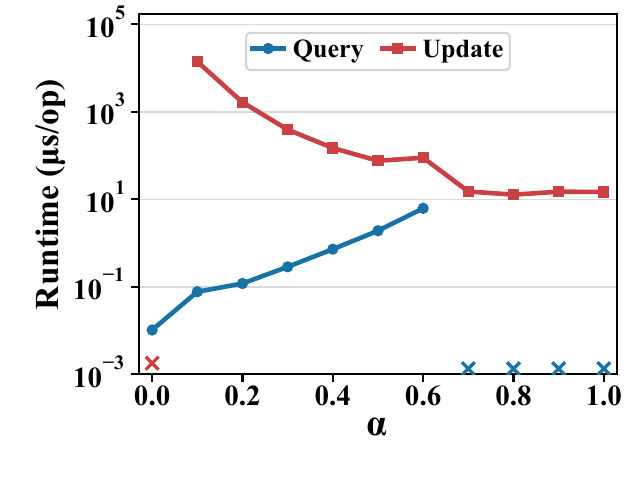}
    \caption{\Geneslarge}
    \label{fig:genes_large_alpha_complexity}
\end{subfigure}
\par\vspace{2pt}
\begin{subfigure}[b]{0.32\columnwidth}
    \includegraphics[width=\linewidth,trim={0 4mm 0 0},clip]{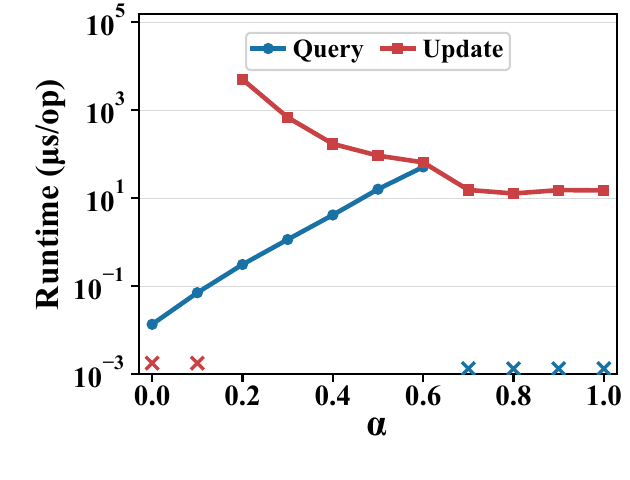}
    \caption{\VIRlarge}
    \label{fig:vir_large_alpha_complexity}
\end{subfigure}\hfill
\begin{subfigure}[b]{0.32\columnwidth}
    \includegraphics[width=\linewidth,trim={0 4mm 0 0},clip]{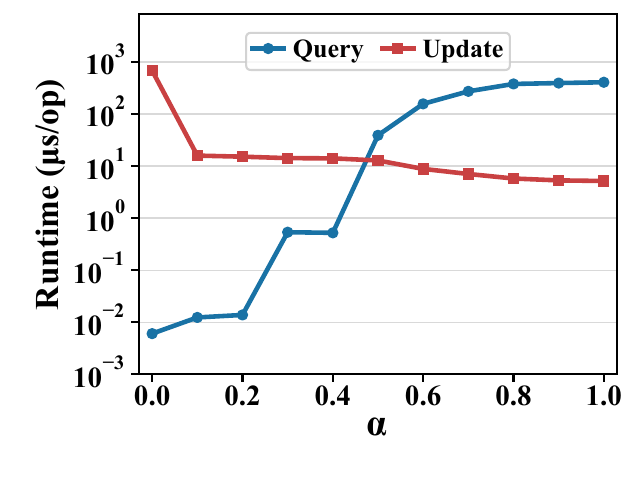}
    \caption{\Amazonlarge}
    \label{fig:amazon_salesrank_alpha_complexity}
\end{subfigure}\hfill
\begin{subfigure}[b]{0.32\columnwidth}
    \includegraphics[width=\linewidth,trim={0 4mm 0 0},clip]{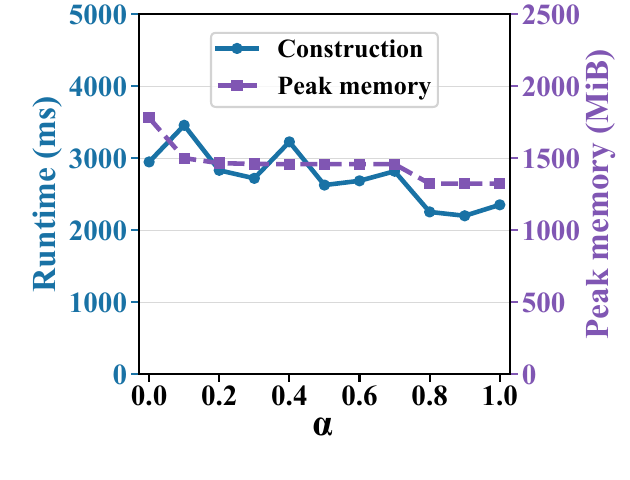}
    \caption{\WebKBsample}
    \label{fig:webkb_alpha_construction_time}
    \label{fig:webkb_alpha_construction_peak_rss}
\end{subfigure}
\caption{Impact of $\alpha$: (a)--(e) average operation time for $\mathcal{W}_\mathbf{Q}$ and $\mathcal{W}_\mathbf{U}$; (f) construction time and space on \WebKBsample. Crosses denote $24$-hour time-outs.}
\label{fig:small_alpha_sweep}
\label{fig:large_alpha_complexity}
\end{figure}

\begin{figure}[htbp]
\renewcommand{\sweepplot}[3]{%
\begin{subfigure}[t]{0.3286\columnwidth}
    \includegraphics[
        width=\linewidth,
        trim={0 4mm 0 0},
        clip
    ]{figs/#1}
    \caption{{\footnotesize #2}}
    \label{#3}
\end{subfigure}%
}
\centering
\captionsetup{skip=3pt}
\captionsetup[subfigure]{skip=0pt}

\sweepplot{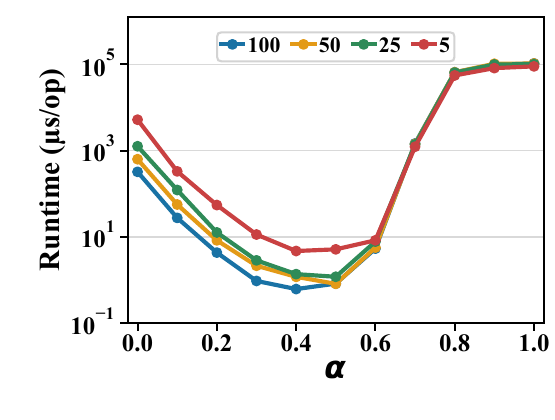}{\WebKBlarge}{fig:webkb_large_alpha_time}\hfill
\sweepplot{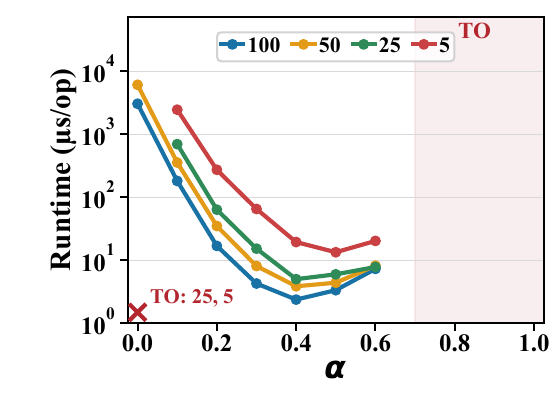}{\Geneslarge}{fig:genes_large_alpha_time}\hfill
\sweepplot{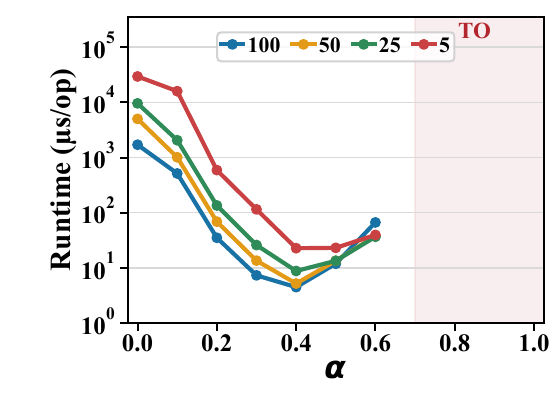}{\VIRlarge}{fig:vir_large_alpha_time}

\par\vspace{2pt}

\sweepplot{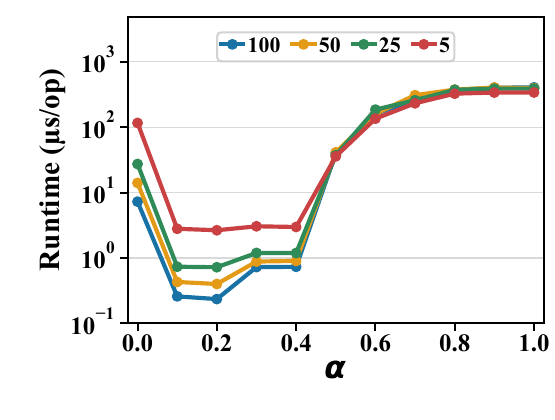}{\Amazonlarge}{fig:amazon_salesrank_alpha_time}\hfill
\sweepplot{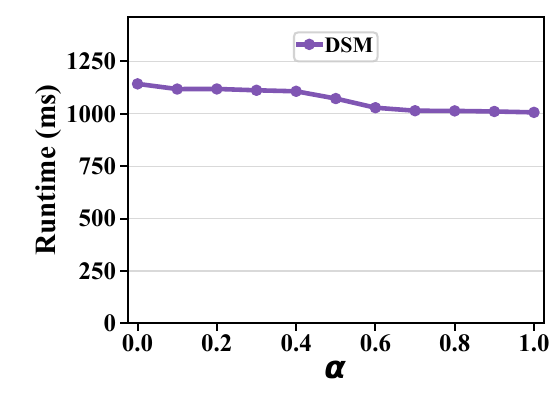}{Construction time}{fig:amazon_salesrank_alpha_buildtime}\hfill
\sweepplot{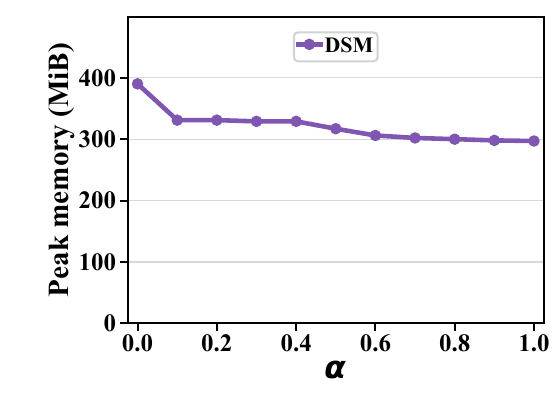}{Space}{fig:amazon_salesrank_alpha_space}

\caption{Impact of $\alpha$: (a)--(d) average operation runtime for $\mathcal{W}_{100}$, $\mathcal{W}_{50}$, $\mathcal{W}_{25}$, and $\mathcal{W}_{5}$; (e)--(f) construction runtime and space on \Amazonlarge. TO denotes a $24$-hour time-out.}
\label{fig:large_mixed_alpha_sweep}
\vspace{-2mm}
\end{figure}

\vspace{-2mm}
\subsubsection{Impact of $\alpha$.}
\Crefrange{fig:alpha_webkb_query_time}{fig:amazon_salesrank_alpha_complexity} show the impact of $\alpha$ on average query time and update time in the query-only workload $\mathcal{W}_{\mathbf{Q}}$ and the update-only workload $\mathcal{W}_{\mathbf{U}}$. The query time increases with $\alpha$ in line with the $\widetilde{\mathcal{O}}(n^{\alpha})$ bound of \Cref{thm:dsm_ds}, whereas the update time decreases with $\alpha$ in line with the $\widetilde{\mathcal{O}}(n^{1-\alpha})$ bound of \Cref{thm:dsm_ds}. \Cref{fig:webkb_alpha_construction_time} shows that $\alpha$ has a small impact on construction time and space on \WebKBsample, as expected by \Cref{thm:dsm_ds}. 

\Crefrange{fig:webkb_large_alpha_time}{fig:amazon_salesrank_alpha_time} show the impact of $\alpha$ on average operation time (over all query and update operations) under different workloads: $\mathcal{W}_{100}$, $\mathcal{W}_{50}$, $\mathcal{W}_{25}$, and $\mathcal{W}_{5}$. The value of $\alpha$ that minimizes the average operation time is $\alpha=0.2$ on \Amazonlarge and 
$0.4$ on all other datasets. \Cref{fig:amazon_salesrank_alpha_buildtime,fig:amazon_salesrank_alpha_space} show the impact of $\alpha$ on construction time and space. Again, the impact of $\alpha$ is very small, in line with \Cref{thm:dsm_ds}. Results for \News{} and
\Genessample{} are in Appendix~\ref{app:news_genes_additional}.

\providecommand{\nsweepmergedplot}[5]{%
\begin{subfigure}[t]{0.155\textwidth}
    \centering
    \includegraphics[width=\linewidth]{figs/#1}
    \caption{#2}
    \label{#5}
\end{subfigure}%
}
\begin{figure*}[t]
\centering
\captionsetup{skip=3pt}
\captionsetup[subfigure]{skip=0pt}

\renewcommand{\thesubfigure}{(\alphalph{\value{subfigure}})}
\captionsetup[subfigure]{labelformat=simple,font={footnotesize,bf},labelfont=bf,
    justification=centering,singlelinecheck=false,skip=2pt}
\nsweepmergedplot{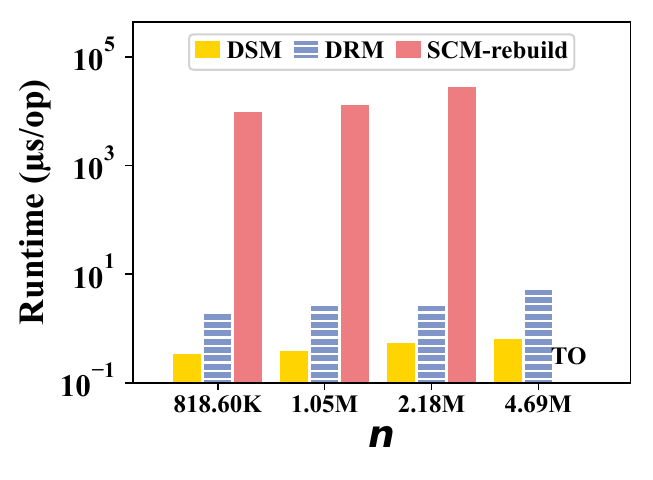}{\WebKBsample{}}{Avg. operation time}{100}{fig:webkb_n_q100_time}\hfill
\nsweepmergedplot{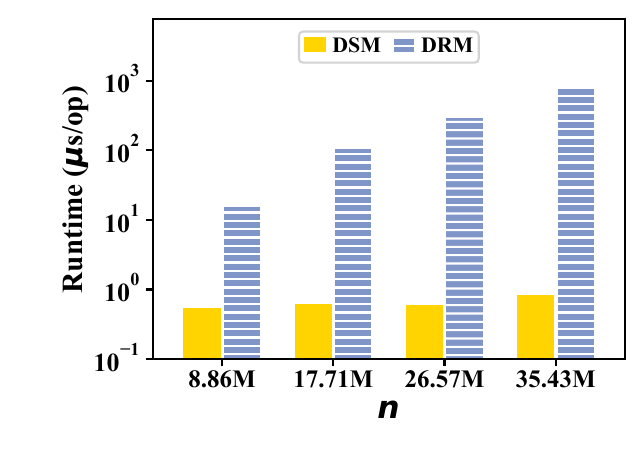}{\WebKBlarge{}}{Avg. operation time}{100}{fig:webkb_large_n_time}\hfill
\nsweepmergedplot{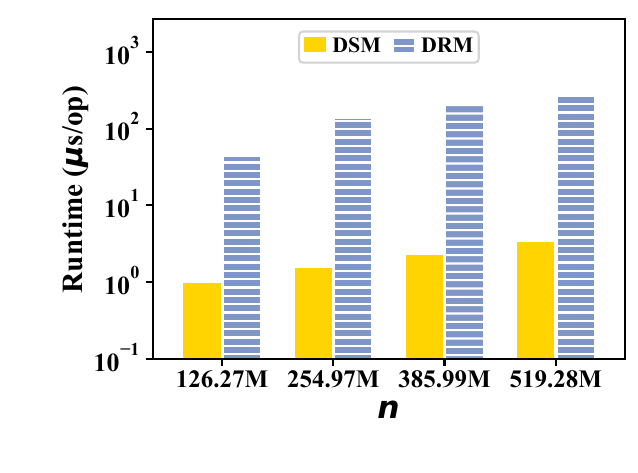}{\Geneslarge{}}{Avg. operation time}{100}{fig:genes_large_n_time}\hfill
\nsweepmergedplot{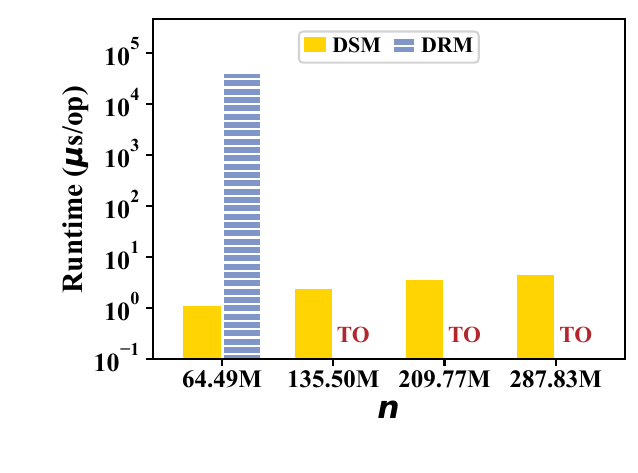}{\VIRlarge{}}{Avg. operation time}{100}{fig:vir_large_n_time}\hfill
\nsweepmergedplot{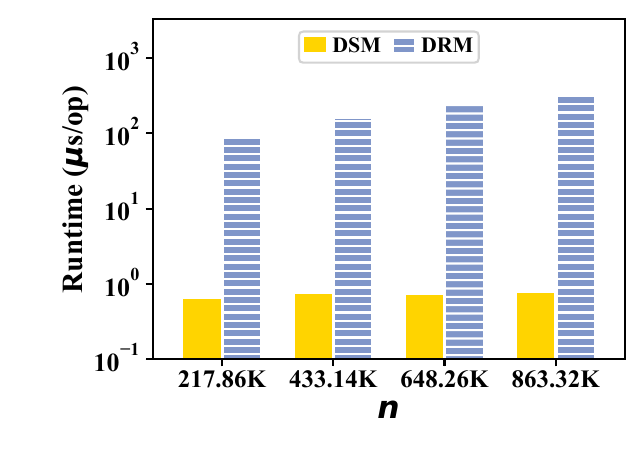}{\Amazonlarge{}}{Avg. operation time}{100}{fig:amazon_salesrank_n_time}\hfill
\nsweepmergedplot{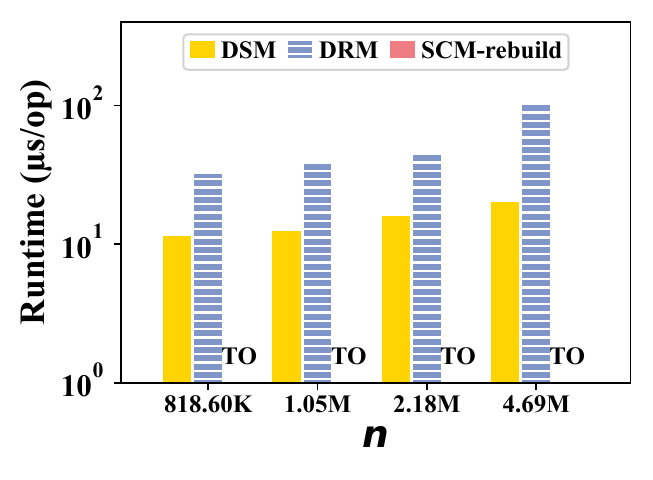}{\WebKBsample{}}{Avg. operation time}{1/100}{fig:webkb_n_q001_time}
\par\medskip

\nsweepmergedplot{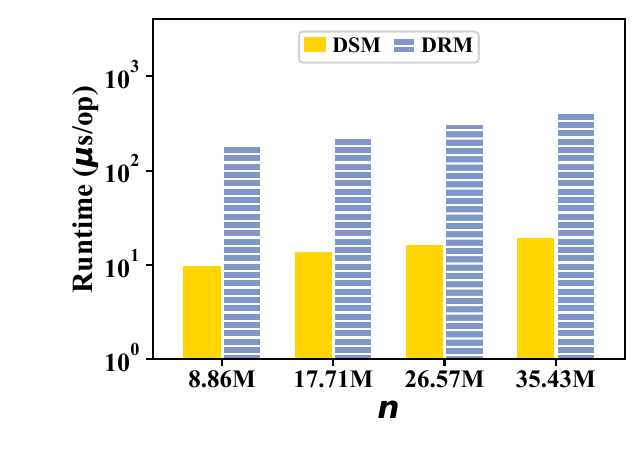}{\WebKBlarge{}}{Avg. operation time}{1/100}{fig:webkb_large_n_q001_time}\hfill
\nsweepmergedplot{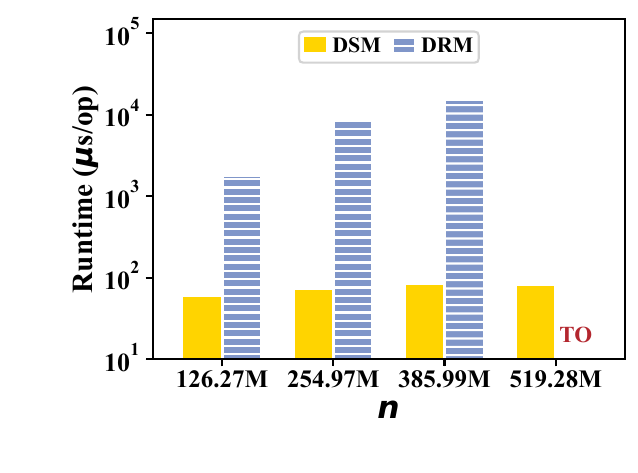}{\Geneslarge{}}{Avg. operation time}{1/100}{fig:genes_large_n_q001_time}\hfill
\nsweepmergedplot{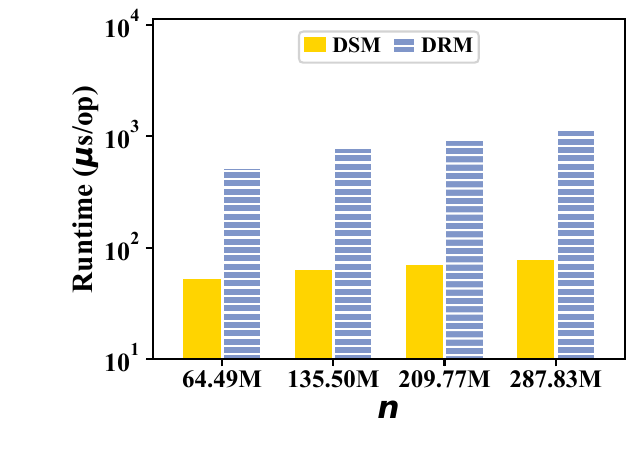}{\VIRlarge{}}{Avg. operation time}{1/100}{fig:vir_large_n_q001_time}\hfill
\nsweepmergedplot{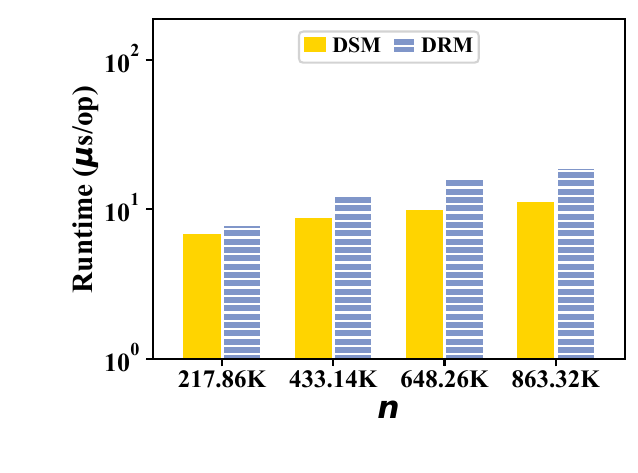}{\Amazonlarge{}}{Avg. operation time}{1/100}{fig:amazon_salesrank_n_q001_time}\hfill
\nsweepmergedplot{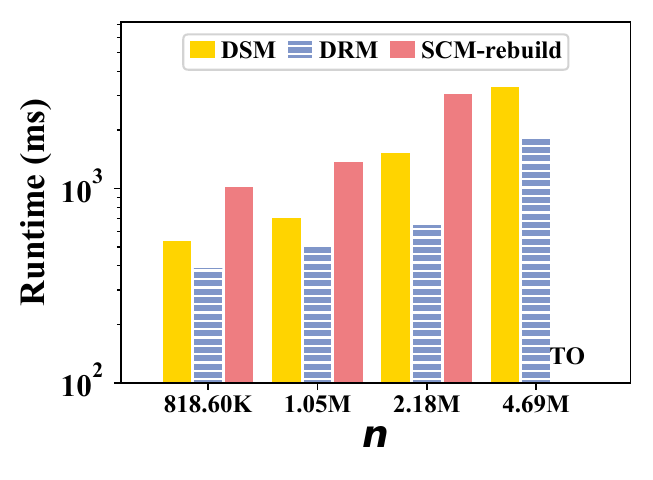}{\WebKBsample{}}{Construction time}{100}{fig:webkb_n_q100_buildtime}\hfill
\nsweepmergedplot{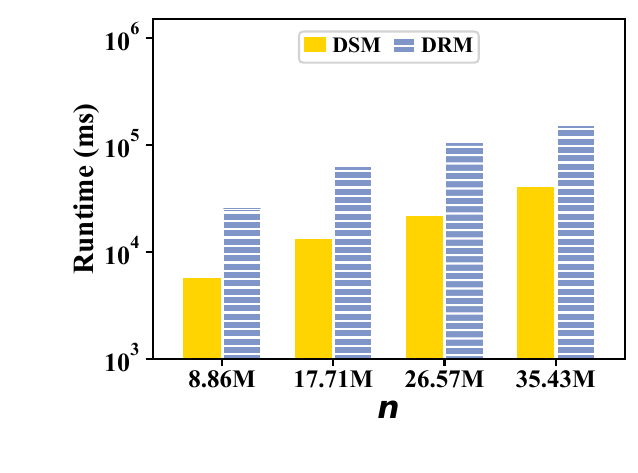}{\WebKBlarge{}}{Construction time}{100}{fig:webkb_large_n_buildtime}
\par\medskip

\nsweepmergedplot{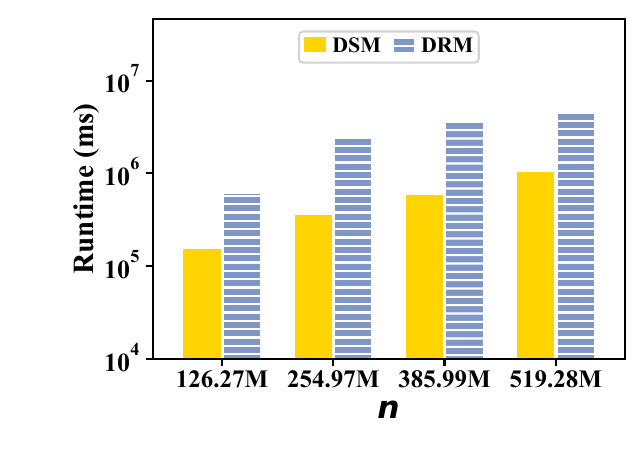}{\Geneslarge{}}{Construction time}{100}{fig:genes_large_n_buildtime}\hfill
\nsweepmergedplot{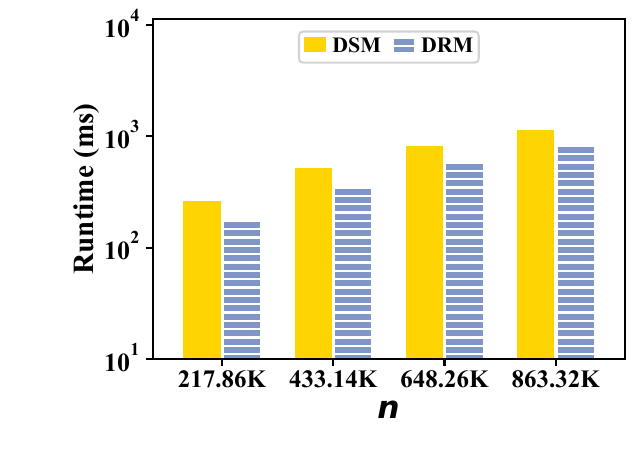}{\Amazonlarge{}}{Construction time}{100}{fig:amazon_salesrank_n_buildtime}\hfill
\nsweepmergedplot{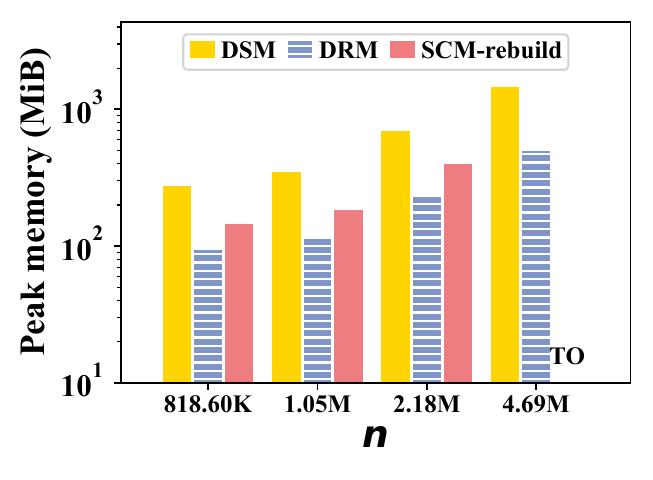}{\WebKBsample{}}{Space}{100}{fig:webkb_n_q100_space}\hfill
\nsweepmergedplot{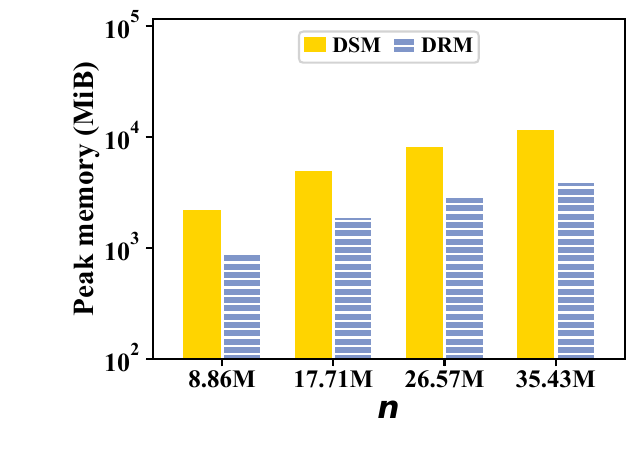}{\WebKBlarge{}}{Space}{100}{fig:webkb_large_n_space}\hfill
\nsweepmergedplot{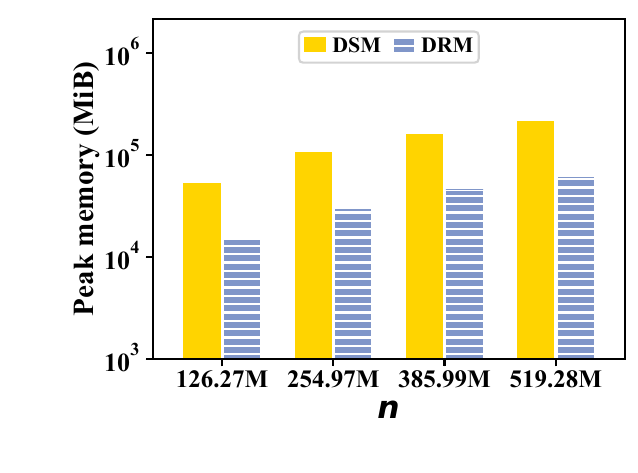}{\Geneslarge{}}{Space}{100}{fig:genes_large_n_space}\hfill
\nsweepmergedplot{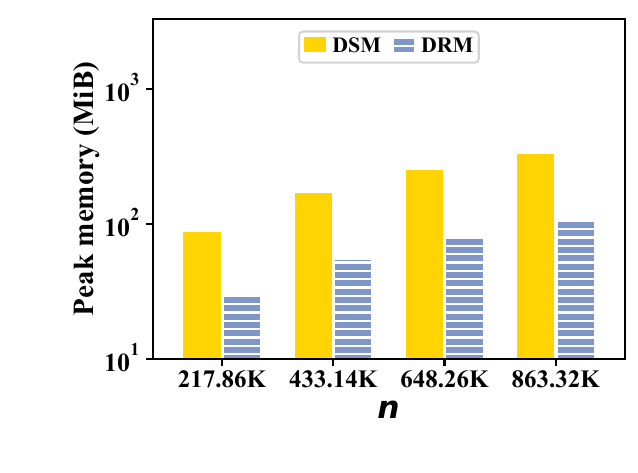}{\Amazonlarge{}}{Space}{100}{fig:amazon_salesrank_n_space}
\caption{Impact of $n$. (a)--(e) report average operation time on
$\mathcal{W}_{100}$ and (f)--(j) on $\mathcal{W}_{1/100}$.
(k)--(n) report construction time and (o)--(r) construction space
on $\mathcal{W}_{100}$. TO indicates a $24$-hour run time-out.}
\label{fig:n_sweep_merged}
\Description{Eighteen charts in three rows of six.
The first ten show average operation time on five datasets under
two workloads. Four construction-time and four space charts
show WebKB-sam, WebKB, Genes, and Amazon under the query-heavy workload.}
\end{figure*}

\vspace{-2mm}
\subsubsection{Impact of $n$.} 
We show that \DSM  is \emph{up to $4$ orders of
magnitude faster} than the competitors and 
\emph{substantially faster to construct on large datasets}. Its space grows linearly with $n$ and is about $3$ times larger than \DRM{}. Thus, \emph{\DSM{} offers an attractive time/space trade-off}.  Appendixes~\ref{app:news_genes_additional} and ~\ref{app:n_sweep_additional} contain additional results. 

\Crefrange{fig:webkb_n_q100_time}{fig:amazon_salesrank_n_time} show that our approach is \emph{up to four orders of magnitude faster} than both \textsc{SCM-rebuild} and \DRM{} under the query-heavy workload $\mathcal{W}_{100}$. Even on the smaller \WebKBsample{} dataset (see \Cref{fig:webkb_n_q100_time}), \textsc{SCM-rebuild} is impractical, requiring more than $24$ hours for $n\geq 4.69\cdot 10^6$. This motivates the need for dynamic data structures for Subtree Mode. \Crefrange{fig:webkb_n_q100_time}{fig:amazon_salesrank_n_time} also show the  
significant benefit of our data structure that exploits the tree structure of the \DSM problem over solutions via the more general \DRM problem. Specifically, \DSM is \emph{up to four orders of magnitude} faster than \DRM. The increasing operation times with $n$ for \DSM are consistent with \Cref{eq:t-alpha}.  

\Crefrange{fig:webkb_n_q001_time}{fig:amazon_salesrank_n_q001_time} show that our approach is \emph{up to $185$} times faster than \DRM{}, even when updates dominate, as in $\mathcal{W}_{1/100}$. It is faster on the larger \Geneslarge{} and \VIRlarge{} datasets; by up to $70\%$ faster on \Amazonlarge{} and up to $185$ times on \Geneslarge{}. On the other hand, \textsc{SCM-rebuild} is again impractical, even on the smaller \WebKBsample{} dataset, where it timed-out after $24$ hours (see \Cref{fig:webkb_n_q001_time}).  

\Crefrange{fig:webkb_n_q100_buildtime}{fig:amazon_salesrank_n_buildtime} 
show the impact of $n$ on construction time under
$\mathcal{W}_{100}$ (the results for $ \mathcal{W}_{1/100}$ are analogous; see Appendix~\ref{app:n_sweep_additional}). On the small \WebKBsample{} dataset, 
\DSM{} is on average about \emph{two times faster} to construct than  
\textsc{SCM-rebuild}.
On the large \WebKBlarge{} and \Geneslarge{} datasets, \DSM{} outperforms \DRM{} by $4.3$ and 
$5.6$ times on average, respectively.  Importantly, the advantage of \DSM{}
over \DRM{} increases with $n$. These results are consistent with the construction-time bounds of $\cO(n)$ for \DSM{} and $\cO(n^{4/3})$ for \DRM{}. \textsc{SCM-rebuild} did not terminate within $24$ hours in the large datasets.     

\Crefrange{fig:webkb_n_q100_space}{fig:genes_large_n_space}
show the impact of $n$ on space under
$\mathcal{W}_{100}$ (the results for $\mathcal{W}_{1/100}$
are analogous; see Appendix~\ref{app:n_sweep_additional}). As expected, the space of all approaches scales linearly with $n$. 
On the small \WebKBsample{} dataset, \DRM{} and
\textsc{SCM-rebuild} use on average $66\%$ and 
$45\%$ less space than \DSM{},
respectively. On the large datasets, \DSM{} uses a reasonable amount of space, which is approximately three times more than that of \DRM{}.

\subsubsection{Impact of $|\colors|$.}  \DSM{} is often \emph{orders of magnitude faster} than both competitors and \emph{substantially faster to construct on large datasets}. Its space is insensitive to $|\colors|$, as expected, but it is more than the competitors. Thus, our approach offers a desirable time/space trade-off. Additional results are in Appendixes~\ref{app:news_genes_additional} and~\ref{app:delta_sweep_build_space}.

\Crefrange{fig:webkb_delta_q100_time}{fig:amazon_salesrank_delta_time}
show that \DSM{} is substantially faster than both competitors 
under the query-heavy workload $\mathcal{W}_{100}$ and \Crefrange{fig:webkb_delta_q001_time}{fig:amazon_salesrank_delta_q001_time} that it remains faster than \DRM{} under $\mathcal{W}_{1/100}$ (e.g., by up to $25.4$ times on the string-derived datasets and up to $80\%$  on \Amazonlarge{}). On the small \WebKBsample{} dataset, \DSM{} is on average
\emph{more than four orders of magnitude faster} 
than \textsc{SCM-rebuild} and $4.7$ times faster than \DRM{}. On \WebKBlarge{} and \Amazonlarge{}, \DSM{} is \emph{up to $600$ times} faster than \DRM{}.  On the largest \Geneslarge{} and \VIRlarge{} datasets, 
\DRM{} \emph{does not terminate within $24$ hours} 
for any $|\colors|$. 

\providecommand{\deltasweepmergedplot}[5]{%
\begin{subfigure}[t]{0.155\textwidth}
    \centering
    \includegraphics[width=\linewidth]{figs/#1}
    \caption{#2}
    \label{#5}
\end{subfigure}%
}
\begin{figure*}[t]
\centering
\captionsetup{skip=3pt}
\captionsetup[subfigure]{skip=0pt}
\setlength{\parskip}{0pt}
\renewcommand{\thesubfigure}{(\alphalph{\value{subfigure}})}
\captionsetup[subfigure]{labelformat=simple,font={footnotesize,bf},labelfont=bf,
    justification=centering,singlelinecheck=false,skip=2pt}

\deltasweepmergedplot{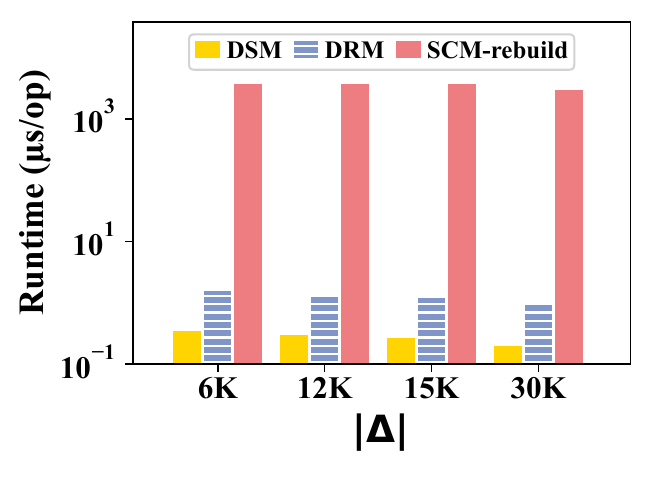}{\WebKBsample{}}{Avg. operation time}{100}{fig:webkb_delta_q100_time}\hfill
\deltasweepmergedplot{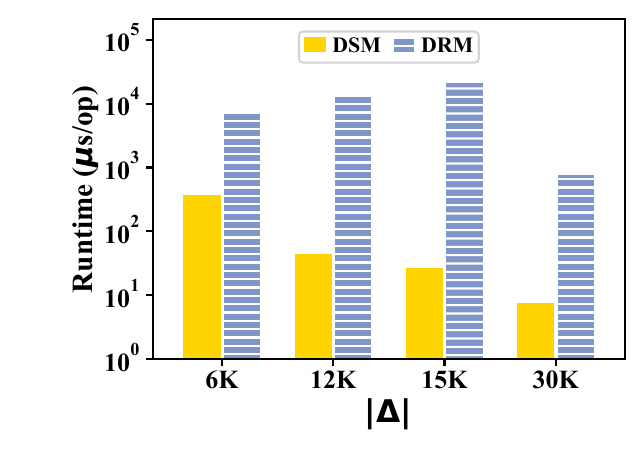}{\WebKBlarge{}}{Avg. operation time}{100}{fig:webkb_large_delta_time}\hfill
\deltasweepmergedplot{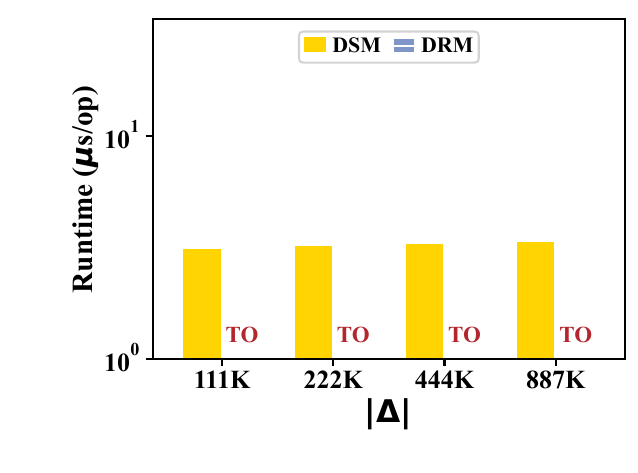}{\Geneslarge{}}{Avg. operation time}{100}{fig:genes_large_delta_time}\hfill
\deltasweepmergedplot{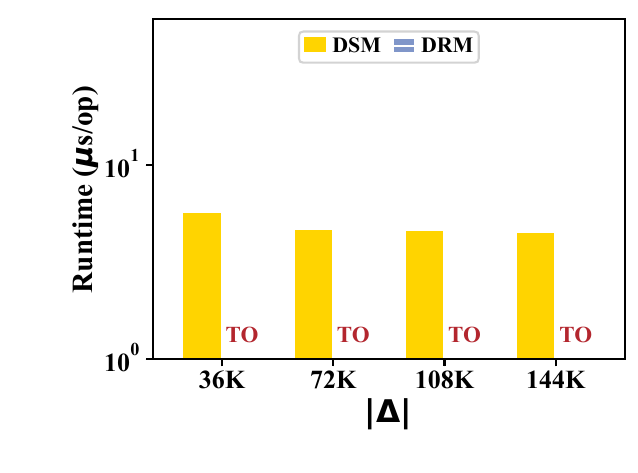}{\VIRlarge{}}{Avg. operation time}{100}{fig:vir_large_delta_time}\hfill
\deltasweepmergedplot{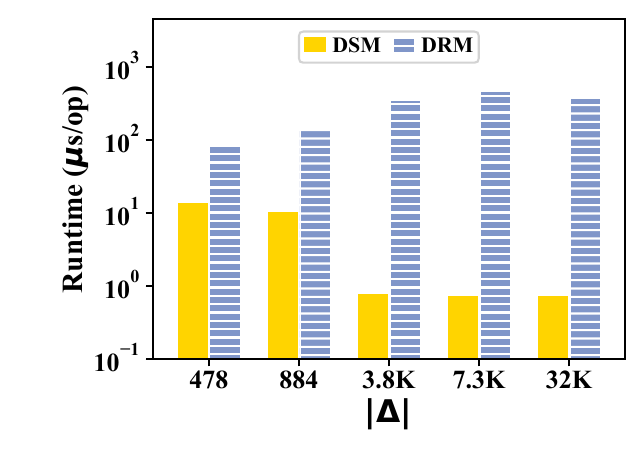}{\Amazonlarge{}}{Avg. operation time}{100}{fig:amazon_salesrank_delta_time}\hfill
\deltasweepmergedplot{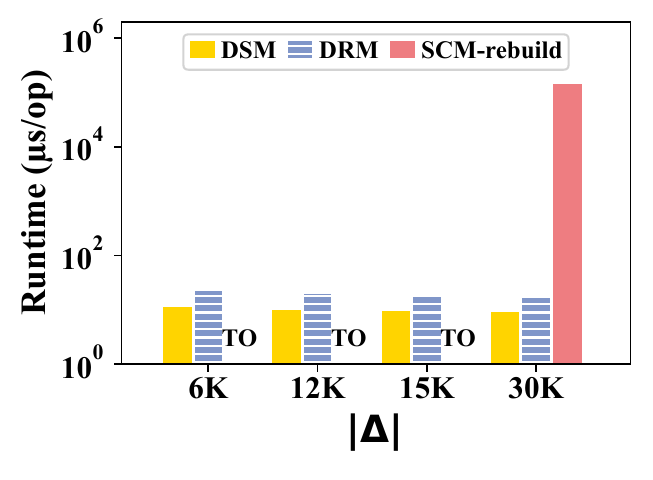}{\WebKBsample{}}{Avg. operation time}{1/100}{fig:webkb_delta_q001_time}
\par\medskip

\deltasweepmergedplot{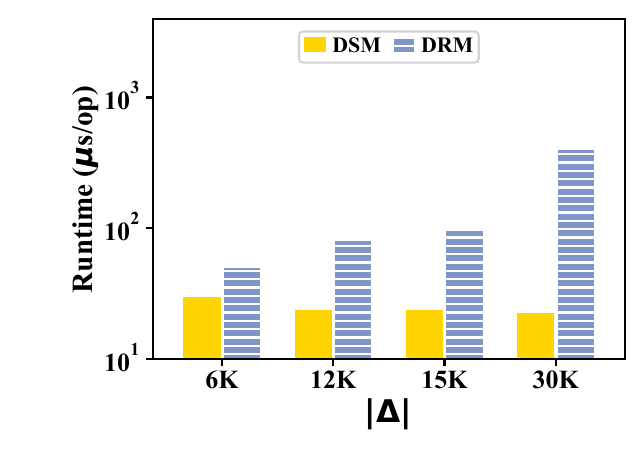}{\WebKBlarge{}}{Avg. operation time}{1/100}{fig:webkb_large_delta_q001_time}\hfill
\deltasweepmergedplot{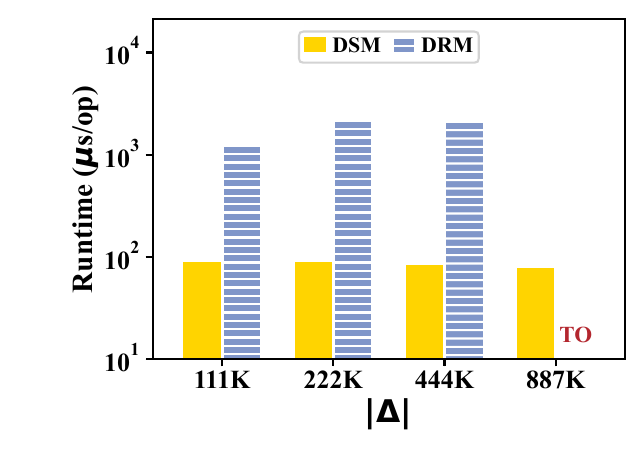}{\Geneslarge{}}{Avg. operation time}{1/100}{fig:genes_large_delta_q001_time}\hfill
\deltasweepmergedplot{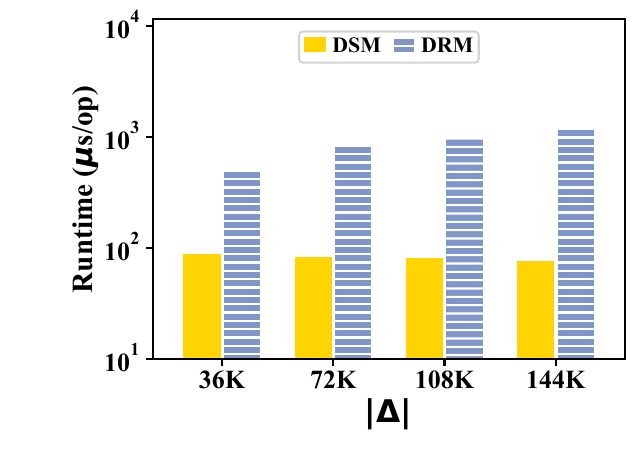}{\VIRlarge{}}{Avg. operation time}{1/100}{fig:vir_large_delta_q001_time}\hfill
\deltasweepmergedplot{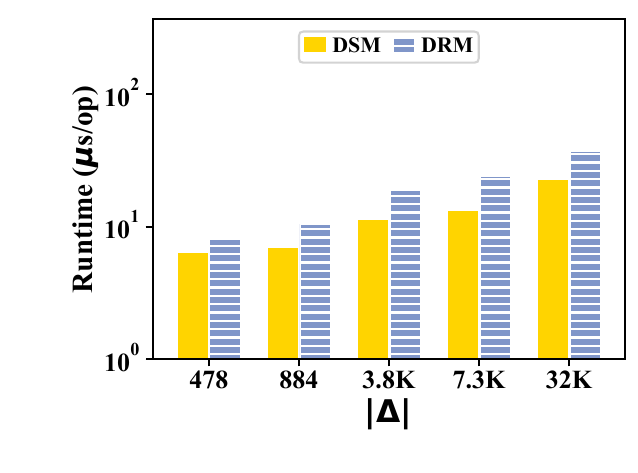}{\Amazonlarge{}}{Avg. operation time}{1/100}{fig:amazon_salesrank_delta_q001_time}\hfill
\deltasweepmergedplot{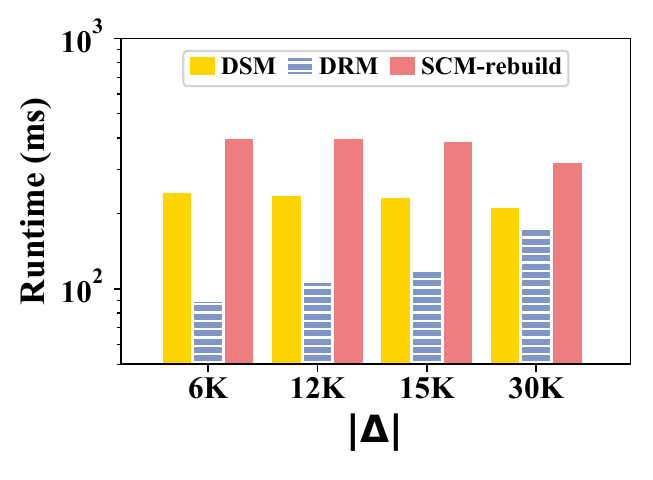}{\WebKBsample{}}{Construction time}{100}{fig:webkb_delta_q100_buildtime}\hfill
\deltasweepmergedplot{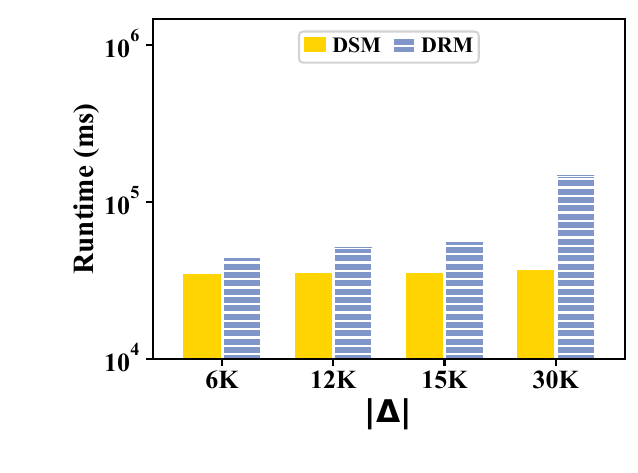}{\WebKBlarge{}}{Construction time}{100}{fig:webkb_large_delta_buildtime}
\par\medskip

\deltasweepmergedplot{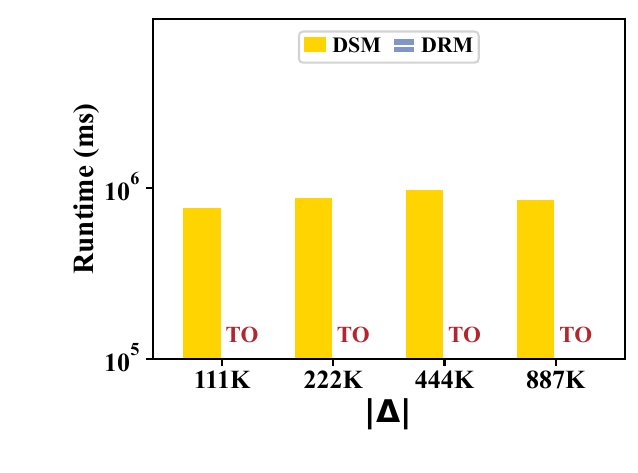}{\Geneslarge{}}{Construction time}{100}{fig:genes_large_delta_buildtime}\hfill
\deltasweepmergedplot{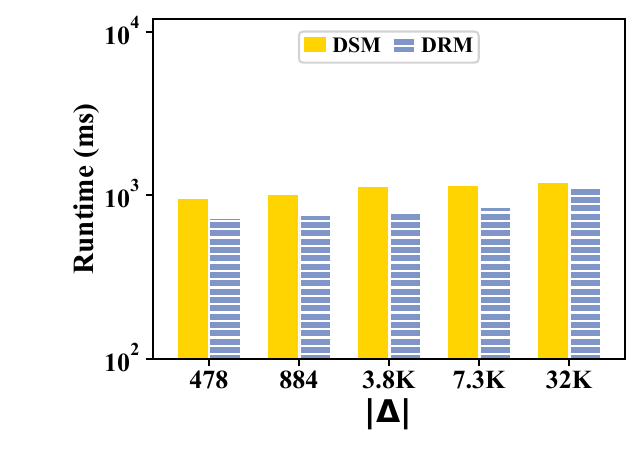}{\Amazonlarge{}}{Construction time}{100}{fig:amazon_salesrank_delta_buildtime}\hfill
\deltasweepmergedplot{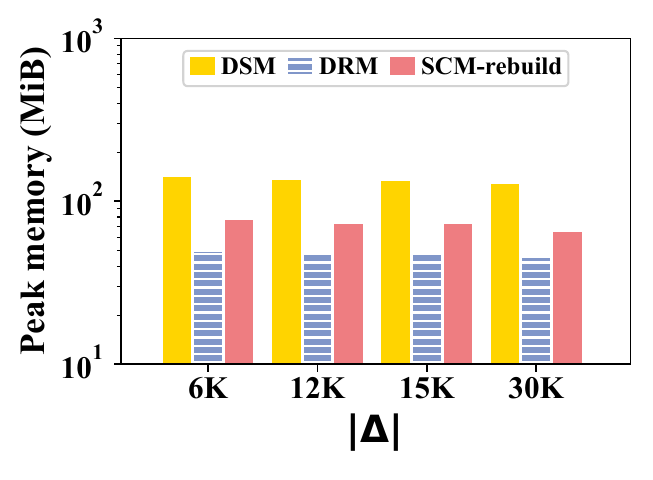}{\WebKBsample{}}{Space}{100}{fig:webkb_delta_q100_space}\hfill
\deltasweepmergedplot{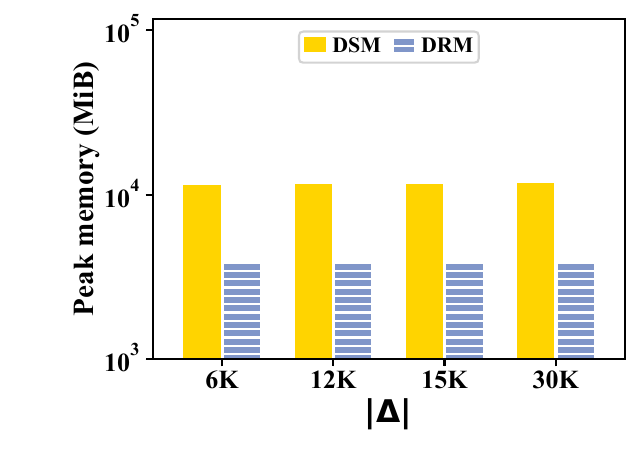}{\WebKBlarge{}}{Space}{100}{fig:webkb_large_delta_space}\hfill
\deltasweepmergedplot{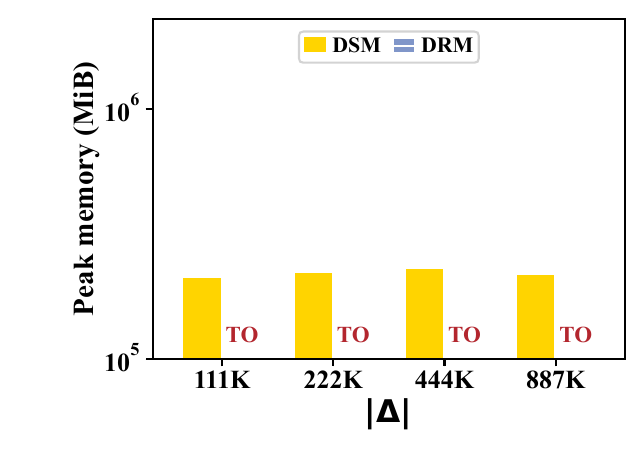}{\Geneslarge{}}{Space}{100}{fig:genes_large_delta_space}\hfill
\deltasweepmergedplot{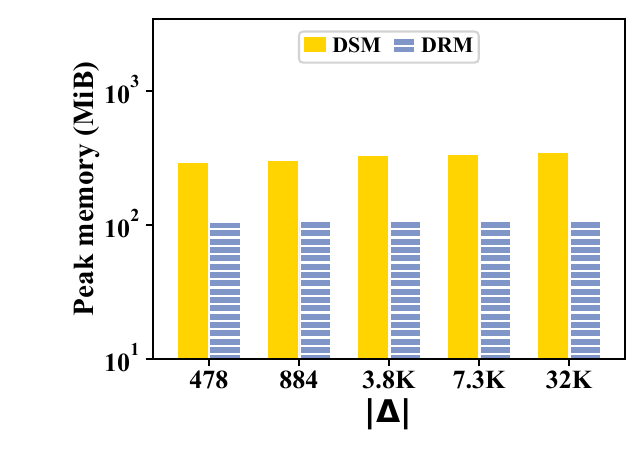}{\Amazonlarge{}}{Space}{100}{fig:amazon_salesrank_delta_space}

\caption{Impact of $|\colors|$. (a)--(e) report average operation time on
$\mathcal{W}_{100}$ and (f)--(j) on $\mathcal{W}_{1/100}$.
(k)--(n) report construction time and (o)--(r) construction space
on $\mathcal{W}_{100}$. TO indicates a $24$-hour run time-out.}
\label{fig:delta_sweep_merged}
\Description{Eighteen charts in three rows of six
compare DSM and the baselines as color count varies on fixed trees.
The first ten show average operation time on five datasets under
two workloads. Four construction-time and four space charts
show WebKB-sam, WebKB, Genes, and Amazon under the query-heavy workload.}
\end{figure*}

\Crefrange{fig:webkb_delta_q100_buildtime}{fig:amazon_salesrank_delta_buildtime} 
show the impact of $|\colors|$ on construction time under
$\mathcal{W}_{100}$. On the small \WebKBsample{} dataset,  
\DSM{} is $64\%$ faster on average to construct than 
\textsc{SCM-rebuild} but it is $2$ times slower on average than \DRM{} (and $31\%$ slower than \DRM on \Amazonlarge). On the larger datasets, however, \DSM{} significantly outperforms \DRM{};  it was $2$ times faster on average on \WebKBlarge{},  
while \DRM \emph{did not terminate within $24$ hours} in \Geneslarge{}. These results are in line with the construction time bounds, $\cO(n)$ for \DSM vs. $\cO(n^{4/3})$ for \DRM. 

\Crefrange{fig:webkb_delta_q100_space}{fig:genes_large_delta_space}
show the impact of $|\colors|$ on space under
$\mathcal{W}_{100}$, which is small as expected by the space complexities of the approaches. On \WebKBsample{}, \DRM{} and 
\textsc{SCM-rebuild} use on average $64.5\%$ and
$46.2\%$ less space than \DSM{}, respectively. 
On \WebKBlarge{}, \DSM{} uses about three times more space than that of \DRM.

\vspace{-4mm}
\subsection{\FDSM Data Structure}
We report results for \Files; additional results are in
Appendix~\ref{app:fdsm_webkb_additional}.

\vspace{+2mm}
\subsubsection{\SM Queries and Updates.}~We evaluate \FDSM under the workload $\mathcal{W}_{100}$, which only has subtree mode queries and updates. This is to compare it against \DSM and \DRM, which do not support add/delete leaf and link-cut update operations, in addition to \textsc{SCM-rebuild}. For the experiments using \Files, we assign colors to tree  leaves so that each color occurs
approximately equally often in the leaves. Starting with $|\colors|=100$, we merge colors to  
obtain $|\colors|\in\{5,10,25,50,100\}$ and use these values in the experiments varying $\colors$. In the experiments varying $n$, we use $|\colors|=50$. 

\vspace{+1mm}
\noindent{\bf Impact of $n$.}~\Cref{fig:few_colors_n_q100_time} shows that
\FDSM is \emph{more than $4$ times} (respectively, \emph{over two orders of magnitude}) faster than \DRM (respectively, \textsc{SCM-rebuild}) on average under $\mathcal{W}_{100}$. The operation time of
\FDSM increases only slightly with $n$, consistent
with the theoretical bounds of
\Cref{thm:fdsm_few_colors}. \DSM is approximately $5$ times
faster than \FDSM, as it is specifically designed for only query and update operations (unlike \FDSM). 

\Cref{fig:few_colors_n_q100_construction_time} shows
the impact of $n$ on construction time under
$\mathcal{W}_{100}$. Our \FDSM data structure is
on average about \emph{two times faster} to construct
than \DSM and \emph{more than four times faster}
than \textsc{SCM-rebuild}. However, \DRM is about
$1.4$ times faster to construct than \FDSM on average.
The construction time of all methods increases in line with their respective bounds. 

\Cref{fig:few_colors_n_q100_space} shows the impact
of $n$ on space under $\mathcal{W}_{100}$.
Interestingly, \FDSM uses \emph{$52\%$ less 
space} than \DSM on average and $9\%$ less
space than \textsc{SCM-rebuild}. On the other hand, \DRM uses $43\%$ less space on average than \FDSM. As expected, the space of all approaches scales
linearly with $n$.

\vspace{+1mm}
\noindent{\bf Impact of $|\colors|$.} \Cref{fig:few_colors_c_q100_time} shows that
\FDSM is \emph{over two orders of magnitude} 
(respectively, \emph{more than $3$ times}) 
faster than \textsc{SCM-rebuild} (respectively, \DRM) 
on average under $\mathcal{W}_{100}$.
The average operation time of \FDSM increases by only
$54\%$ as $|\colors|$ increases $20$ times, 
showing a much weaker dependence on $|\colors|$
than the linear dependence in its worst-case query-time bound of
$\cO(|\colors|\log^2 n)$ in 
\Cref{thm:fdsm_few_colors}. 
\DSM is about $5$ times faster than
\FDSM, as it is designed for query
and update operations on a fixed tree. 

\Cref{fig:few_colors_c_q100_construction_time}
shows that \FDSM is 
\emph{more than $4$ (respectively, $2$) times} faster to construct than \textsc{SCM-rebuild} (respectively, \DSM) on average 
 under $\mathcal{W}_{100}$. \DRM takes on average
$21\%$ less time to be constructed than \FDSM.
Importantly, the construction time of \FDSM
varies by less than $3\%$ as $|\colors|$ increases
$20$ times, consistent with its
$\cO(n\log n)$ construction-time bound.

\Cref{fig:few_colors_c_q100_space}
shows that \FDSM uses on average
\emph{$21\%$ (respectively, $52\%$) less space} 
than \textsc{SCM-rebuild} (respectively, \DSM)
under $\mathcal{W}_{100}$. \DRM uses on average
$40\%$ less space than \FDSM.
Importantly, the space consumption of \FDSM
varies by less than $0.3\%$ as $|\colors|$ increases
$20$ times, consistent with its
$\cO(n)$ space bound.

\vspace{-2mm}
\subsubsection{Leaf additions and deletions}

We evaluate \FDSM using workloads consisting 
of leaf additions and deletions, in which a new colored leaf is added and then 
immediately deleted. 
\begin{figure}[htbp]
\centering
\captionsetup{skip=3pt}
\captionsetup[subfigure]{skip=0pt}

\sweepplot{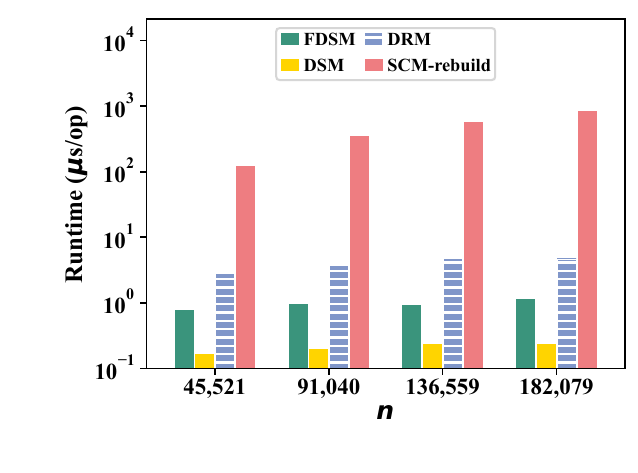}
{Avg. operation time}{fig:few_colors_n_q100_time}\hfill
\sweepplot{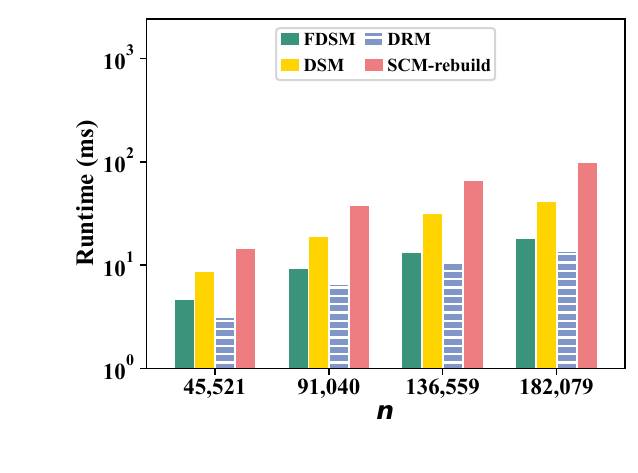}
{Construction time}{fig:few_colors_n_q100_construction_time}\hfill
\sweepplot{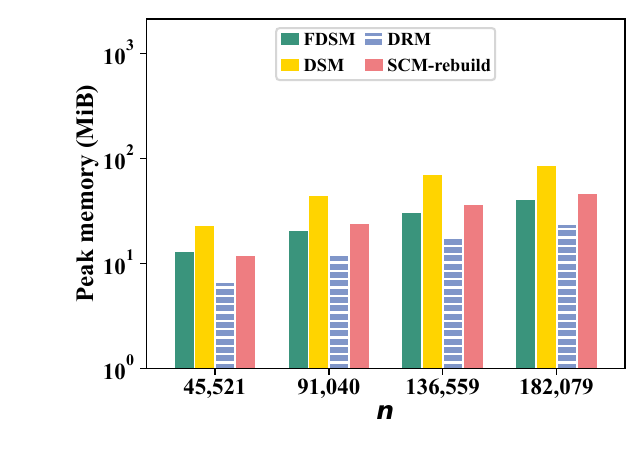}
{Space}{fig:few_colors_n_q100_space}
\caption{Impact of $n$ on \Files for $\mathcal{W}_{100}$.}
\label{fig:few_colors_n_q100}
\end{figure}
\begin{figure}[htbp]
\centering
\captionsetup{skip=3pt}
\captionsetup[subfigure]{skip=0pt}

\sweepplot{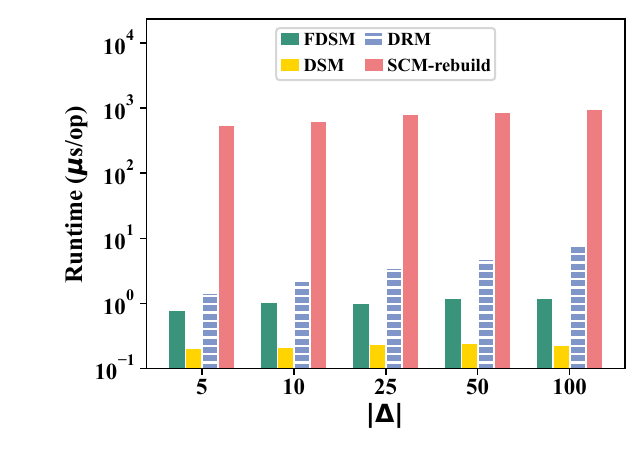}
{Avg. operation time}{fig:few_colors_c_q100_time}\hfill
\sweepplot{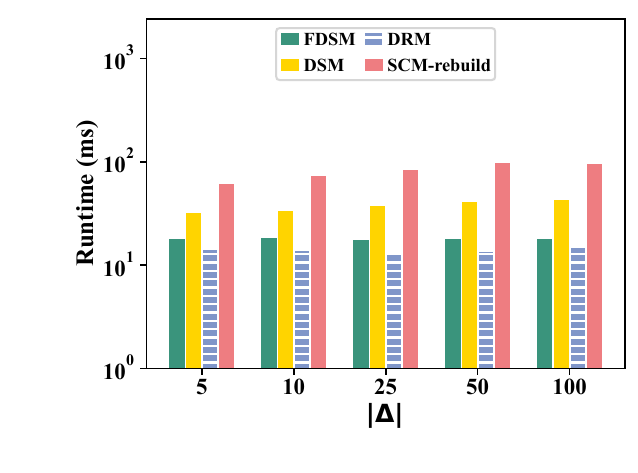}
{Construction time}{fig:few_colors_c_q100_construction_time}\hfill
\sweepplot{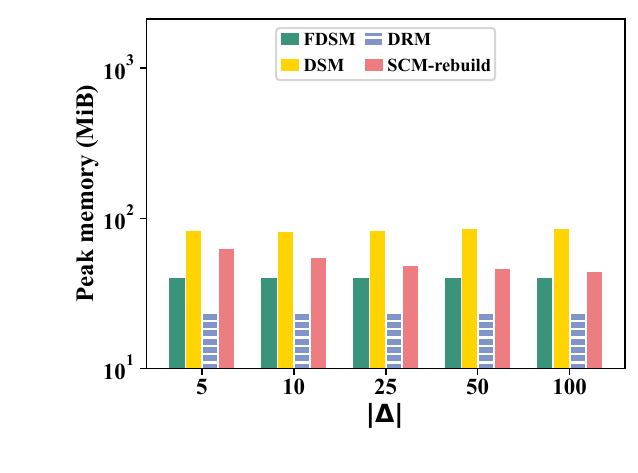}
{Space}{fig:few_colors_c_q100_space}
\caption{Impact of $|\colors|$ on \Files{} for $\mathcal{W}_{100}$.}
\label{fig:few_colors_c_q100}
\centering
\captionsetup{skip=3pt}
\captionsetup[subfigure]{skip=0pt}



\fdsmopplot{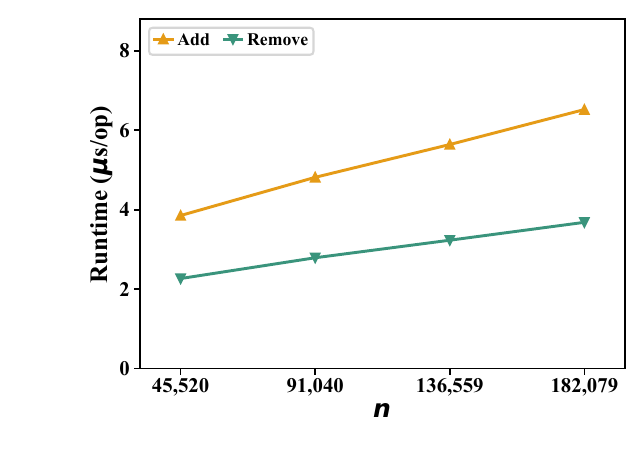}
{Avg. operation time}{fig:leaf_pair_n_time}\hfill
\fdsmopplot{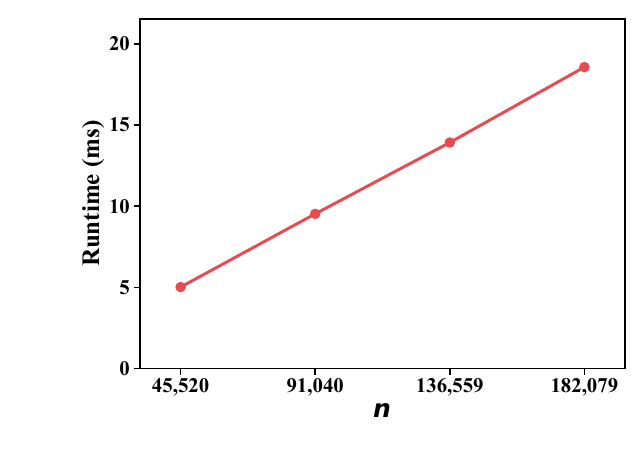}
{Construction time}{fig:leaf_pair_n_construction_time}\hfill
\fdsmopplot{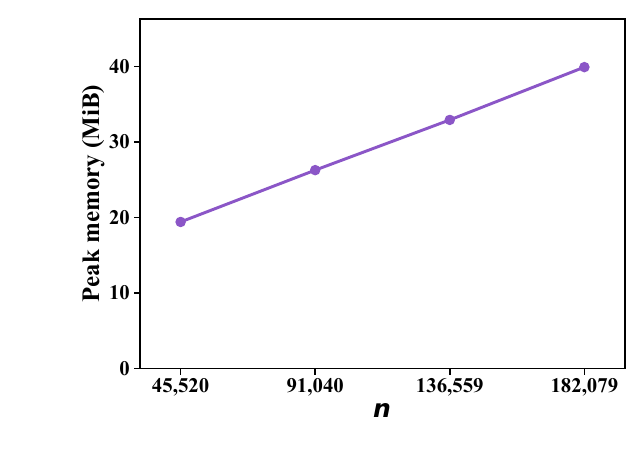}
{Space}{fig:leaf_pair_n_space}\hfill


\fdsmopplot{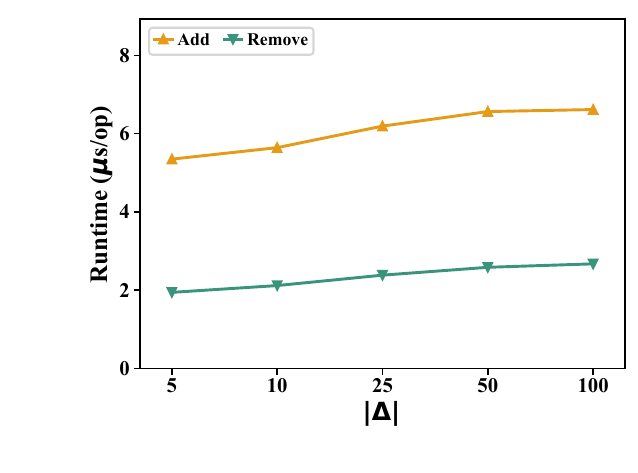}
{Avg. operation time}{fig:leaf_pair_C_time}\hfill
\fdsmopplot{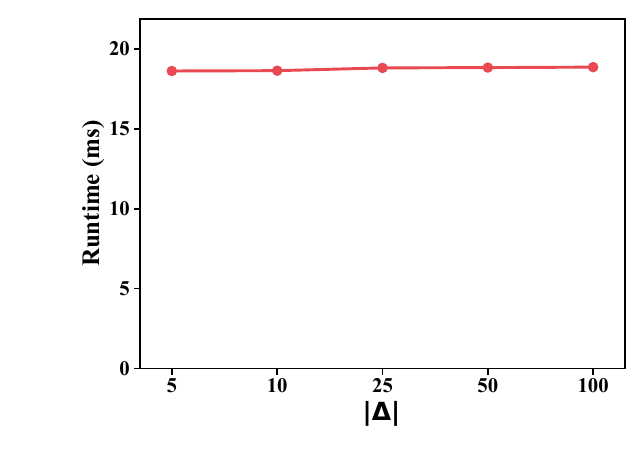}
{Construction time}{fig:leaf_pair_C_construction_time}\hfill
\fdsmopplot{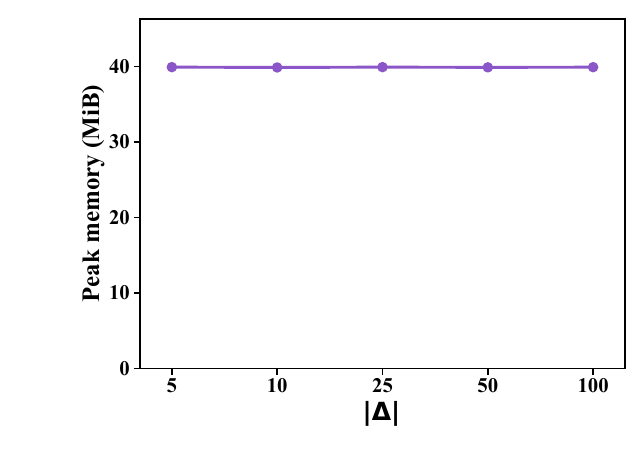}
{Space}{fig:leaf_pair_C_space}

\caption{Impact of $n$ and $|\colors|$ on \FDSM
leaf insertions and deletions: (a)--(c)
varying $n$ with $|\colors|=50$;
(d)--(f) varying $|\colors|$ with $n=182079$.}
\label{fig:fdsm_leaf_updates}
\vspace{-2mm}
\end{figure}

\vspace{+1mm}
\noindent{\bf Impact of $n$.}~
We use $4$ nested instances of the \Files dataset with $|\colors|=50$, replaying the same sequence of \num{83221} leaf addition/deletion pairs at every size. Attachment points are sampled uniformly from the original nodes  of the smallest instance, which are also present in every larger instance. The attachment nodes, insertion colors, operation order, and workload length are fixed across sizes.

\Cref{fig:leaf_pair_n_time} shows that
the average time of leaf additions (respectively, deletions)
increases by only \emph{about $69\%$ (respectively, $62\%$)}
as $n$ increases $4$ times. Leaf additions take approximately
$1.7$ times longer than deletions on average. \Cref{fig:leaf_pair_n_construction_time} shows that
the construction time of \FDSM increases by
approximately $3.7$ times as $n$ increases
$4$ times. 
\Cref{fig:leaf_pair_n_space} shows that
the space consumption of \FDSM increases overall
by approximately $2$ times as $n$ increases
$4$ times. These results are consistent with the bounds in \Cref{thm:fdsm_few_colors}. 

\vspace{+1mm}
\noindent{\bf Impact of $|\colors|$.}~
We use the \Files dataset and 
perform \num{83221} leaf addition/deletion pairs
per color setting. Attachment points alternate between
randomly selected initial leaves and internal nodes.

\Cref{fig:leaf_pair_C_time} shows that
the average time of leaf additions (respectively,
deletions) increases by only
\emph{$24\%$ (respectively, $37\%$)}
as $|\colors|$ increases $20$ times.  
\Cref{fig:leaf_pair_C_construction_time}
shows that the construction time of \FDSM
increases by only $1.3\%$ as $|\colors|$ increases
$20$ times. 
\Cref{fig:leaf_pair_C_space} shows that
the space consumption of \FDSM varies by less
than $0.2\%$ as $|\colors|$ increases $20$ times. These results are 
consistent with the bounds 
in \Cref{thm:fdsm_few_colors}.

\subsubsection{Link-Cut Updates (Subtree Moves)} 
We evaluate \FDSM using workloads comprised 
of link-cut updates (subtree moves), generated randomly while
preserving the validity of the evolving tree.

\vspace{+1mm}
\noindent{\bf Impact of $n$.}~We use $4$ nested trees with 
$|\colors|=50$, performing the same sequence of \num{166442}
subtree moves at every size, with node IDs remapped appropriately.
Each move detaches a non-root subtree and
reattaches it under another node outside that subtree.
The moves are generated randomly on the evolving tree,
ensuring that every operation is valid. 
\Cref{fig:subtree_move_n_time} shows that the average operation time increases as $n$ increases, as expected by the respective bound.  \Cref{fig:subtree_move_n_construction_time}
shows that the construction time 
increases by approximately $3.7$ times as $n$
increases $4$ times, consistent with the worst-case 
$\cO(n\log n)$ bound.
\Cref{fig:subtree_move_n_space} shows that
the space for \FDSM increases in line with the $\cO(n)$ space bound in 
\Cref{thm:fdsm_few_colors}.

\vspace{+1mm}
\noindent{\bf Impact of $|\colors|$.}~
We use the \Files dataset and a workload comprised of 
\num{166442} subtree moves. 
\Cref{fig:subtree_move_C_time} shows that
the average time of subtree moves increases by
only \emph{about $58\%$} as $|\colors|$ increases
$20$ times, so $|\colors|$ has a smaller impact than that implied by the 
$\cO(|\colors|\log^2 n)$ bound in
\Cref{thm:fdsm_few_colors}. 
\Cref{fig:subtree_move_C_construction_time}
shows that the construction time of \FDSM
varies by less than $2.1\%$ as $|\colors|$
increases $20$ times, consistent with its
$\cO(n\log n)$ construction-time bound. 
\Cref{fig:subtree_move_C_space} shows that
the space of \FDSM varies by
less than $0.2\%$ as $|\colors|$ increases
$20$ times, consistent with its
$\cO(n)$ space bound.




\begin{figure}
\centering
\captionsetup{skip=3pt}
\captionsetup[subfigure]{skip=0pt}

\fdsmopplot{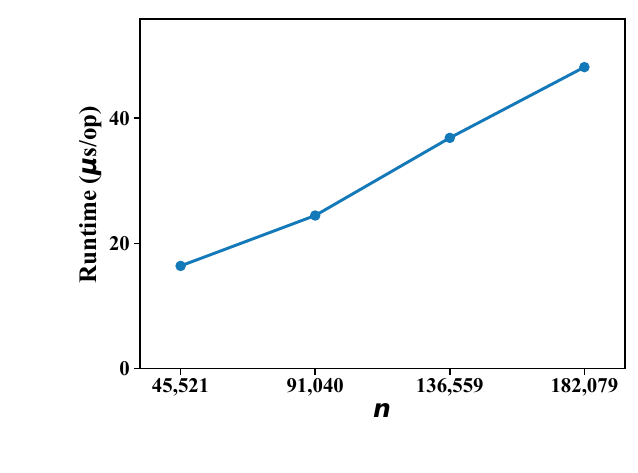}
{Move time}{fig:subtree_move_n_time}\hfill
\fdsmopplot{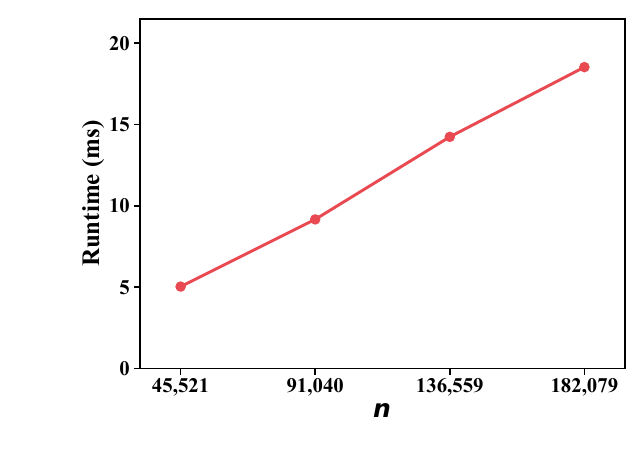}
{Construction time}{fig:subtree_move_n_construction_time}\hfill
\fdsmopplot{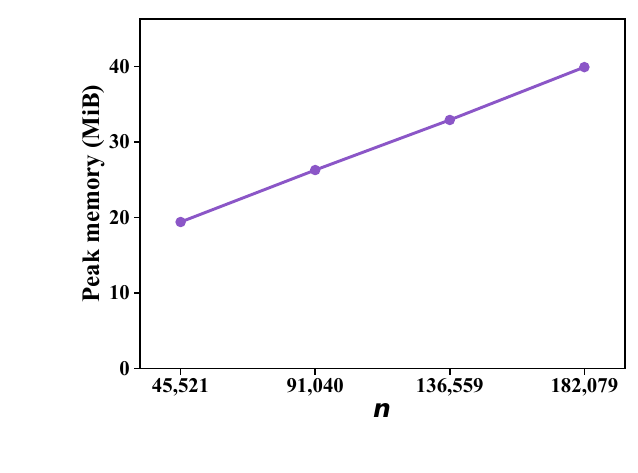}
{Space}{fig:subtree_move_n_space}\hfill

\fdsmopplot{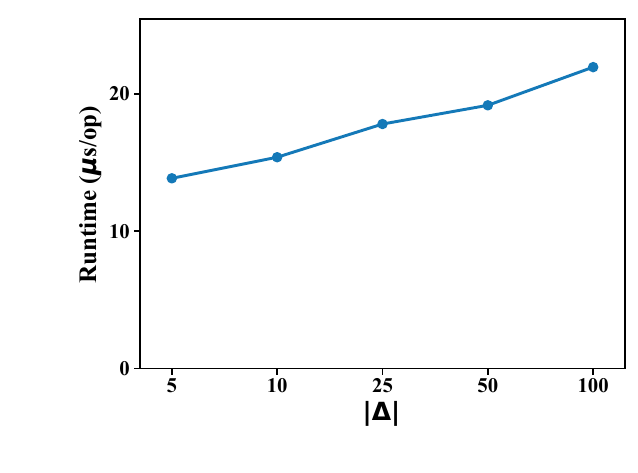}
{Move time}{fig:subtree_move_C_time}\hfill
\fdsmopplot{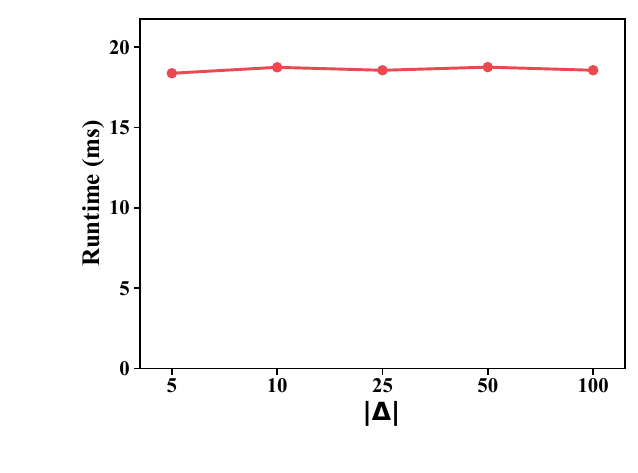}
{Construction time}{fig:subtree_move_C_construction_time}\hfill
\fdsmopplot{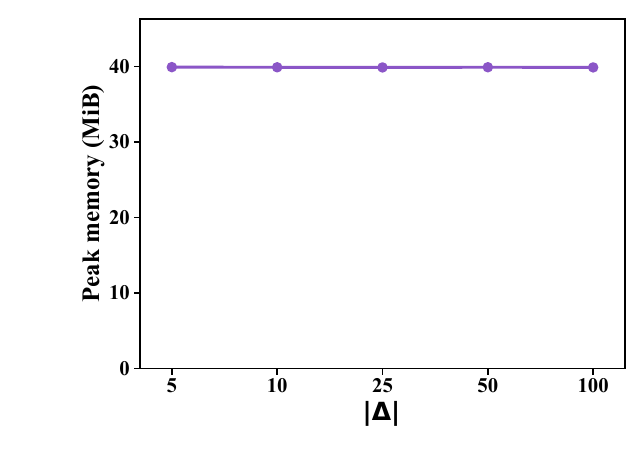}
{Space}{fig:subtree_move_C_space}

\caption{Impact of $n$ and $|\colors|$ on \FDSM
subtree moves: (a)--(c) varying $n$ with
$|\colors|=50$; (d)--(f) varying $|\colors|$
with $n=182079$.}
\label{fig:fdsm_subtree_moves}
\end{figure}

\vspace{-3mm}
\subsection{Case Studies}
\label{sec:case_studies}

We evaluate our data structures using
\emph{real changes in an e-commerce platform
and a software repository}. 
We replay these changes in their original order,
together with subtree-mode queries. The construction time and space results  are omitted for space reasons.

\begin{figure}
\centering
\captionsetup{skip=3pt}
\captionsetup[subfigure]{skip=0pt}
\begin{subfigure}[t]{0.18\columnwidth}
\centering
\includegraphics[width=\linewidth]{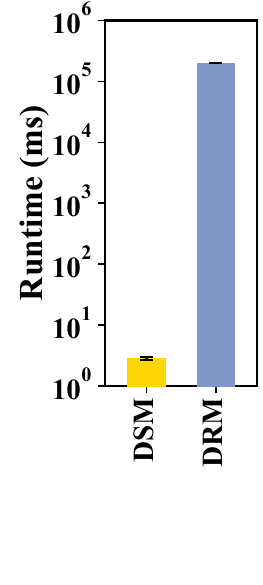}
\caption{\footnotesize Workload}
\label{fig:amazon_case_study_runtime}
\end{subfigure}\hfill
\begin{subfigure}[t]{0.18\columnwidth}
\centering
\includegraphics[width=\linewidth]{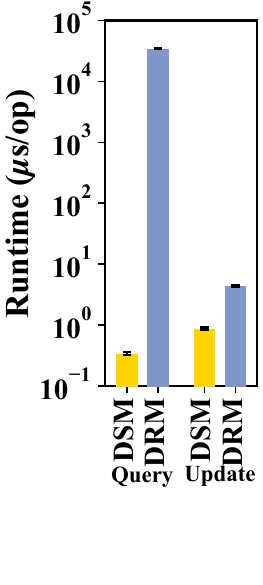}
\caption{\footnotesize Operations}
\label{fig:amazon_case_study_operation_runtime}
\end{subfigure}\hfill
\begin{subfigure}[t]{0.18\columnwidth}
\centering
\includegraphics[width=\linewidth]{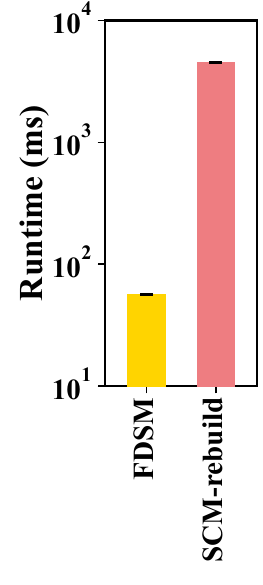}
\caption{\footnotesize Workload}
\label{fig:files_case_study_runtime}
\end{subfigure}\hfill
\begin{subfigure}[t]{0.18\columnwidth}
\centering
\includegraphics[width=\linewidth]{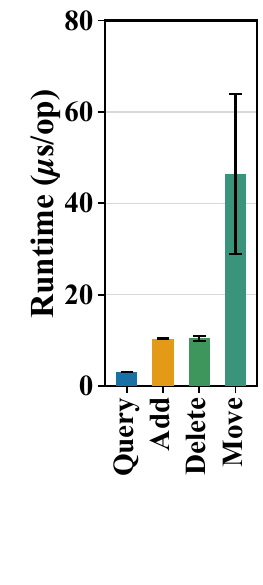}
\caption{\footnotesize Operations}
\label{fig:files_case_study_operation_runtime}
\end{subfigure}
\caption{(a) Total query and update time and (b) average query and update
time for \DSM{} and \DRM{} on \Amazonlarge{}; (c) total workload time and
(d) average operation time on \Files{}. Error bars denote standard deviations over five repetitions.}
\label{fig:case_study_runtime}
\vspace{-2mm}
\end{figure}

\vspace{+1mm}
\noindent{\bf E-Commerce Product Taxonomy.}~
We use the \Amazonlarge{} dataset to evaluate whether \DSM{} maintains meaningful category summaries under real rating updates. Unlike the scalability experiments, we assign products to five rating tiers derived from their cumulative average ratings and replay subsequent reviews chronologically. The replay contains \num{1000} tier-changing updates and \num{5845} queries to the affected categories and their ancestors, up to the root.  
\Cref{fig:amazon_case_study_runtime} shows that
\DSM{} executes the queries and updates
\emph{about $\num{71000}$ times faster} than \DRM{}.   
\Cref{fig:amazon_case_study_operation_runtime} shows that
\emph{queries and updates take only about $0.34$ and
$0.87$ microseconds on average}, respectively.
The mode or its frequency changes more than
$\num{3000}$ times during the replay, demonstrating
the importance of efficiently maintaining
subtree-mode information as the data evolve. 

\vspace{+1mm}
\noindent{\bf An Evolving Software Repository.}~
We use the \Files{} dataset to evaluate whether
\FDSM{} efficiently maintains directory summaries
as a real software repository evolves.
We construct the initial tree from the LLVM repository snapshot of 10 November 2025, representing directories as internal nodes and files as leaves, with edges following the directory hierarchy. Unlike the scalability experiments, we color each file leaf by one of $14$ extension-based categories, including source files, headers, LLVM intermediate representations, documentation, and scripts. Files without extensions form a separate category, while unclassified extensions are grouped as ``other''. The same category mapping is used throughout the replay. Starting from this snapshot, we replay \num{1000} consecutive LLVM commits: file and directory additions and deletions are replayed as structural updates, and cross-directory moves as link-cut updates; after each commit, we query affected directories, their ancestors, five fixed major project directories, and the repository root. The workload contains \num{13501} operations, including \num{12806} queries and \num{695} updates.  

\Cref{fig:files_case_study_runtime} shows that
\FDSM{} processes the entire workload
\emph{about $80$ times faster} than
\textsc{SCM-rebuild}. \Cref{fig:files_case_study_operation_runtime} shows that
\emph{queries, leaf additions, and deletions take
only about $3$, $10$, and $10$ microseconds on average},
respectively, while subtree moves take about
$46$ microseconds.
The mode or its frequency changes more than
\num{1000} times across \num{194} directories,
demonstrating the importance of solving \SM in the dynamic setting. 

\section{Conclusion}\label{sec:conclusion}
Our work initiated the study of dynamic subtree mode, proposing efficient
data structures for \DSM and \FDSM with conditionally near-optimal
bounds.  It also raises a few direct open questions. First, our conditional optimality for \DSM and hardness results
for \FDSM motivate studying approximate \FDSM, which reports a
sufficiently frequent color instead of an exact subtree mode. Other
directions include incremental \FDSM,
allowing only node additions for append-only applications, and
reporting the top-$K$ most frequent colors in a subtree.


\ifarxiv
\section*{AI Usage Declaration}
The authors used Anthropic's Claude Opus 5.5 via the chat interface.
The core ideas, algorithms, and proofs presented in this paper were developed entirely by the authors. The authors wrote the initial implementation and subsequently used the model to assist with code optimization. The authors also wrote the initial manuscript and subsequently used the model to help identify imprecise statements and minor errors, and to improve clarity and presentation. The model also wrote the TikZ code for \Cref{fig:subtree-mode-operations,fig:omv-reduction} based on the authors' description. The authors reviewed all AI-assisted revisions and take full responsibility for the final content of the paper and the accompanying code.
\fi

\FloatBarrier{}
\ifarxiv
\bibliographystyle{alphaurl}
\else
\bibliographystyle{ACM-Reference-Format}
\fi
\bibliography{references}


\clearpage
\onecolumn
\appendix
\printomittedproofs

\newcommand{\appendixsmallplot}[3]{%
\begin{subfigure}[b]{0.48\columnwidth}
    \includegraphics[width=\linewidth]{figs/#1}
    \caption{#2}
    \label{#3}
\end{subfigure}%
}

\section{Top-1 Document Retrieval with Document Additions}\label{sec:1DR}

In the Top-1 Document Retrieval problem~\cite{DBLP:conf/soda/Muthukrishnan02}, we are asked to preprocess a collection $\mathcal{S}$ of documents, so that given a query pattern $P$, we can output the document in which $P$ occurs most frequently. Here, we explain how to solve this problem under document additions~\cite{DBLP:journals/siamcomp/NavarroN17}. We identify the color $i \in \colors$ with the document $S_i$, so that $|\colors| = |\mathcal{S}|$.
We build the generalized suffix tree of $\mathcal{S}$ from left to right with Ukkonen's algorithm~\cite{DBLP:journals/algorithmica/Ukkonen95}.
Appending a character $c$ to $S_i$ triggers zero or more updates, each of which (i) attaches a new leaf colored $i$ through a new edge starting with $c$, or (ii) splits an edge with a new internal node and attaches a new leaf colored $i$ to it through a $c$-edge.
Subdividing an edge inserts a unary internal node and takes $\cO(\log n)$ time by the Euler-sequence extension in the proof of \Cref{thm:fdsm_few_colors}. Attaching the new colored leaf then takes $\cO(\log^2 n)$ time. Thus each structural update takes $\cO(\log^2 n)$ time, where $n$ denotes the total length of $\mathcal{S}$ including its sentinels.
Since Ukkonen's algorithm creates $\cO(n)$ nodes in $\cO(n\log n)$ time, maintaining the augmented structure takes $\cO(n\log^2 n)$ time in total and $\cO(n)$ space.
Once $S_i$ and its sentinel $\$_i$ have been appended, every suffix of $S_i$ is represented by a leaf, so we can spell any pattern $P$ of length $m$, reach its locus $v$, and query the subtree mode of $v$, which is the top-1 document for $P$~\cite{static-paper}, in $\cO(m\log n+|\colors|\log^2 n)$ time (\Cref{thm:fdsm_few_colors}). If the locus is implicit on an edge, we query its lower endpoint, which has the same leaf descendants.
For $|\colors|=\cO(1)$, this matches, up to $\log^{\cO(1)} n$ factors,
the query and update bounds of the $\cO(n)$-space data structure of~\cite{DBLP:journals/siamcomp/NavarroN17},
which, however, solves the Top-$K$ counterpart and is significantly more involved.

\section{Additional Material for Experimental Evaluation}\label{appendixC}

\subsection{Details About the Datasets}
\label{app:dataset-details}

To construct \Amazonlarge{}, we retain products with a supplied category path and a rating before 1 January 2014. When a product has multiple paths, we select the deepest one, breaking ties lexicographically. Shared category prefixes become internal nodes, and products become leaves. Product colors are derived from the smallest positive supplied sales rank, grouped into bands of width \num{1000}. This gives \num{3802} colors, including a separate color for products without a supplied rank. Because ranks are category-specific, these colors serve as benchmark labels rather than a universal measure of sales volume.

\subsection{Additional Results for \DSM on \News and \Genessample}
\label{app:news_genes_additional}

\providecommand{\smallresultrowplot}[4][0.155\textwidth]{%
\begin{subfigure}[t]{#1}
\centering
\includegraphics[width=\linewidth]{figs/#2}
\caption{#3}\label{#4}
\end{subfigure}%
}

\begin{figure}[!htbp]
\centering
\captionsetup[subfigure]{skip=1pt,font={footnotesize,bf},labelfont=bf}
\smallresultrowplot[0.235\textwidth]{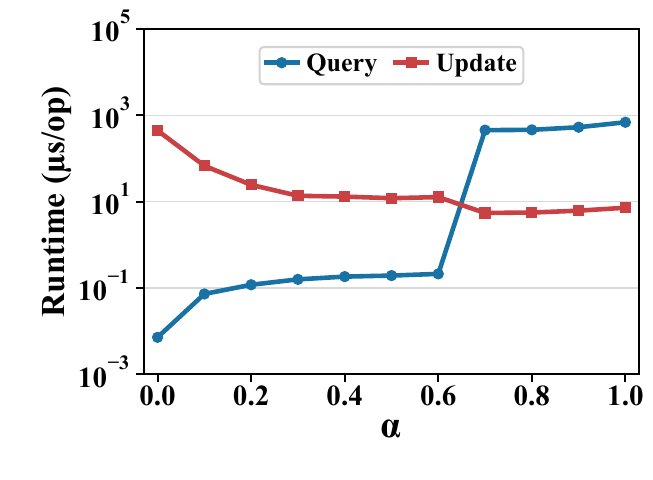}{\News}{fig:news_alpha_query_update_time}\hfill
\smallresultrowplot[0.235\textwidth]{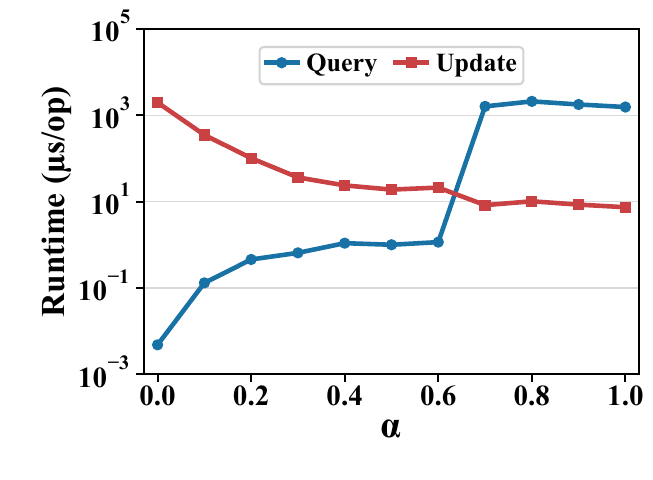}{\Genessample}{fig:genes_alpha_query_update_time}\hfill
\smallresultrowplot[0.235\textwidth]{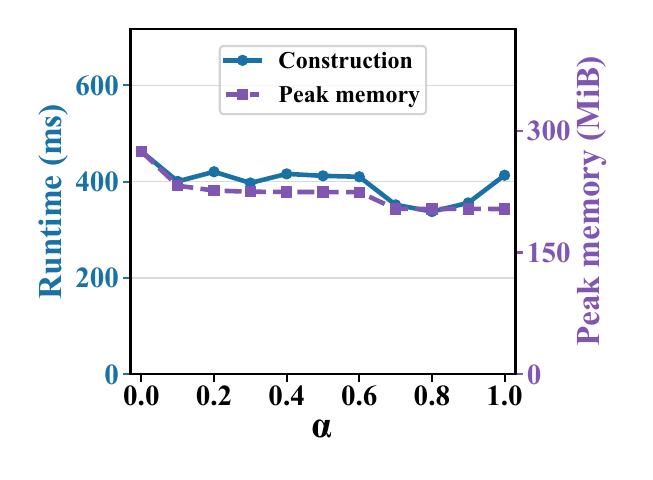}{\News}{fig:news_alpha_construction_memory}\hfill
\smallresultrowplot[0.235\textwidth]{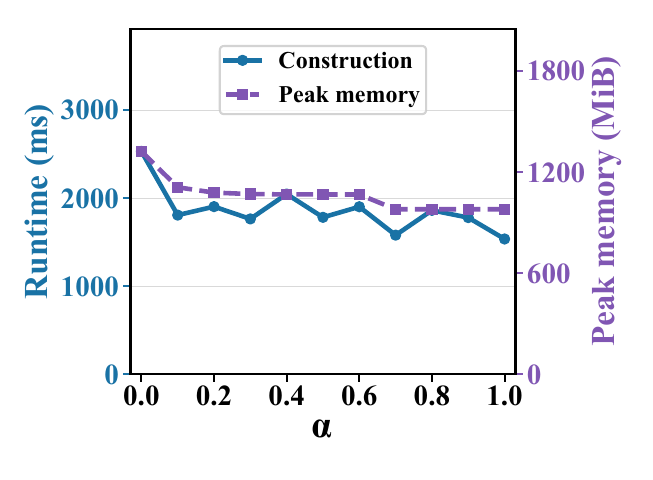}{\Genessample}{fig:genes_alpha_construction_memory}
\caption{Impact of $\alpha$ on \News and \Genessample.
(a)--(b) query and update time; (c)--(d) construction time
(left axis) and space (right axis).}
\label{fig:small_alpha_news_genes}
\end{figure}

\Cref{fig:small_alpha_news_genes} shows the additional results
for varying $\alpha$ on \News{} and \Genessample{}.

\begin{figure}[!htbp]
\centering
\captionsetup[subfigure]{skip=1pt,font={footnotesize,bf},labelfont=bf}
\smallresultrowplot{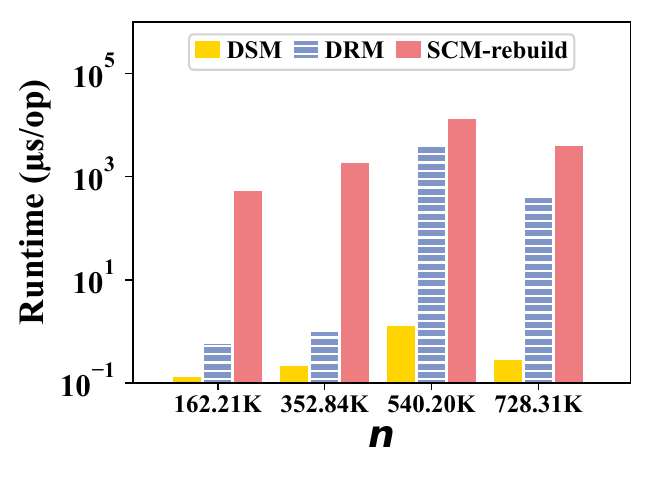}{\News}{fig:news_n_q100_time}\hfill
\smallresultrowplot{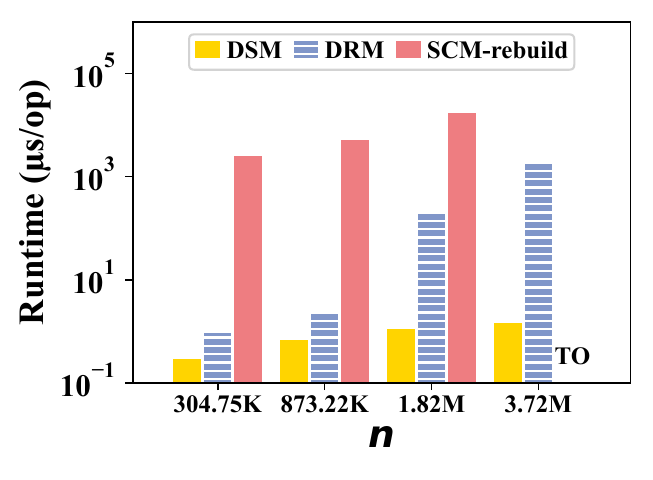}{\Genessample}{fig:genes_n_q100_time}\hfill
\smallresultrowplot{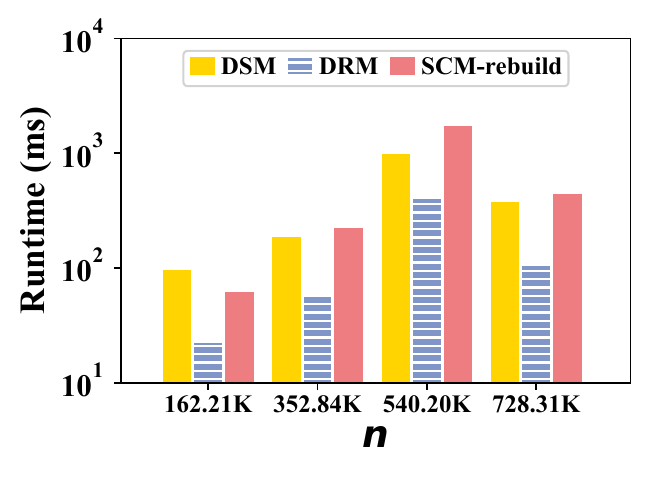}{\News}{fig:news_n_q100_buildtime}\hfill
\smallresultrowplot{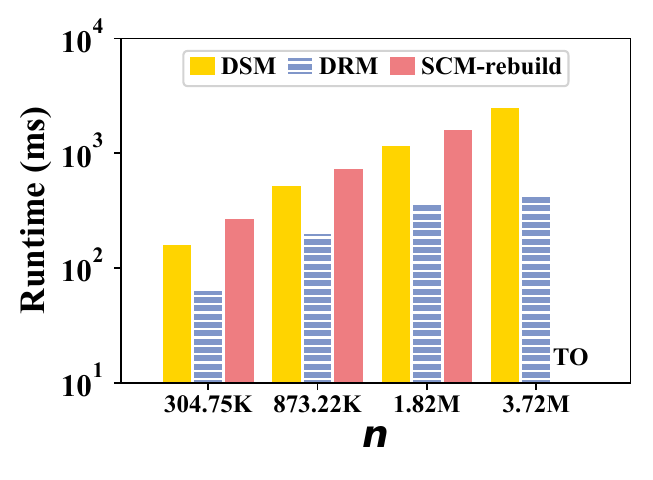}{\Genessample}{fig:genes_n_q100_buildtime}\hfill
\smallresultrowplot{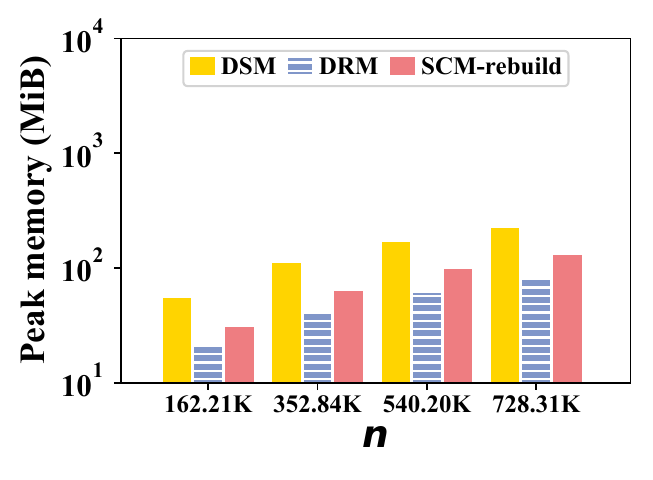}{\News}{fig:news_n_q100_space}\hfill
\smallresultrowplot{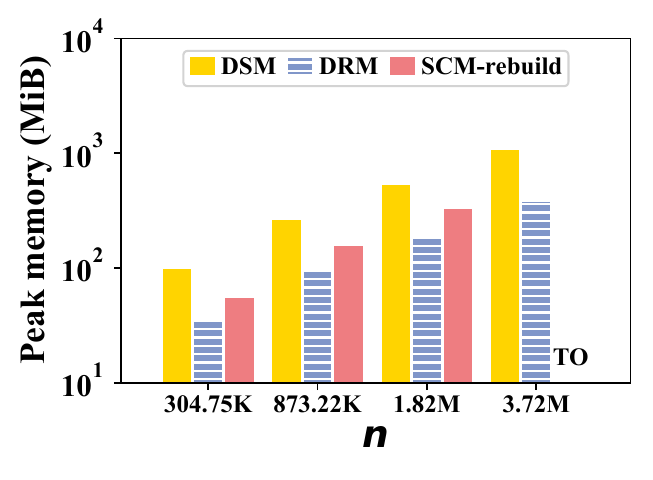}{\Genessample}{fig:genes_n_q100_space}
\caption{Impact of $n$ on \News and \Genessample under $\mathcal{W}_{100}$.
(a)--(b) average operation time; (c)--(d) construction time;
(e)--(f) space. TO denotes a $24$-hour time-out.}
\label{fig:n_news_genes_q100}
\end{figure}

\Cref{fig:n_news_genes_q100} shows the additional results
for varying $n$ on \News{} and \Genessample{} under $\mathcal{W}_{100}$.

\begin{figure}[!htbp]
\centering
\captionsetup[subfigure]{skip=1pt,font={footnotesize,bf},labelfont=bf}
\smallresultrowplot{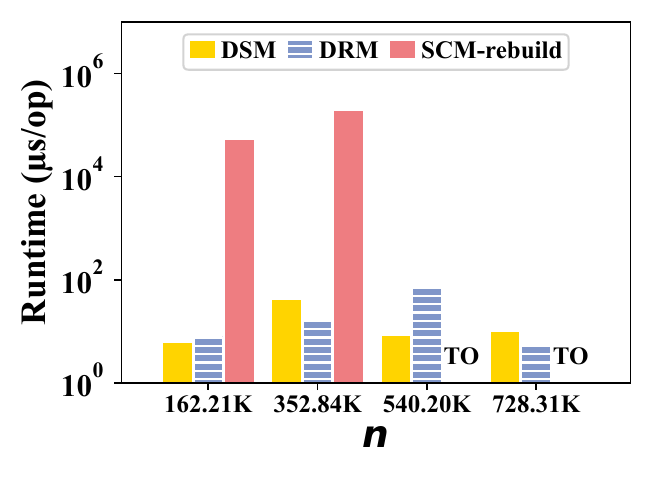}{\News}{fig:news_n_q001_time}\hfill
\smallresultrowplot{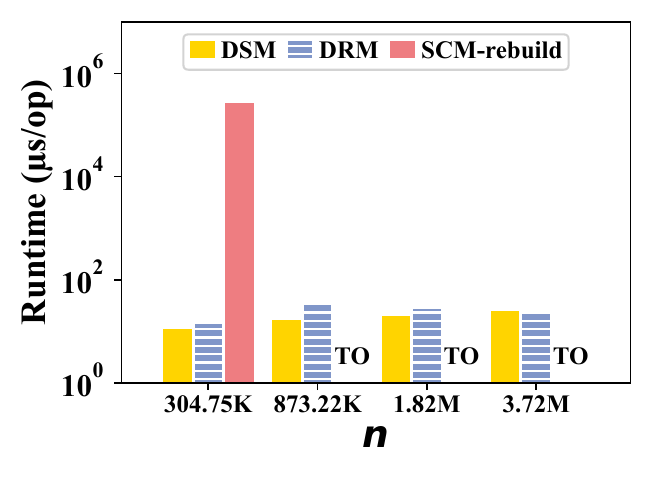}{\Genessample}{fig:genes_n_q001_time}\hfill
\smallresultrowplot{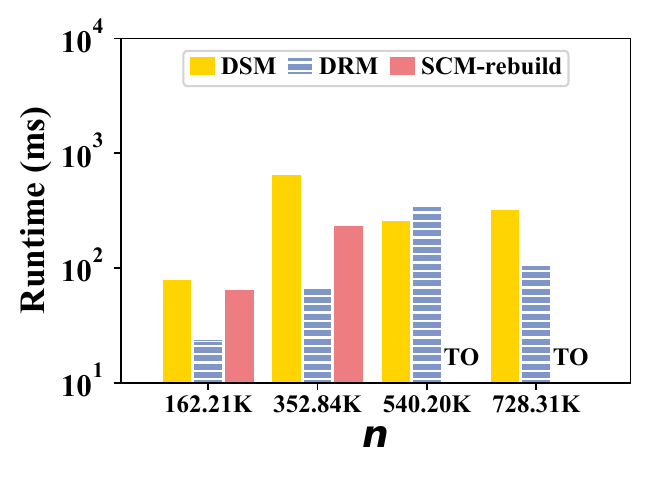}{\News}{fig:news_n_q001_buildtime}\hfill
\smallresultrowplot{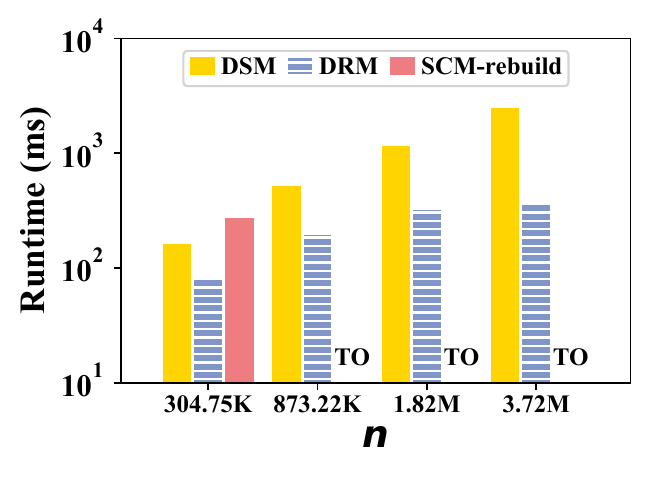}{\Genessample}{fig:genes_n_q001_buildtime}\hfill
\smallresultrowplot{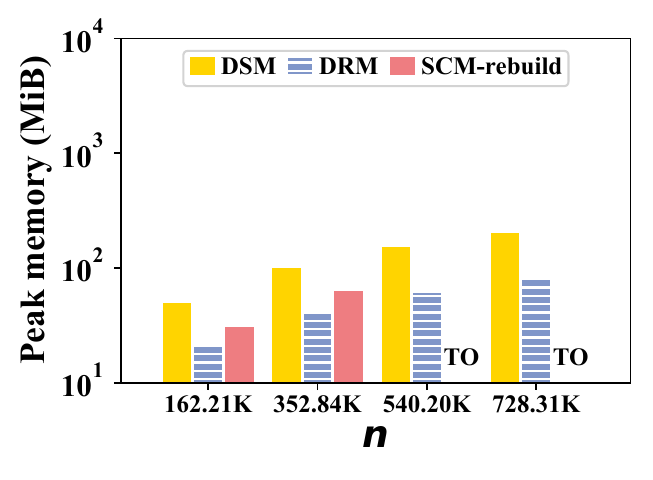}{\News}{fig:news_n_q001_space}\hfill
\smallresultrowplot{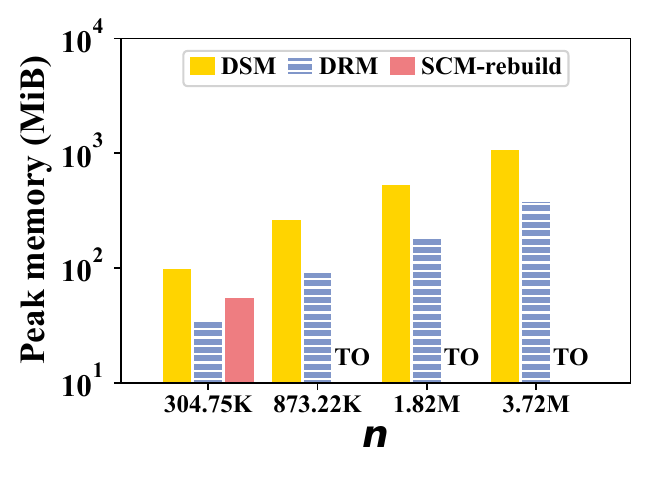}{\Genessample}{fig:genes_n_q001_space}
\caption{Impact of $n$ on \News and \Genessample under $\mathcal{W}_{1/100}$.
(a)--(b) average operation time; (c)--(d) construction time;
(e)--(f) space. TO denotes a $24$-hour time-out.}
\label{fig:n_news_genes_q001}
\end{figure}

\Cref{fig:n_news_genes_q001} shows the additional results
for varying $n$ on \News{} and \Genessample{} under $\mathcal{W}_{1/100}$.

\begin{figure}[!htbp]
\centering
\captionsetup[subfigure]{skip=1pt,font={footnotesize,bf},labelfont=bf}
\smallresultrowplot{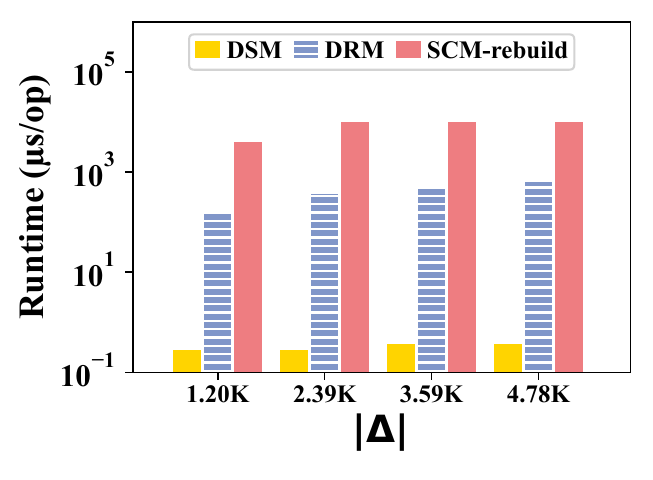}{\News}{fig:news_delta_q100_time}\hfill
\smallresultrowplot{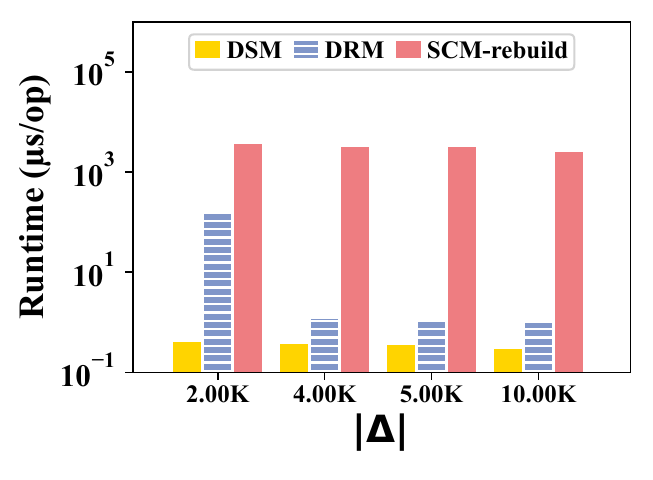}{\Genessample}{fig:genes_delta_q100_time}\hfill
\smallresultrowplot{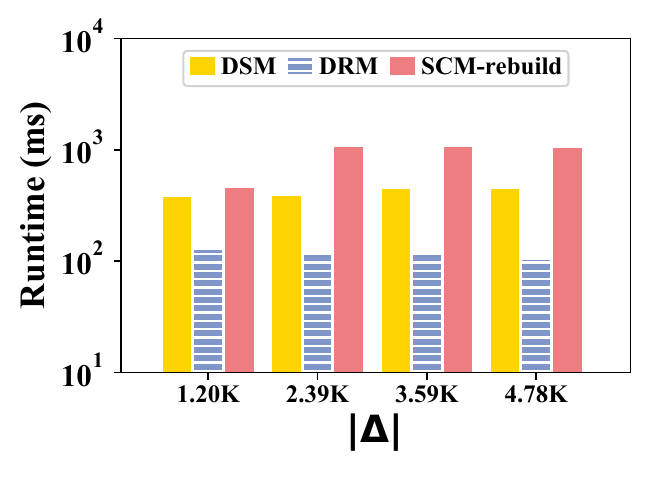}{\News}{fig:news_delta_q100_buildtime}\hfill
\smallresultrowplot{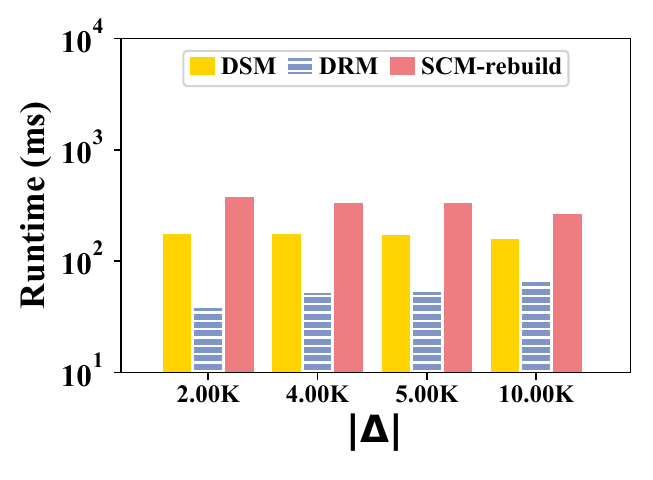}{\Genessample}{fig:genes_delta_q100_buildtime}\hfill
\smallresultrowplot{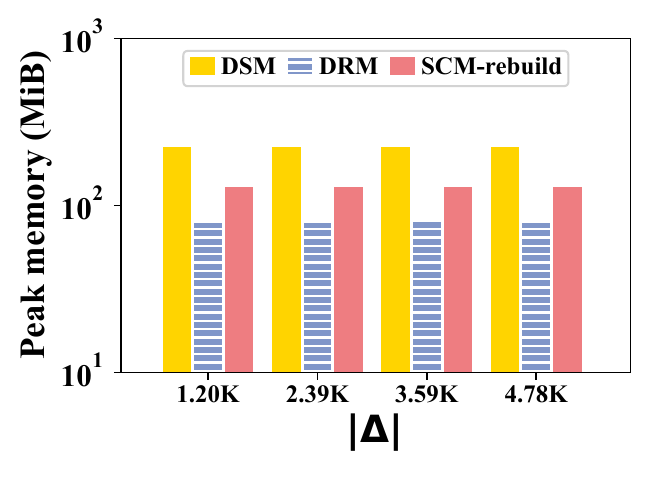}{\News}{fig:news_delta_q100_space}\hfill
\smallresultrowplot{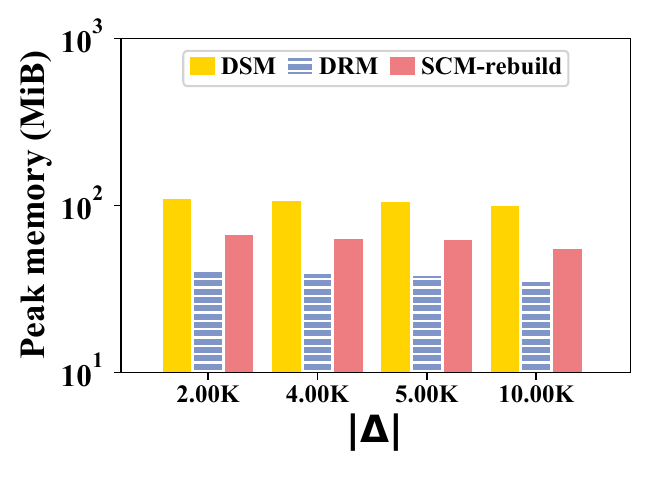}{\Genessample}{fig:genes_delta_q100_space}
\caption{Impact of $|\Delta|$ on \News and \Genessample under $\mathcal{W}_{100}$.
(a)--(b) average operation time; (c)--(d) construction time;
(e)--(f) space.}
\label{fig:delta_news_genes_q100}
\end{figure}

\Cref{fig:delta_news_genes_q100} shows the additional results
for varying $|\Delta|$ on \News{} and \Genessample{} under $\mathcal{W}_{100}$.

\begin{figure}[!htbp]
\centering
\captionsetup[subfigure]{skip=1pt,font={footnotesize,bf},labelfont=bf}
\smallresultrowplot{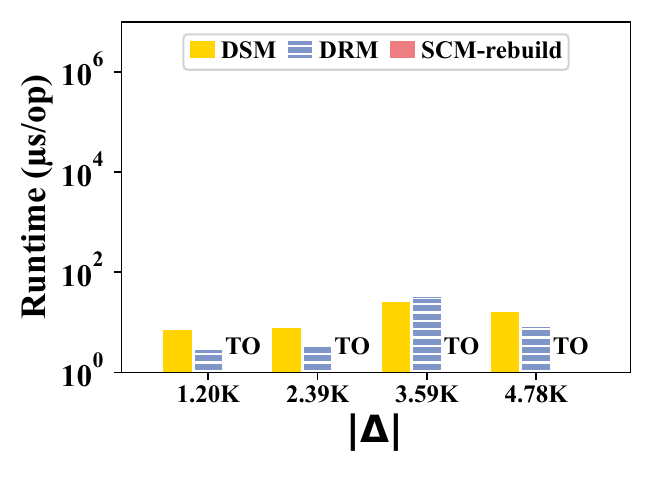}{\News}{fig:news_delta_q001_time}\hfill
\smallresultrowplot{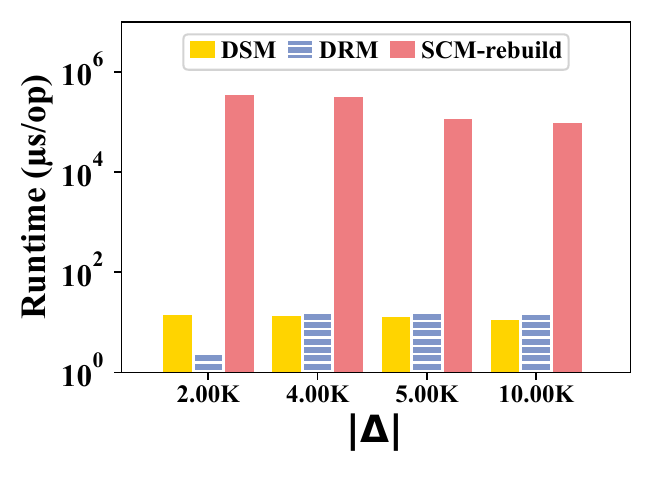}{\Genessample}{fig:genes_delta_q001_time}\hfill
\smallresultrowplot{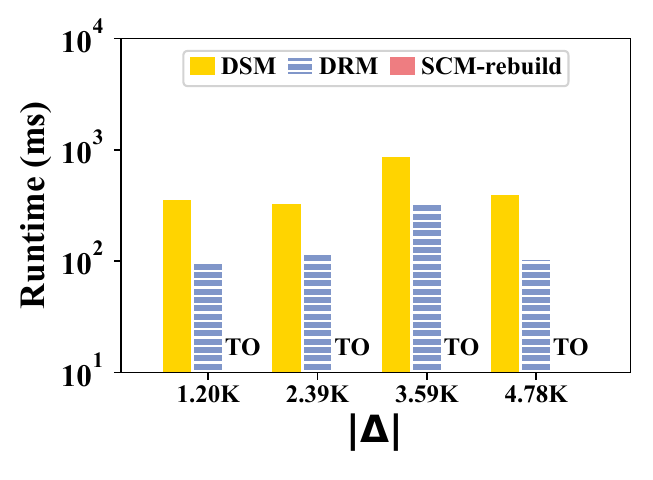}{\News}{fig:news_delta_q001_buildtime}\hfill
\smallresultrowplot{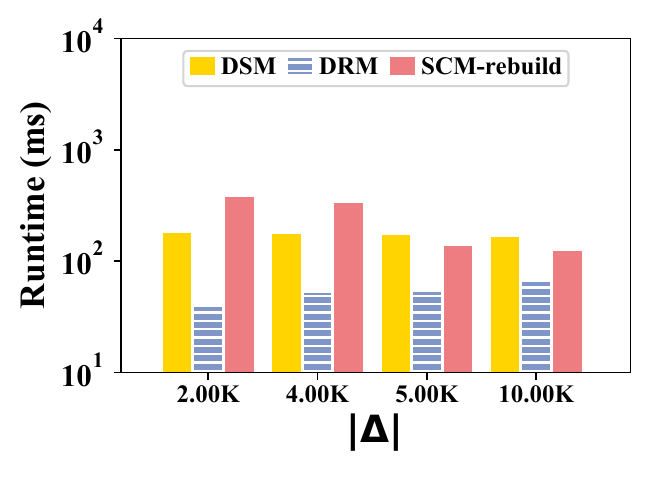}{\Genessample}{fig:genes_delta_q001_buildtime}\hfill
\smallresultrowplot{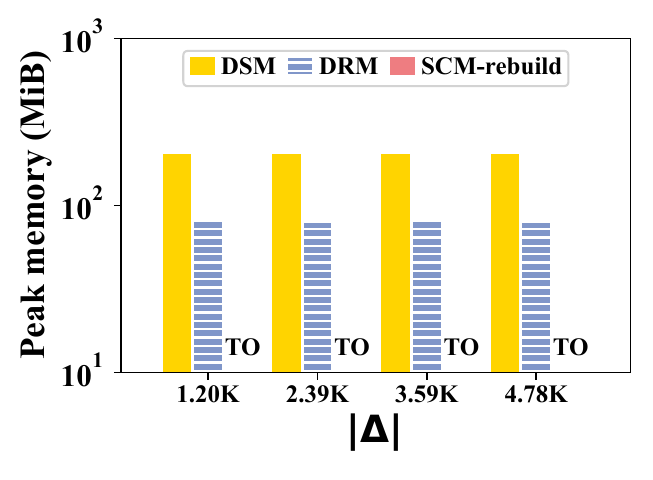}{\News}{fig:news_delta_q001_space}\hfill
\smallresultrowplot{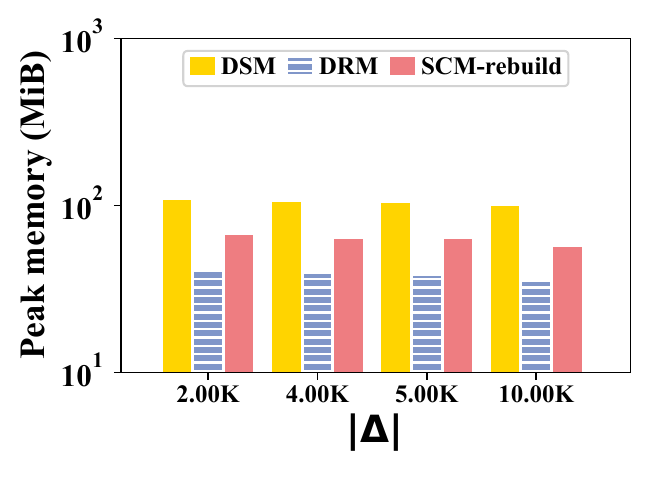}{\Genessample}{fig:genes_delta_q001_space}
\caption{Impact of $|\Delta|$ on \News and \Genessample under $\mathcal{W}_{1/100}$.
(a)--(b) average operation time; (c)--(d) construction time;
(e)--(f) space. TO denotes a $24$-hour time-out.}
\label{fig:delta_news_genes_q001}
\end{figure}

\Cref{fig:delta_news_genes_q001} shows the additional results
for varying $|\Delta|$ on \News{} and \Genessample{} under $\mathcal{W}_{1/100}$.

\subsection{Additional Results for \DSM for the Impact of $n$}
\label{app:n_sweep_additional}

\Cref{fig:n_sweep_build_space_appendix} reports the additional
construction-time and space results omitted from the main text.

\providecommand{\nsweepmergedplot}[5]{%
\begin{subfigure}[t]{0.155\textwidth}
    \centering
    \includegraphics[width=\linewidth]{figs/#1}
    \caption{#2}
    \label{#5}
\end{subfigure}%
}
\begin{figure*}[t]
\centering
\captionsetup{skip=3pt}
\captionsetup[subfigure]{skip=0pt}

\renewcommand{\thesubfigure}{(\alphalph{\value{subfigure}})}
\captionsetup[subfigure]{labelformat=simple,font={footnotesize,bf},labelfont=bf,
    justification=centering,singlelinecheck=false,skip=2pt}
\nsweepmergedplot{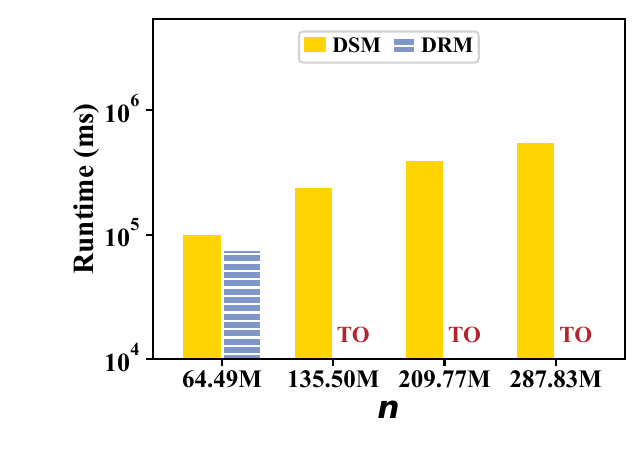}{\VIRlarge{}}{Construction time}{100}{fig:vir_large_n_buildtime}\hfill
\nsweepmergedplot{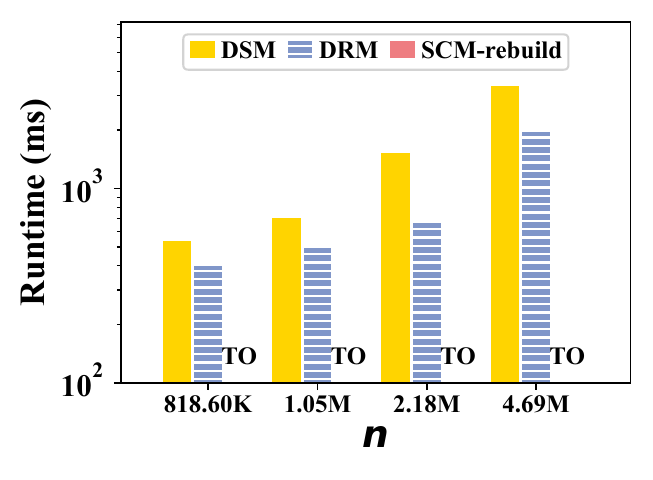}{\WebKBsample{}}{Construction time}{1/100}{fig:webkb_n_q001_buildtime}\hfill
\nsweepmergedplot{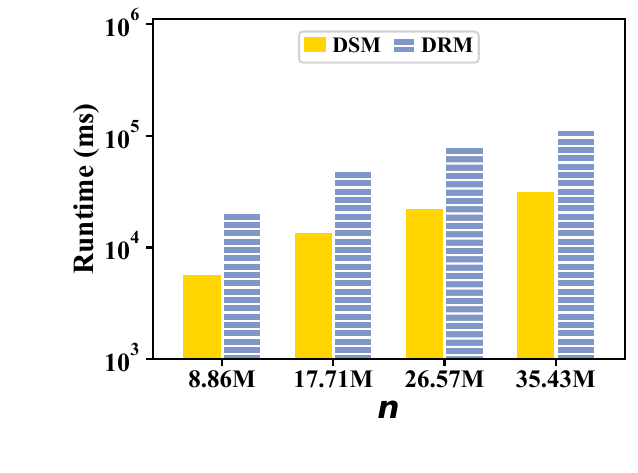}{\WebKBlarge{}}{Construction time}{1/100}{fig:webkb_large_n_q001_buildtime}\hfill
\nsweepmergedplot{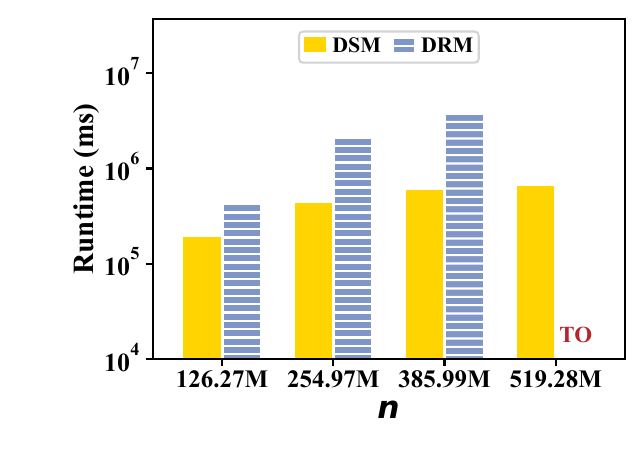}{\Geneslarge{}}{Construction time}{1/100}{fig:genes_large_n_q001_buildtime}\hfill
\nsweepmergedplot{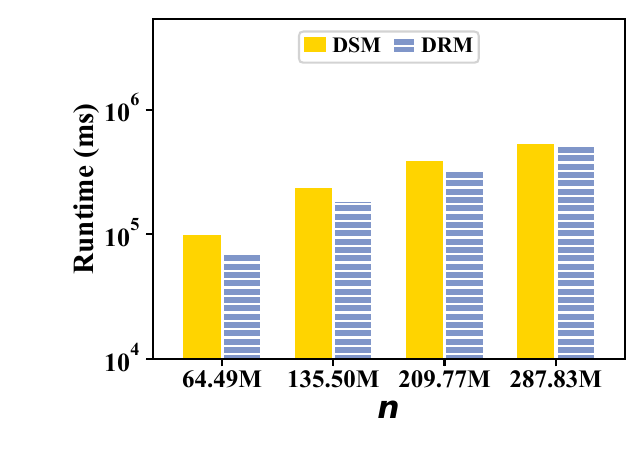}{\VIRlarge{}}{Construction time}{1/100}{fig:vir_large_n_q001_buildtime}\hfill
\nsweepmergedplot{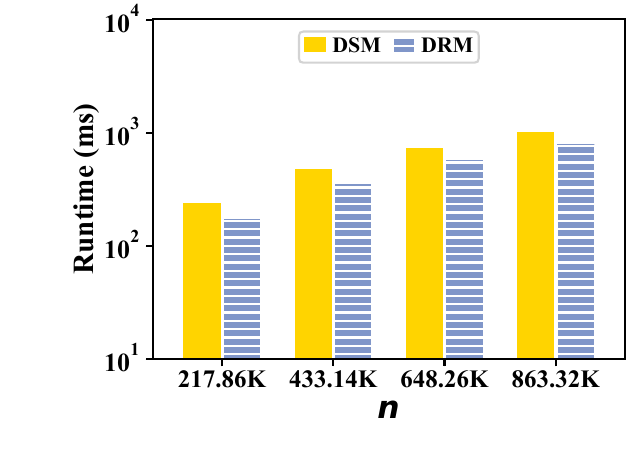}{\Amazonlarge{}}{Construction time}{1/100}{fig:amazon_salesrank_n_q001_buildtime}
\par\medskip
\nsweepmergedplot{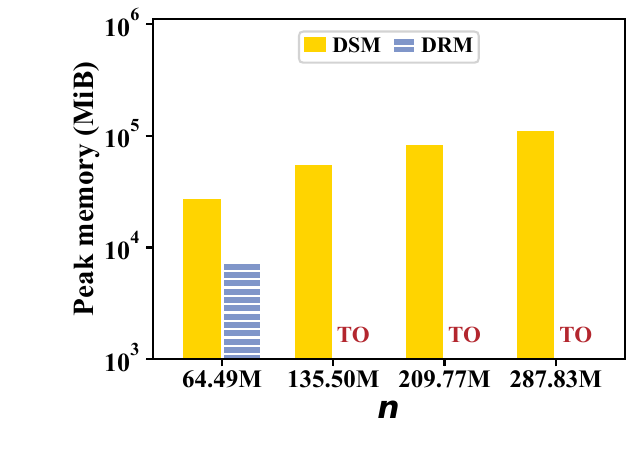}{\VIRlarge{}}{Space}{100}{fig:vir_large_n_space}\hfill
\nsweepmergedplot{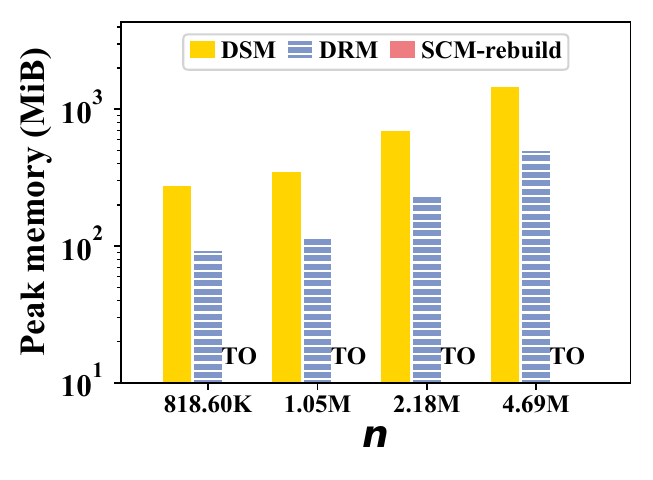}{\WebKBsample{}}{Space}{1/100}{fig:webkb_n_q001_space}\hfill
\nsweepmergedplot{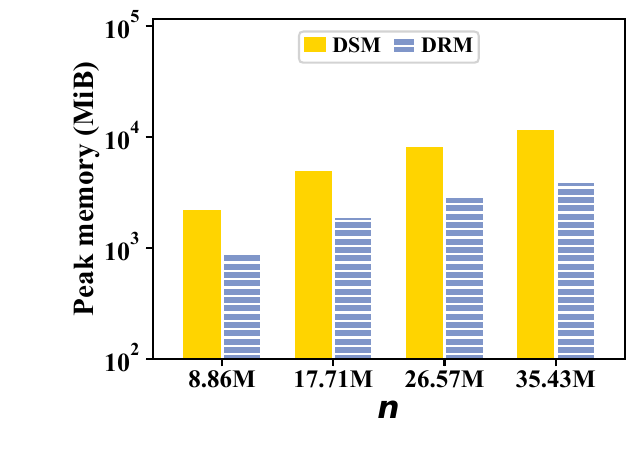}{\WebKBlarge{}}{Space}{1/100}{fig:webkb_large_n_q001_space}\hfill
\nsweepmergedplot{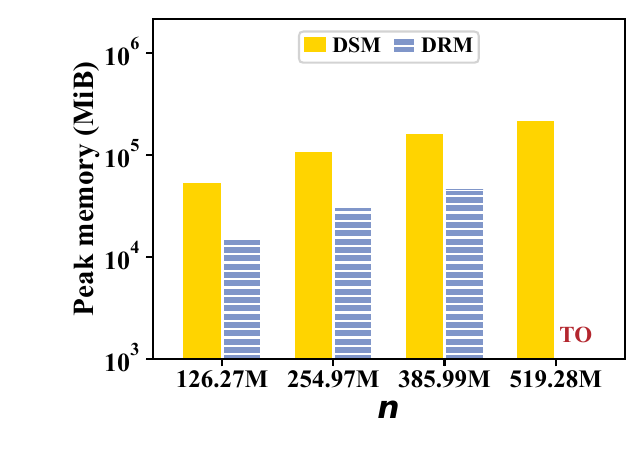}{\Geneslarge{}}{Space}{1/100}{fig:genes_large_n_q001_space}\hfill
\nsweepmergedplot{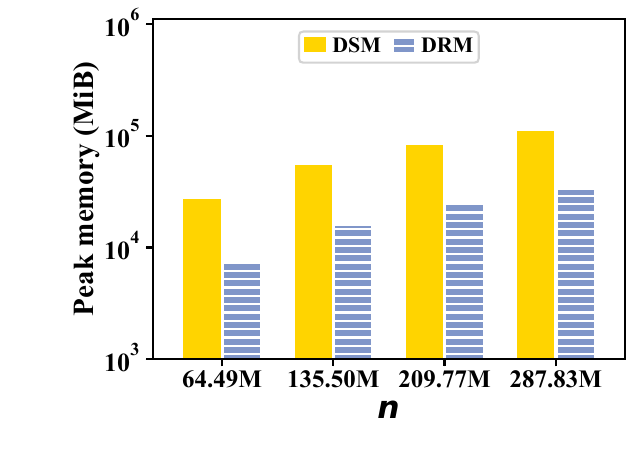}{\VIRlarge{}}{Space}{1/100}{fig:vir_large_n_q001_space}\hfill
\nsweepmergedplot{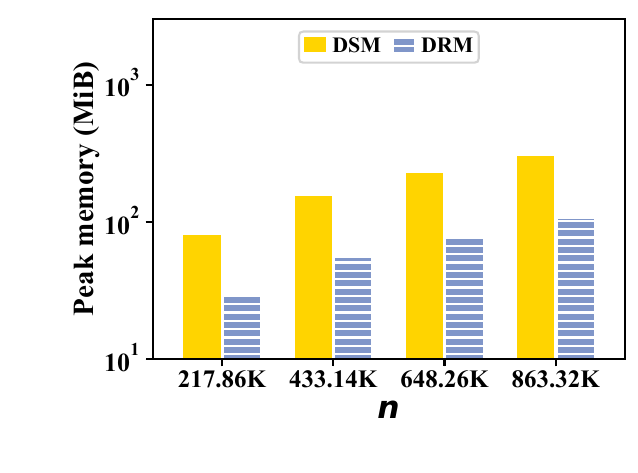}{\Amazonlarge{}}{Space}{1/100}{fig:amazon_salesrank_n_q001_space}
\caption{Additional results for the impact of $n$.
(a)--(f) report construction time and (g)--(l) construction space.
(a) and (g) use $\mathcal{W}_{100}$; the remaining panels use
$\mathcal{W}_{1/100}$. TO indicates a $24$-hour run time-out.}
\label{fig:n_sweep_build_space_appendix}
\Description{Twelve charts in two rows of six. Six show construction time
and six show construction space: Vir under the query-heavy workload,
followed by all five datasets under the update-heavy workload.
All size axes show actual total node counts.}
\end{figure*}

\subsection{Additional Results for the Impact of Color Count}
\label{app:delta_sweep_build_space}
\Cref{fig:delta_sweep_build_space_appendix} reports the construction and
space results omitted from the main text.
\providecommand{\auditdeltaplot}[3]{%
\begin{subfigure}[t]{0.155\textwidth}
  \centering
  \includegraphics[width=\linewidth]{figs/#1}
  \caption{#2}\label{#3}
\end{subfigure}%
}
\begin{figure*}[t]
\centering
\captionsetup[subfigure]{font={footnotesize,bf},labelfont=bf,
justification=centering,singlelinecheck=false,skip=2pt}
\auditdeltaplot{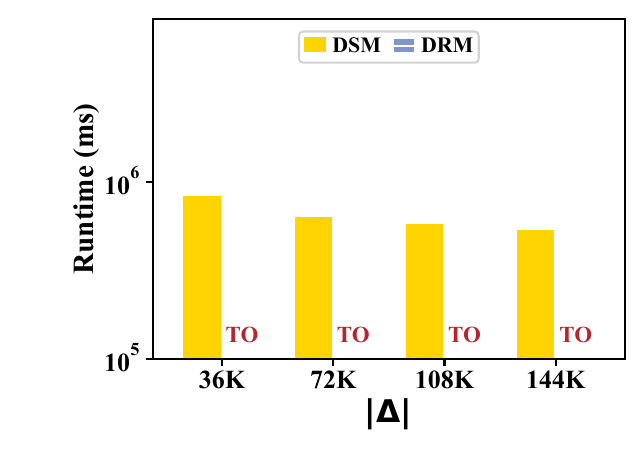}{\VIRlarge{}}{fig:vir_large_delta_buildtime}\hfill
\auditdeltaplot{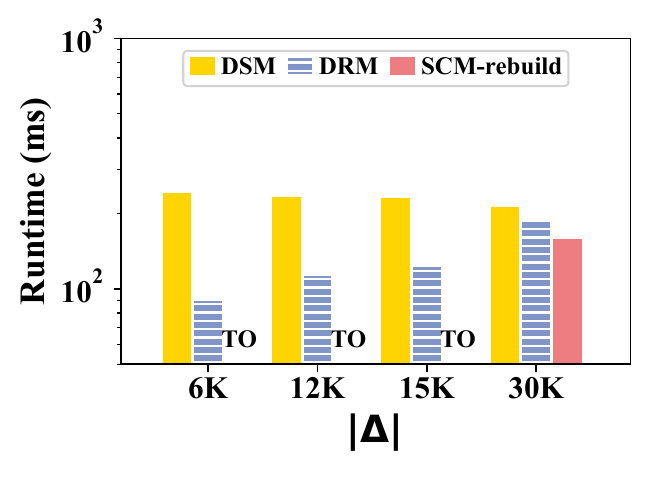}{\WebKBsample{}}{fig:webkb_delta_q001_buildtime}\hfill
\auditdeltaplot{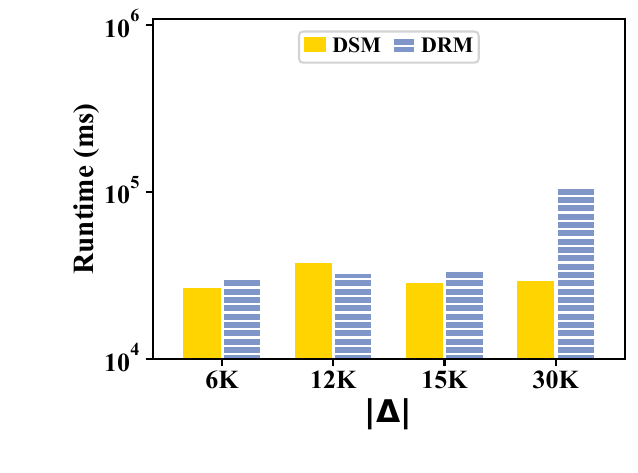}{\WebKBlarge{}}{fig:webkb_large_delta_q001_buildtime}\hfill
\auditdeltaplot{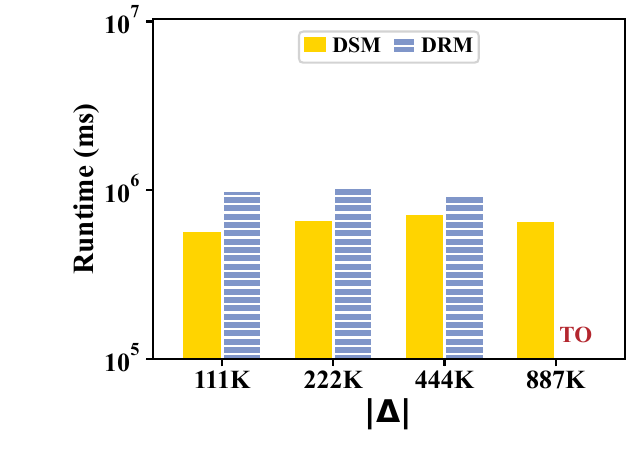}{\Geneslarge{}}{fig:genes_large_delta_q001_buildtime}\hfill
\auditdeltaplot{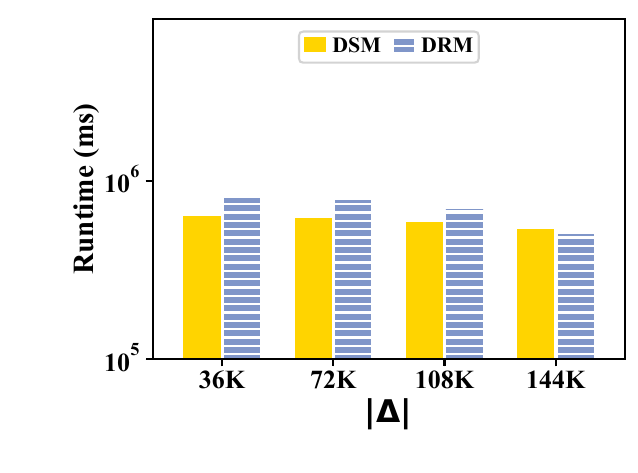}{\VIRlarge{}}{fig:vir_large_delta_q001_buildtime}\hfill
\auditdeltaplot{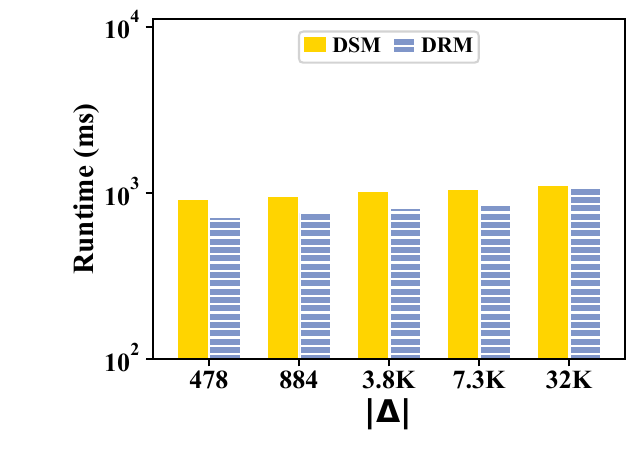}{\Amazonlarge{}}{fig:amazon_salesrank_delta_q001_buildtime}

\par\medskip
\auditdeltaplot{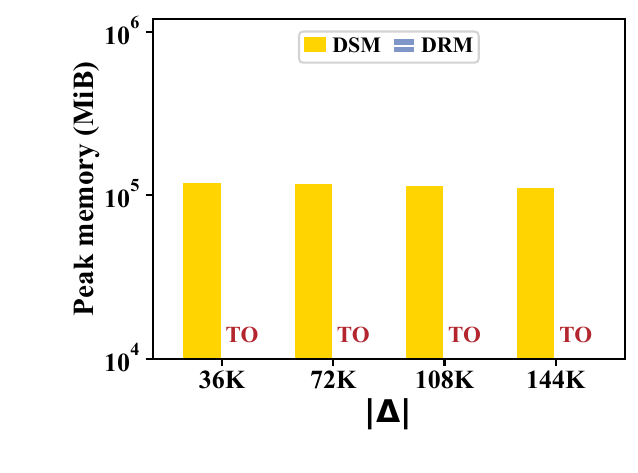}{\VIRlarge{}}{fig:vir_large_delta_space}\hfill
\auditdeltaplot{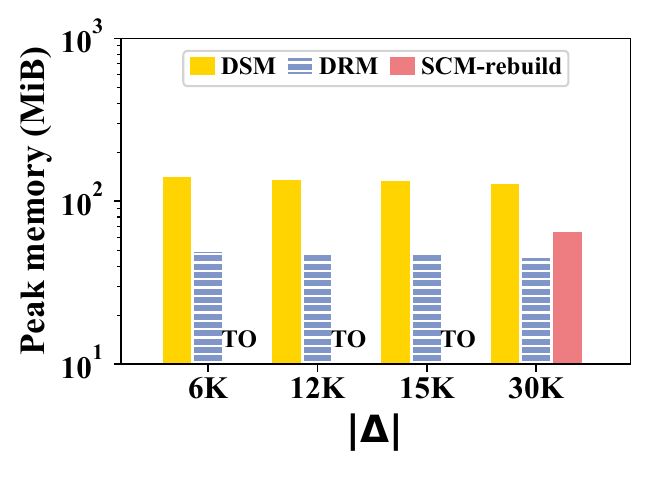}{\WebKBsample{}}{fig:webkb_delta_q001_space}\hfill
\auditdeltaplot{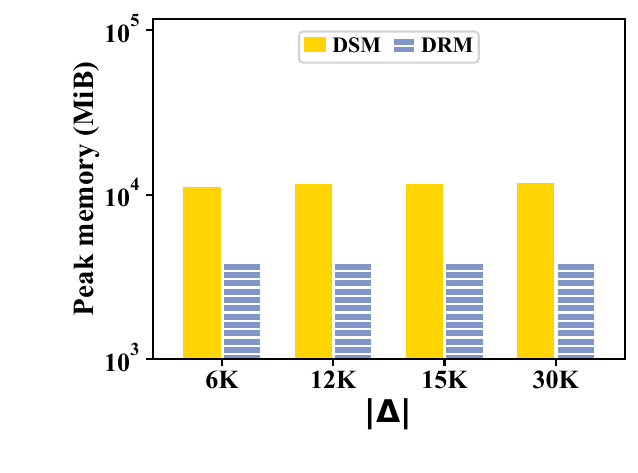}{\WebKBlarge{}}{fig:webkb_large_delta_q001_space}\hfill
\auditdeltaplot{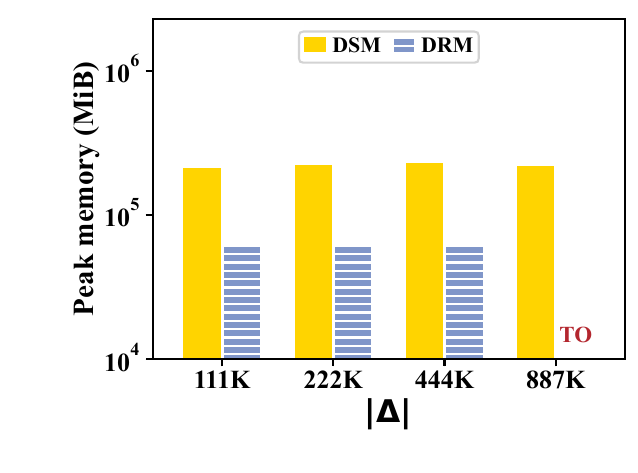}{\Geneslarge{}}{fig:genes_large_delta_q001_space}\hfill
\auditdeltaplot{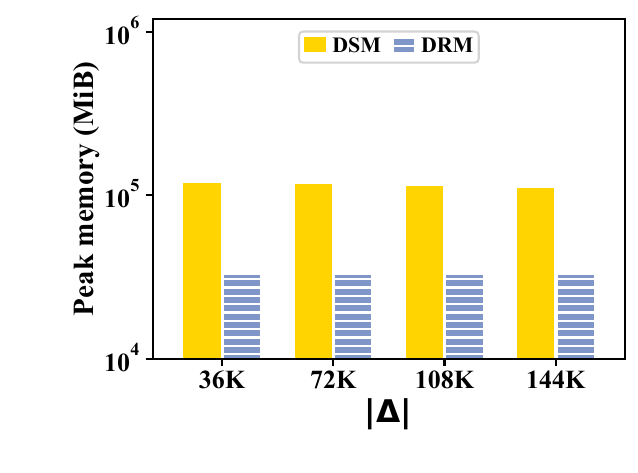}{\VIRlarge{}}{fig:vir_large_delta_q001_space}\hfill
\auditdeltaplot{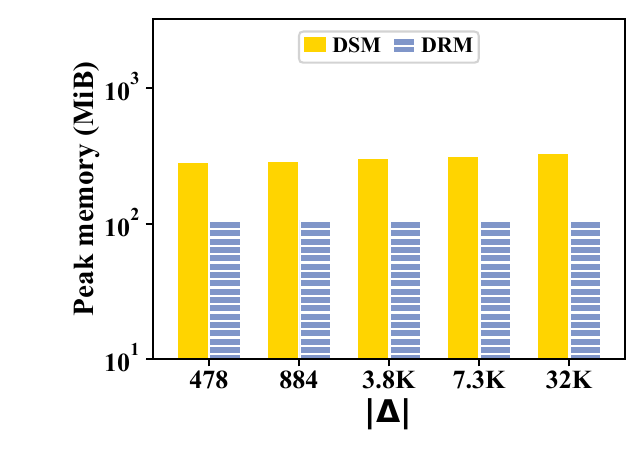}{\Amazonlarge{}}{fig:amazon_salesrank_delta_q001_space}

\caption{Additional results for the impact of $|\Delta|$.
(a)--(f) report construction time and (g)--(l) construction space.
(a) and (g) use $\mathcal{W}_{100}$; the other panels use
$\mathcal{W}_{1/100}$. TO indicates a $24$-hour run time-out.}
\label{fig:delta_sweep_build_space_appendix}
\Description{Twelve charts in two rows. The first row shows construction
time and the second construction peak memory. Each row contains Vir under
the query-heavy workload, followed by WebKB-sam, WebKB, Genes, Vir, and
Amazon under the update-heavy workload.}
\end{figure*}

\subsection{Additional Results for \FDSM{} on \WebKBlarge{}}
\label{app:fdsm_webkb_additional}

\providecommand{\webkbappendixplot}[3]{%
\begin{subfigure}[t]{0.155\textwidth}
\centering\includegraphics[width=\linewidth]{figs/#1}
\caption{#2}\label{#3}
\end{subfigure}%
}

\begin{figure}[!htbp]
\centering
\captionsetup{skip=3pt}
\renewcommand{\thesubfigure}{(\alph{subfigure})}
\captionsetup[subfigure]{labelformat=simple,font={footnotesize,bf},labelfont=bf,skip=1pt,justification=centering,singlelinecheck=false}
\webkbappendixplot{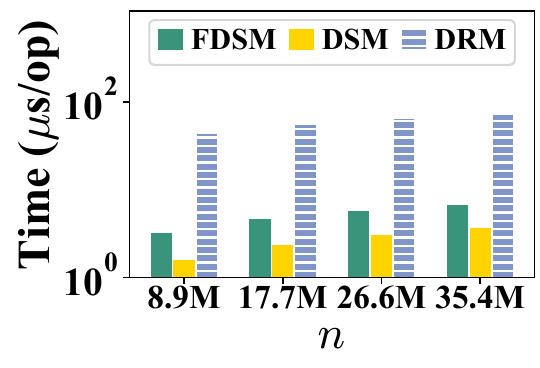}{Avg. op. time}{fig:fdsm_webkb_n_q100_time}\hfill
\webkbappendixplot{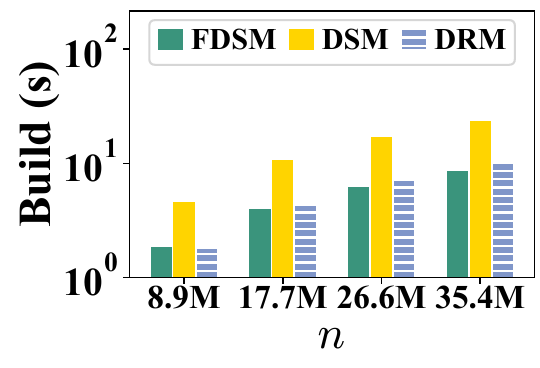}{Construction time}{fig:webkb_n_q100_construction_time}\hfill
\webkbappendixplot{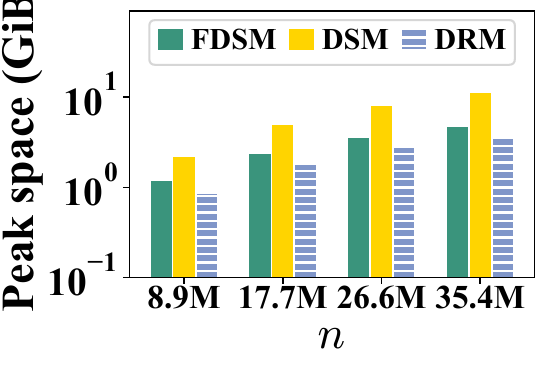}{Space}{fig:fdsm_webkb_n_q100_space}\hfill\webkbappendixplot{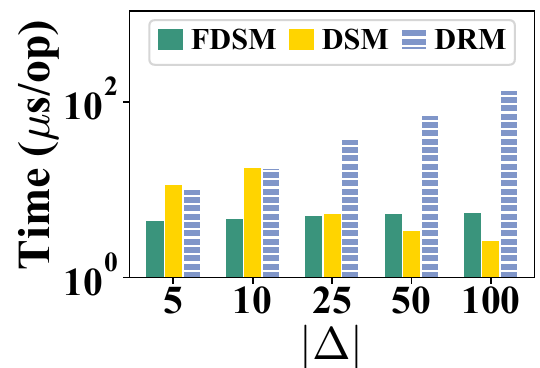}{Avg. op. time}{fig:webkb_C_q100_time}\hfill
\webkbappendixplot{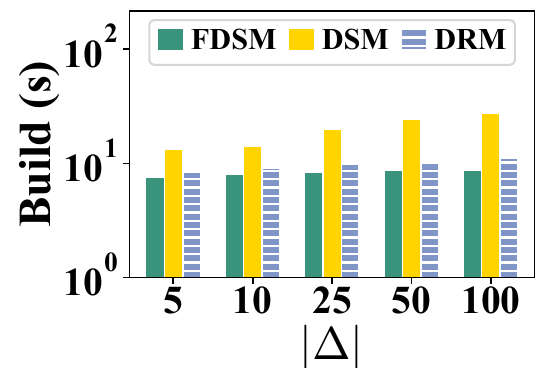}{Construction time}{fig:webkb_C_q100_construction_time}\hfill
\webkbappendixplot{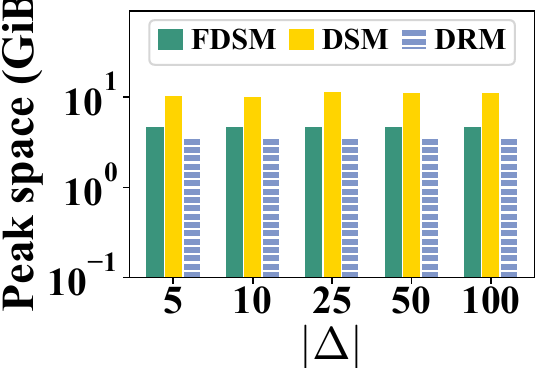}{Space}{fig:webkb_C_q100_space}
\caption{Query/recoloring experiments on \WebKBlarge{} under $\mathcal{W}_{100}$, comparing \FDSM{}, \DSM{}, and \DRM{}. (a)--(c) varying $n$; (d)--(f) varying $|\colors|$.}
\label{fig:webkb_query_recolor}
\label{fig:webkb_n_q100}
\label{fig:webkb_C_q100}
\Description{Six charts in one row. The left three vary the number of nodes; the right three vary the number of colors. Each group shows operation time, construction time, and peak space.}
\end{figure}

\begin{figure}[!htbp]
\centering
\captionsetup{skip=3pt}
\renewcommand{\thesubfigure}{(\alph{subfigure})}
\captionsetup[subfigure]{labelformat=simple,font={footnotesize,bf},labelfont=bf,skip=1pt,justification=centering,singlelinecheck=false}
\webkbappendixplot{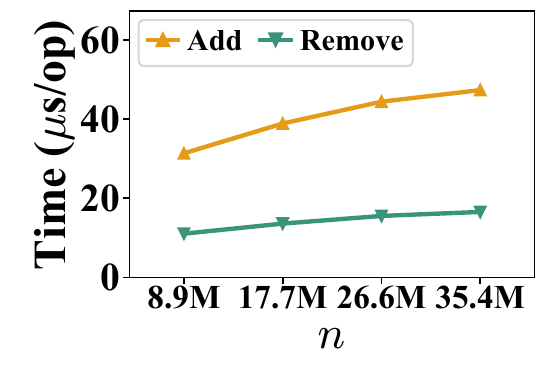}{Avg. op. time}{fig:webkb_leaf_pair_n_0}\hfill
\webkbappendixplot{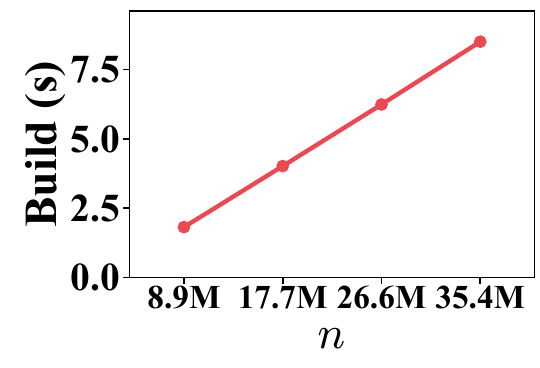}{Construction time}{fig:webkb_leaf_pair_n_1}\hfill
\webkbappendixplot{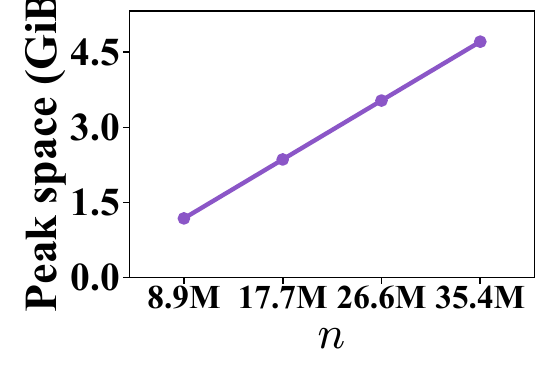}{Space}{fig:webkb_leaf_pair_n_2}\hfill
\webkbappendixplot{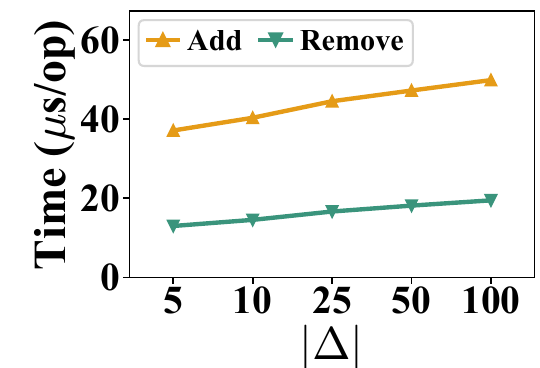}{Avg. op. time}{fig:webkb_leaf_pair_C_0}\hfill
\webkbappendixplot{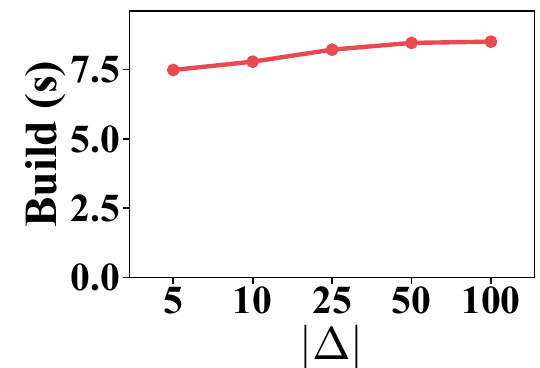}{Construction time}{fig:webkb_leaf_pair_C_1}\hfill
\webkbappendixplot{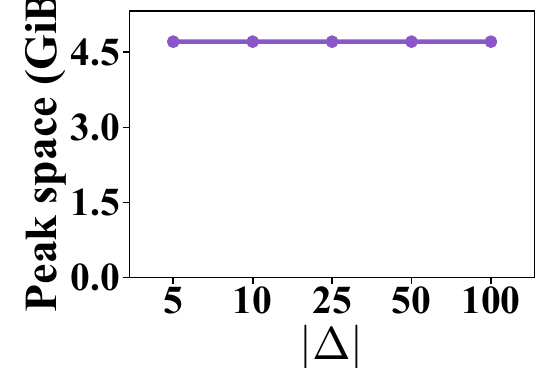}{Space}{fig:webkb_leaf_pair_C_2}
\caption{Leaf additions and deletions with \FDSM{} on \WebKBlarge{}. (a)--(c) varying $n$; (d)--(f) varying $|\colors|$.}
\label{fig:webkb_leaf_pair}
\Description{Six charts in one row. The left three vary the number of nodes; the right three vary the number of colors. Each group shows operation time, construction time, and peak space.}
\end{figure}

\begin{figure}[!htbp]
\centering
\captionsetup{skip=3pt}
\renewcommand{\thesubfigure}{(\alph{subfigure})}
\captionsetup[subfigure]{labelformat=simple,font={footnotesize,bf},labelfont=bf,skip=1pt,justification=centering,singlelinecheck=false}
\webkbappendixplot{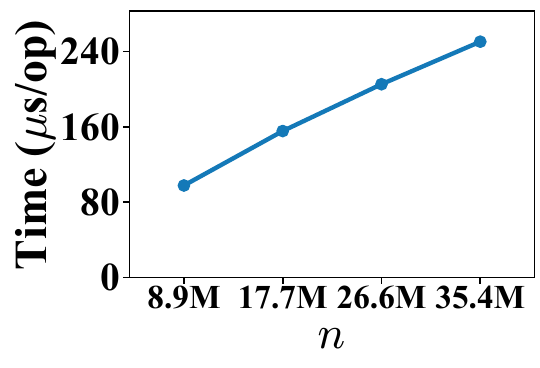}{Move time}{fig:webkb_subtree_move_n_0}\hfill
\webkbappendixplot{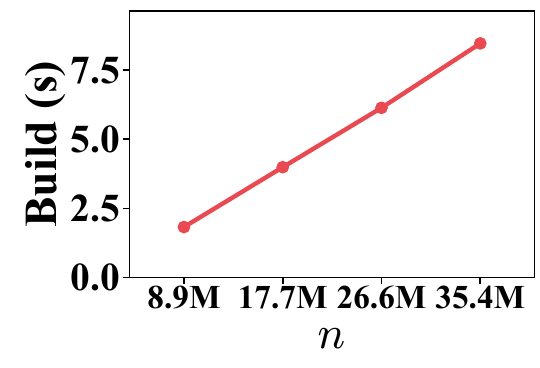}{Construction time}{fig:webkb_subtree_move_n_1}\hfill
\webkbappendixplot{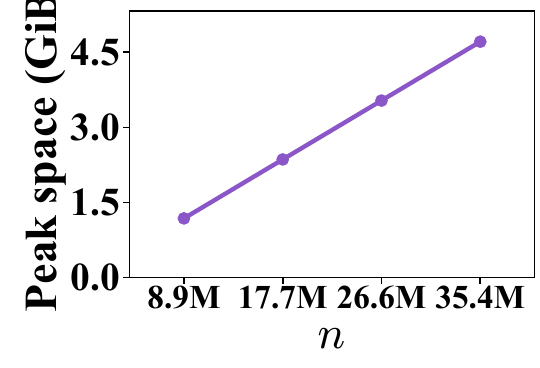}{Space}{fig:webkb_subtree_move_n_2}\hfill
\webkbappendixplot{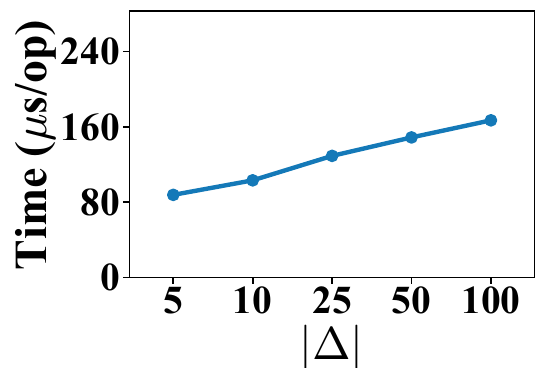}{Move time}{fig:webkb_subtree_move_C_0}\hfill
\webkbappendixplot{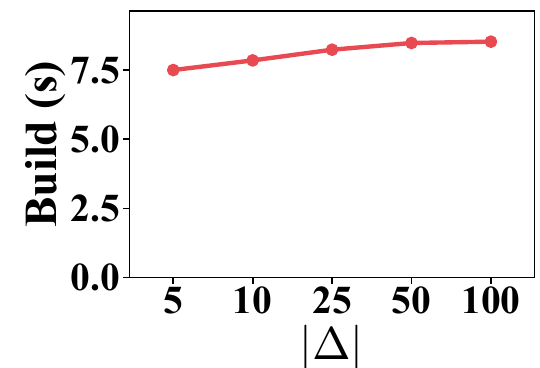}{Construction time}{fig:webkb_subtree_move_C_1}\hfill
\webkbappendixplot{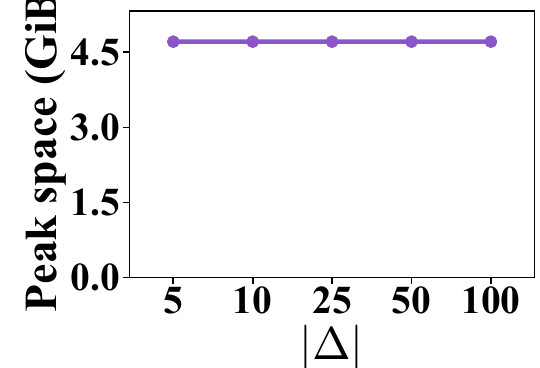}{Space}{fig:webkb_subtree_move_C_2}
\caption{Subtree moves with \FDSM{} on \WebKBlarge{}. (a)--(c) varying $n$; (d)--(f) varying $|\colors|$.}
\label{fig:webkb_subtree_move}
\Description{Six charts in one row. The left three vary the number of nodes; the right three vary the number of colors. Each group shows operation time, construction time, and peak space.}
\end{figure}

\Cref{fig:webkb_query_recolor} reports query/recoloring results under $\mathcal{W}_{100}$;
\Cref{fig:webkb_leaf_pair,fig:webkb_subtree_move} report structural updates.
Within each figure, (a)--(c) vary $n$ with $|\colors|=50$, and
(d)--(f) vary $|\colors|$ with $n=\num{35429531}$.
The $n$-axis tick labels use millions (M); the exact sizes are
\num{8857383}, \num{17714766}, \num{26572148}, and \num{35429531}.

%
%

%
%

\ifarxiv\else
\section*{AI Usage Declaration}
The authors used Anthropic's Claude Opus 5.5 via the chat interface.
The core ideas, algorithms, and proofs presented in this paper were developed entirely by the authors. The authors wrote the initial implementation and subsequently used the model to assist with code optimization. The authors also wrote the initial manuscript and subsequently used the model to help identify imprecise statements and minor errors, and to improve clarity and presentation. The model also wrote the TikZ code for \Cref{fig:subtree-mode-operations,fig:omv-reduction} based on the authors' description. The authors reviewed all AI-assisted revisions and take full responsibility for the final content of the paper and the accompanying code.
\fi

\end{document}